\documentclass[twocolumn]{aastex62}
\usepackage{hyperref}
\usepackage{amsmath}
\usepackage{booktabs ,float}
\usepackage{gensymb}
\usepackage{xspace}
\usepackage{xcolor}
\usepackage{chngcntr}
\usepackage{colortbl}
\usepackage{ctable}
\usepackage[all]{nowidow}
\newcommand{\dprime}{^{\prime\prime}\mkern-1.2mu}

\definecolor{highlight}{HTML}{CB1F47}

\newcommand{\deemph}[1]{{\color{black!60}#1}}

\graphicspath{{./}{figures/}}

\received{\today}
\revised{\today}
\accepted{\today}

\shorttitle{Gaia CVs}
\shortauthors{Abrahams et al.}

\begin{document}

\title{\bf Informing the Cataclysmic Variable Donor Sequence from\\ Gaia DR2 Color-Magnitude and Inferred Variability Metrics} 
\correspondingauthor{Ellianna S. Abrahams}
\email{ellianna@berkeley.edu}

\author[0000-0002-9879-1183]{Ellianna S. Abrahams}
\thanks{NSF Graduate Research Fellow}
\thanks{Two Sigma Ph.D. Fellow}
\affiliation{Department of Astrophysics, University of California, Berkeley, CA 94720-3411, USA}
\affiliation{Department of Statistics, University of California, Berkeley, CA 94720-3860, USA}

\author[0000-0002-7777-216X]{Joshua S. Bloom}
\affiliation{Department of Astrophysics, University of California, Berkeley, CA 94720-3411, USA}

\author{Nami Mowlavi}
\affiliation{Department of Astronomy, University of Geneva, Ch. des Maillettes 51, 1290 Versoix, Switzerland}

\author[0000-0003-4373-7777]{Paula Szkody}
\affil{University of Washington,
Department of Astronomy, Box 351580,
Seattle, WA 98195, USA}

\author[0000-0003-4996-9069]{Hans-Walter Rix}
\affiliation{Max Planck Institute for Astronomy, D-69117 Heidelberg, Germany}

\author[0000-0001-8962-8010]{Jean-Paul Ventura}
\affiliation{Department of Astrophysics, American Museum of Natural History, New York, NY 10024, USA}
\affiliation{Department of Physics and Astronomy, Hunter College, City University of New York, New York, NY 10065, USA}

\author{Thomas G. Brink}
\affiliation{Department of Astrophysics, University of California, Berkeley, CA 94720-3411, USA}

\author[0000-0003-3460-0103]{Alexei V. Filippenko}
\thanks{Miller Senior Fellow}
\affiliation{Department of Astrophysics, University of California, Berkeley, CA 94720-3411, USA}
\affiliation{Miller Institute for Basic Research in Science, University of California, Berkeley, CA  94720, USA}

\begin{abstract}
  Short-period cataclysmic variables (spCVs), with orbital periods below the period gap \hbox{($P_{\rm orb} < 2$\,hr)}, offer insight into the evolutionary models of CVs and can serve as strong emitters of detectable gravitational waves (GWs) for next-generation space-based GW observatories. To identify new spCV candidates, we crossmatch a catalog of known CVs with periods from 70\,min to 8\,hr to sources with well-measured parallaxes in the {\it Gaia} second data release (DR2). We uncover and fit a surprisingly (apparently) monotonic relationship between the color--absolute-magnitude diagram (CMD) position and $P_{\rm orb}$ of these CVs, revealed in DR2. To supplement the CMD-$P_{\rm orb}$ relation we also develop a method for identifying sources with large photometric variability, a characteristic trait of spCVs. Though {\it Gaia} is inherently a time-domain survey, the DR2 contains only a small fraction of sources with photometric light curves. Using such light curves, however, we construct a machine-learned regression model to predict physically informative variability metrics for sources in the CMD locus of known spCVs based solely on time-averaged observational covariates present in DR2. Using this approach we identify 3,253 candidate spCVs, of which $\sim95$\% are previously unknown. Inspection of archival SDSS spectra of these candidates suggests that $>$82\% are likely to be spCVs. This is a noticeably higher recovery rate than the typical recovery ($\sim$30\%) in previous light-curve searches, which bias toward flaring and active systems. We obtain optical spectra of nine previously uncharacterized systems with the Shane telescope at Lick Observatory and confirm that all objects are CV systems. We measure $P_{\rm orb}$ for seven systems using archival {\it Gaia} and Palomar Transient Factory light curves, all of which are spCVs and three of which do not have previous $P_{\rm orb}$ measurements.  We use the CMD-$P_{\rm orb}$ relation to infer the detectability of these systems to the upcoming {\it LISA} mission, and find that six of them may be coherent {\it LISA} verification binaries, with an estimated SNR $> 5$ in the 4\,yr mission. This paper demonstrates that the time-averaged {\it Gaia} catalog is a powerful tool in the methodical discovery and characterization of even semi-stochastic time-varying objects, making it complementary to missions like ZTF, TESS, and the Vera Rubin LSST in the efficient search for rare and unusual variable systems.
\end{abstract}


\section{Introduction} \label{sec:intro}

\noindent Time-domain surveys have cataloged $\sim 10^6$ variable stars in the Galaxy \citep{jayasinghe_2019}, a small fraction of the Galactic objects observed and recorded by static surveys. Compared to the $>10^9$ stars published with astrometric and photometric measurements by the {\it Gaia} mission in its second data release \citep[DR2;][]{gaiacollaboration_2018}, at least an order of magnitude more variables (with flux changes $>1$\%) would be expected \citep{holl_2018}. For example, the occurrence rates of 0.9\%--2.2\% for eclipsing binaries \citep[EBs;][]{kirk_2016} place lower limits of $\sim 10^6$--$10^7$ EBs alone in the {\it Gaia} catalog considering a 68\% detectability limit of {\it Kepler} EBs \citep{kochoska_2017}. While {\it Gaia} is expected to release the light curves for all sources at the conclusion of its mission, search and classification methods that do not rely on fitting the full set of time-resolved data are necessary to make computations tractable. 

We search the entire {\it Gaia} dataset for some of the most energetically time-variable systems in the Milky Way, cataclysmic variables (CVs). CVs are compact binary systems composed of a white dwarf (WD) member and a low-mass main-sequence (MS) member, and often generate an array of bright, energetic outbursts which can occur on a semi-stochastic or semi-periodic basis. For most of CV evolution the WD primary accretes mass from the Roche-lobe-filling MS donor star, which is thought to be driven out of thermal equilibrium and therefore radius-inflated, via an accretion disk \citep{warner_1995}. Since the 1970s, population studies of CVs have revealed a dearth of systems with orbital periods ($P_{\rm orb}$) of $\sim 2$--3\,hr \citep{livio_1983, ritter_1984, knigge_2011}. It is thought that angular momentum loss (AML) in short-period CVs is dominated by gravitational radiation \citep{paczynski_1981}, while AML in CV systems above the gap is dominated by magnetic braking \citep{verbunt_1981}. \citet{knigge_2006} estimates the bounds of this gap as $2.15\pm0.03$ to $3.18\pm0.04$\,hr, and within this regime, models assume that as the donor star loses mass and transitions to a fully convective interior, the resultant changing surface magnetic fields cause it to deflate to its equilibrium radius. With the donor star no longer filling its Roche lobe, accretion turns off until the system transitions out of the period gap \citep{warner_1995}. This model for CV evolution is known as the disrupted magnetic breaking model, and is well-supported by observational studies \citep{townsley_2003, knigge_2006, townsley_2009}. 

Assuming that gravitational radiation is the only AML mechanism below the period gap, a theoretical minimum $P_{\rm orb} \approx 65$--70\,min can be calculated \citep{kolb_1993, goliasch_2015, kalomeni_2016}. However, the shortest observed periods are noticably longer than this, most recently cited at $79.6 \pm 0.2$ min \citep{mcallister_2019}. This discrepancy can be resolved if an additional source of AML contributes to the orbital evolution below the gap \citep[][hereafter Knigge11]{knigge_2011a}; or, since donor stars are smaller at shorter $P_{\rm orb}$ and therefore fainter, such systems might have yet to be detected.

Owing to the evolution of the magnetic field and the transition of the donor star to its fully convective state \citep{garraffo_2018}, the mass transfer rate in CV systems is 1--2 orders of magnitude faster above the period gap. These rates decrease consistently as the donor star loses mass, thus giving rise to the model that CVs should spend most of their lives below the period gap and that most of these systems should accumulate around the $P_{\rm orb}$ minimum. These models predict that $\sim 99$\% of CV systems should be observed with $P_{\rm orb} < 2$\,hr \citep{kolb_1993, howell_2001}. However, this is not reflected in the observations of CV populations. Instead, only $\sim 83$\% of CVs are found below the period gap in volume-limited studies of nearby populations \citep{pala_2020}. This is thought to be due to brightness-threshold limitations of all-sky surveys and the difficulty of performing a bias-free spectroscopic campaign on faint objects. Historically, the ratio of observed CVs below the period gap was even further from the predictions, but the Sloan Digital Sky Survey (SDSS) filled in many missing systems by probing deeper than previous studies \citep{szkody_2011}. 

Short-period CVs (spCVs; CVs with $P_{\rm orb} < 2$\,hr) are therefore both rare and important astrophysical objects as they offer much-needed empirical insight and constraints to the evolutionary models of CVs. Furthermore, some spCV systems will emit a gravitational wave (GW) signal sufficiently loud to be coherently detectable in the upcoming  {\it Laser Interferometer Space Antenna (LISA)} mission \citep{danzmann_2000}. Such systems, called verification binaries by the mission, would provide important tests of {\it LISA}'s performance. Additionally, measurements of the gravitational radiation in such systems, combined with robust and detailed population studies of the period distribution in large samples, would provide crucial insight into CV evolution.

All-sky scanning missions like the Palomar Transient Factory \citep[PTF;][]{law_2009}, the Catalina Real-Time Transient Survey \citep[CRTS;][]{drake_2014}, the All-Sky Automated Survey \citep[ASAS, and later ASAS-SN;][]{pojmanski_2002, shappee_2014}, and the Zwicky Transient Facility \citep[ZTF;][]{bellm_2014} have intentionally added to the numbers of CV discoveries in recent years, but only release data for a specific subset of bright, resolved sources. Now, with the advent of deep and precise surveys like {\it Gaia}, we can conduct a more automated search for these elusive spCVs across the entire sky with the use of variability metrics. The use of such metrics has been demonstrated in searching for other classes of variables \citep{deason_2017, belokurov_2017, vioque_2020, mowlavi_2020}; see Section \ref{subsec:vari_metrics}.

{\it Gaia}'s mission directive is to build the largest 3-dimensional (3D) map of the Galaxy to date. To measure precise positions, motions, and parallaxes for more than 1.4 billion stars the {\it Gaia} mission \citep{gaiacollaboration_2016} obtains tens to hundreds of measurements for each star over several years \citep{lindegren_2018}, making it an inherently time-domain survey. In addition to the parallax measurements published for most objects, {\it Gaia} will eventually release (DR4) multi-epoch photometric light curves for every star that it has observed during its 5-year nominal mission. However, in the most recent release, {\it Gaia} DR2, only 550,737 light curves were provided. There were 363,969 objects\footnote{This number is not expected to change in the upcoming Early Data Release 3 (EDR3), which will appear on Dec. 3, 2020. The next variable star release is currently being planned for DR3, which is foreseen for 2022.} released with classification into a number of variable star classes \citep[][hereafter Rim18]{rimoldini_2019}. These classifications were made using a set of attributes describing the light-curve statistics and physical measurements of each object as described by \hyperlink{cite.rimoldini_2019}{Rim18}, and include both the interquartile range (IQR) and median absolute deviation (MAD) summarizing the distribution spread of the time-resolved observations. IQR in particular can be used to summarize variability amplitude across multiple classification types \citep[see Figure 9 of][]{gaiacollaboration_2019}.

In addition to releasing parallaxes, the DR2 also provides measurements of time-averaged photometry in three bands for $\sim 1.7$ billion stars \citep{riello_2018}. Information about how the brightness of a stellar object changes over time is inherently encoded in these time-averaged statistics and their uncertainties. A measurement of the standard deviation across a light curve will be biased by outliers, such as flares and outbursts, and will be determined by the amplitude of stellar variation, when well-sampled \citep{mowlavi_2020}. By using the information available in the measurements of average photometric flux and flux uncertainty, we use the {\it static} DR2 catalog to locate highly variable stars and systems from averaged position, parallax distance, and averaged photometry alone, in advance of the future {\it Gaia} release of the time-resolved information for the entire catalog. These methods seek to find an alternative to the IQR and MAD measurements made directly from the light curves in \hyperlink{cite.rimoldini_2019}{Rim18}, by relying instead on the time-averaged measurements that were released across DR2.

In this paper, we report the discovery of $3,253$ new candidate spCV systems as a test case for predicting {\it Gaia} variability from static data. To find these CVs we show a new relationship revealed in the {\it Gaia} data between CV $P_{\rm orb}$, $G_{\rm BP}-G_{\rm RP}$ color, and $M_{\rm G}$. In Section \ref{sec:measure_vari}, we describe our methodology for predicting variability from the physics buried in time-averaged {\it Gaia} photometry and its uncertainty. Section \ref{sec:cv_fit} shows the relationship between CV period and the {\it Gaia} color--absolute-magnitude diagram (CMD) and discuss our model fit to this relationship. In Sections \ref{sec:cv_search} and \ref{sec:previous_char}, we present our technique for selecting candidate CVs from the full {\it Gaia} database and discuss previous characterizations of known sample members. Spectroscopic follow-up observations of nine of the candidate systems are reported in Section \ref{sec:new_spectra}, confirming their CV status. In Section \ref{sec:new_lcs}, we present time-domain follow-up observations of seven systems from available light-curve catalogs, like the {\it Gaia} DR2 Variability Catalog, the {\it Gaia} Alerts Database, and PTF. These results are discussed in Section \ref{sec:discussion}; in particular, we demonstrate that there is color-dependent scatter in the semi-empirical CV donor sequence and we estimate that six of our candidate spCVs are {\it LISA} verification binary candidates. Section \ref{sec:summary} summarizes our work and places this study in context with ongoing inquiry.

\vspace{5px}

\section{Measuring Variability with {\it Gaia} DR2} \label{sec:measure_vari}

\subsection{Variability Metrics} \label{subsec:vari_metrics}

\noindent \citet{gaiacollaboration_2019} measure the photometric dispersion for objects with released {\it Gaia} light curves using the IQR in order to characterize variable stars, but without the full probability distribution function (PDF) of the photometric measurements for the vast majority of {\it Gaia} objects, we do not have access to the IQR. As an alternative, we rely on two variability metrics using the {\it Gaia} averaged photometry and uncertainty. Both metrics make use of the fact that information about the dispersion of each individual photometric measurement is embedded in its uncertainty.

The first metric, as defined by Equation 2 of \citet{deason_2017}, measures the root-mean-square (RMS) dispersion of the flux over the full set of {\it Gaia} observations:

\begin{equation}
    \sigma_f = \sqrt{N_{\rm obs}}\times\frac{\delta f}{f} ,
\label{eqtn:sigma_f}
\end{equation}{}

\noindent where $f$ is the flux in a chosen bandpass, $\delta f$ is the flux uncertainty, and $N_{\rm obs}$ is the number of observations used to calculate the flux. 

\begin{figure}
  \begin{center}
  \includegraphics[width=240px]{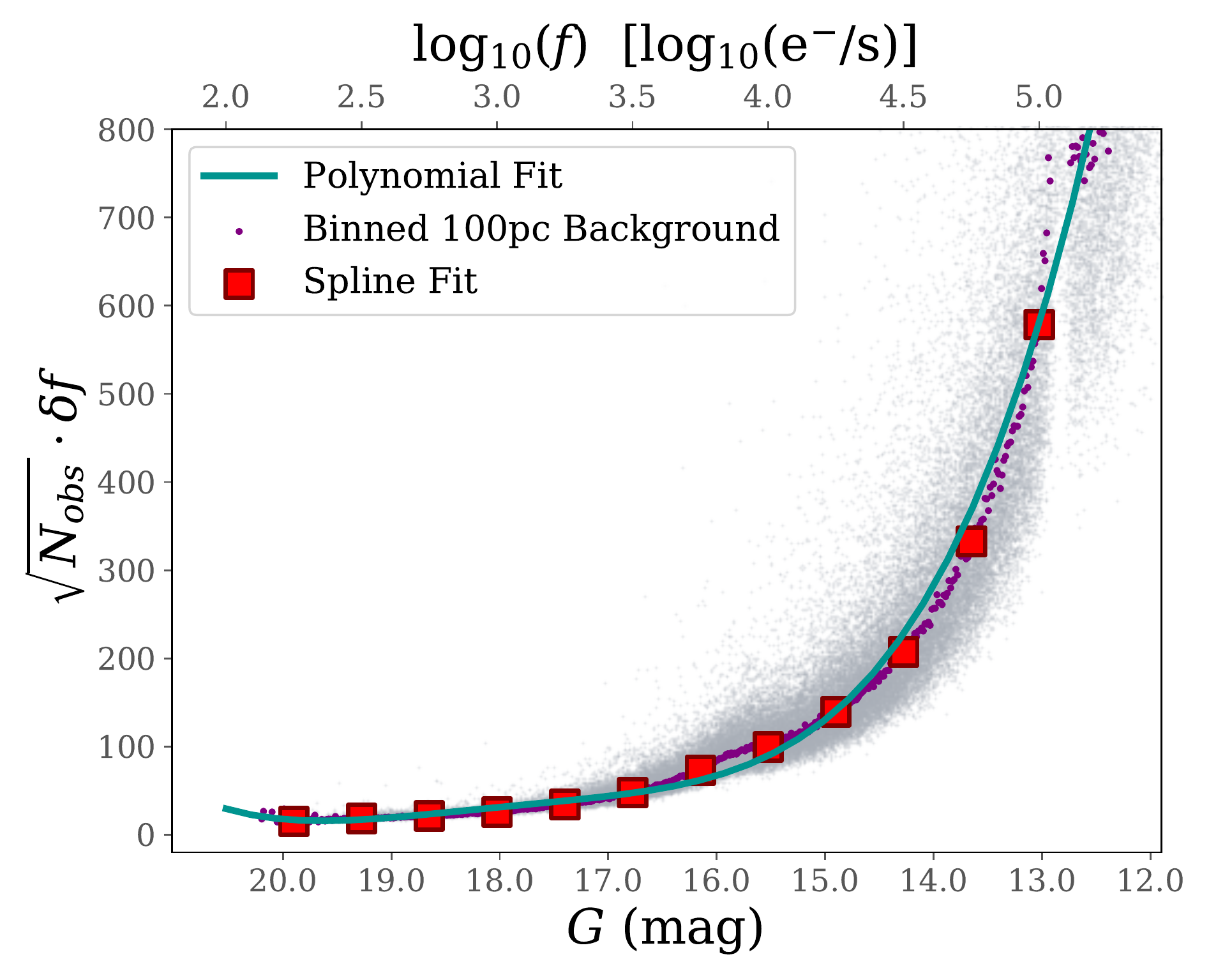}
  \caption{Empirical dispersion versus flux in {\it Gaia} establishing a fiducial background. The purple dots are the binned 100\,pc background sample. Data were binned using a 2D binning of $\sqrt{N_{\rm obs}}\times\delta f$ as a function of $\log_{10}(f)$. We fit a spline function to the data as illustrated in red. While spline functions overfit the data when searching for the true, underlying functional form, we are only interested in how a specific star might compare to the background sample. For this reason we found the spline fit to be more accurate than the fourth-order polynomial fit shown in green. The discontinuity in the DR2 photometry reduction is visible in the 100\,pc background sample,  illustrating the necessity of photometry cuts of $G > 13$\,mag for the use of our variability metrics.}
  \label{fig:mag_bg}
  \end{center}
\end{figure}

\begin{figure*}[]
  \centering
  \includegraphics[width=500px]{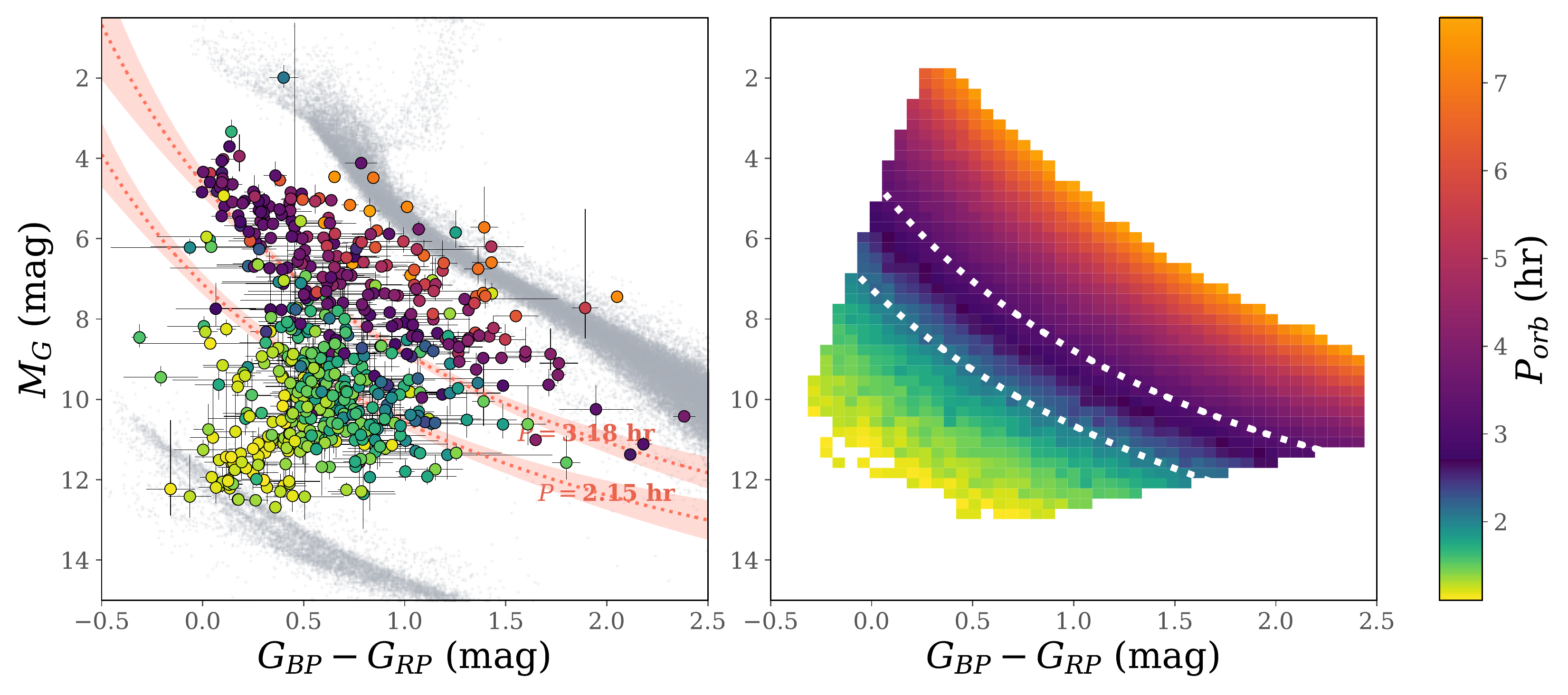}
  \caption{{\it Left panel:} The CMD for the \hyperlink{cite.ritter_2003}{RK16}-{\it Gaia} crossmatch for CVs with $P_{\rm orb} <$ 8\,hr, indicated by the circles colored according to $P_{\rm orb}$. For reference, the 100\,pc quiescent sample is shown here in gray. CVs primarily inhabit the space between the MS and the WD sequence. The relationship between the CMD position and period is apparent on the CMD, with longer period CVs falling closer to the MS and shorter period CVs falling closer to the WD sequence. Lines of constant $P_{\rm orb}$ (from the model fit in Eq. \ref{eqtn:porb_hat}) that enclose the CV period gap are shown as dotted red lines, along with their 5$\sigma$ confidence bands. {\it Right panel:} Full 2D representation of the Eq. \ref{eqtn:porb_hat} model fit for $P_{\rm orb}$. The fit is only reliable within the bounds of the data, and the intrinsic scatter, $\sigma_{\rm int}$, of the fit is indicated by random Gaussian noise. The dotted white lines indicate the bounds of the period gap.}
  \vspace{10pt}
  \label{fig:rk_cmd}
\end{figure*}

This metric has been employed in previous studies of variability populations in {\it Gaia}: on Mira variables \citep{deason_2017}, RR Lyrae stars \citep{belokurov_2017}, pre-main-sequence objects \citep{vioque_2020}, and most recently, large-amplitude variables \citep{mowlavi_2020}. However, since the fluctuations in variable stars are not necessarily Gaussian, this metric does not measure the true dispersion. Furthermore, some of this uncertainty will be due to shot noise, low observation counts, and telescope systematics.

We further develop a second new metric that calculates the deviation of the measured flux uncertainty in a given object from the average flux uncertainty that is associated with the object's magnitude averaged across {\it Gaia} DR2:

\begin{equation}
    \epsilon_f = \sqrt{N_{\rm obs}}\times\frac{\delta f_{\rm object}}{\overline{\delta f_{\rm mag}}},
\label{eqtn:epsilon_f}
\end{equation}{}

\noindent where $\delta f_{\rm object}$ is the flux uncertainty in the object of interest. $\overline{\delta f_{\rm mag}}$ is calculated by fitting a spline to the 2D binning of $\sqrt{N_{\rm obs}}\times\delta f$ as a function of $\mathrm{log}_{10}(f)$ for a random subsample of 100,000 {\it Gaia} objects within 100\,pc (referred to as the 100\,pc background from here) that follow the recommended ``{\it Gaia} Gold'' recipes of \citet{gaiacollaboration_2018a} and \citep{lindegren_2018}. We limit to 100\,pc since {\it Gaia} is more complete at fainter magnitudes at closer search radii, and this radius is still large enough to include variables of many subtypes. This subsample is likely biased, as we only selected objects that had measurements within the \citet{bailer-jones_2018} Catalog (BJ18 herein) and with limits on parallax over error of $\varpi / \delta_{\varpi}>10$ that were observed by the {\it Gaia} mission at least eight times by the DR2 release. We illustrate the functional fit of $f_{\rm mag}$ in Figure \ref{fig:mag_bg}. Owing to magnitude-dependent systematics in the photometric calibration of {\it Gaia} DR2 \citep{evans_2018}, and since this metric is itself dependent on magnitude, we only recommend the use of this metric for $M_{\rm G} > 13$\,mag or $\log_{10}(f_G) < 5.1$.

\subsection{Predicting Variability from Observables} \label{subsec:rf_model}

\noindent In cases of light curves that are nearly approximated by periodic waveforms like sinusoids or sawtooths, Eqs. \ref{eqtn:sigma_f} and \ref{eqtn:epsilon_f} can be used to derive the amplitude and range of $G$ variability \citep{mowlavi_2020}. However, in the case of flaring or semi-stochastic variables, these variability metrics cannot be mapped linearly to light-curve statistics. Combining the metrics with the {\it Gaia} DR2 observables and their uncertainties, we construct a random forest (RF) regression model \citep{breiman_2001}, which is by definition nonparametric, to predict IQR, which is more stable than amplitude or range, for all but a withheld 20\% of the full DR2 Variability Catalog where $M_{\rm G} > 13$\,mag. We simultaneously fit the regression to predict the MAD. While IQR provides a steeper, and therefore more distinguishable, measurement of variability, MAD is closer to the measurement of standard deviation and is therefore easier to predict with $\sigma_f$.

We find that the most important features in predicting IQR$^\prime$ and MAD$^\prime$\footnote{These statistics are primed to indicate that they are estimators, and not measured from time-resolved light curves.} are $\sigma_f$, $\epsilon_f$, $G - G_{\rm BP}$, $G_{\rm BP}-G_{\rm RP}$, and $G - G_{\rm RP}$ colors, and flux uncertainties in each band. Other observables like parallax, position, and proper motion do not have a significant contribution in predicting IQR$^\prime$ and MAD$^\prime$. Our model predicts IQR$^\prime$ with an explained variance score (EVS) of 0.90 (meaning that our model explains 90\% of the underlying dispersion of the input covariates) and a mean-squared error (MSE) of 0.004. We predict MAD$^\prime$ with an EVS of 0.92 and MSE of 0.008. Appendix \ref{app:rf} contains a further discussion of the RF model used.

\section{CV Periodicity on the {\it Gaia} CMD} \label{sec:cv_fit}

\begin{deluxetable*}{ c c c c c }
\tablenum{1}
\tablecaption{CVs from \hyperlink{cite.ritter_2003}{RK16} with a Robust {\it Gaia} Crossmatch
\label{tab:rk16}}
\tablehead{\colhead{{\it Gaia} \texttt{source\_id}} & \colhead{\hyperlink{cite.ritter_2003}{RK16} Name} & \colhead{$P_{\rm orb}$} & \colhead{IQR$^\prime$} & \colhead{MAD$^\prime$} \vspace{-5px} \\ 
\colhead{} & \colhead{} & \colhead{\deemph{[hr]}} & \colhead{} & \colhead{}}
\setlength{\tabcolsep}{12pt}
\startdata 
481788819721445376 & J0557+6832 & 1.2696 & 0.2732 & 0.1887 \\
5171137394568701184 & KN Cet & 1.2716 & 0.2624 & 0.1742 \\
2698490156365025536 & J2141+0507 & 1.272 & 0.2371 & 0.1079 \\
1206925328171879808 & J1605+2405 & 1.296 & 0.3407 & 0.2189 \\
546910213373341184 & J0221+7322 & 1.296 & 0.3679 & 0.2401 \\
1593786703401901952 & J1457+5148 & 1.298 & 0.3515 & 0.2345 \\
6680933641676624384 & V4738 Sgr & 1.3006 & 0.3520 & 0.2041 \\
4610221438876442368 & PR Her & 1.3013 & 0.1219 & 0.0925 \\
1176468611268115200 & J1433+1011 & 1.3018 & 0.1434 & 0.0915 \\
2104562321825510400 & J1853+4203 & 1.3061 & 0.4677 & 0.2921 \\
\multicolumn{5}{c}{...} \\
\enddata
\tablecomments{We show a truncated table of the full \hyperlink{cite.ritter_2003}{RK16}-{\it Gaia} crossmatch of 582 objects, provided for $70$\,min $< P_{\rm orb} < 8$\,hr. The full data are available in a machine-readable format in the electronic version of this paper.}
\end{deluxetable*}

\noindent \citet{townsley_2002} predict from evolutionary models that CVs should be identifiable from their position on the CMD. This has been observationally confirmed in {\it Gaia} DR2 using a variety of CV samples \citep{pala_2020, abril_2020}. Using the astrometric and time-averaged photometric DR2 catalog, we explore the location of CVs in the {\it Gaia} CMD and find that periodicity is a function of CMD position. 

We crossmatched the latest Ritter \& Kolb Catalog of Cataclysmic Variables \citep[RK16;][]{ritter_2003}\footnote{Catalog Update: RKcat7.24, 2016} to {\it Gaia} DR2. We limit our \hyperlink{cite.ritter_2003}{RK16} crossmatch to the 1,335 objects that have $P_{\rm orb}$ under 8\,hr following the assumption of \hyperlink{cite.knigge_2011a}{Knigge11} that all CVs are ``born" $\sim 6$\,hr above the gap.  We crossmatch each \hyperlink{cite.ritter_2003}{RK16} object to a 20$\dprime$ radius in {\it Gaia}, and then propagate the subsets of {\it Gaia} 20$\dprime$ crossmatches to the J2000 epoch using their proper-motion measurements. Since we use this crossmatch to define the limits used in Section \ref{sec:cv_search}, we made strict cuts on {\it Gaia} quality flags, even though this will remove some of the most variable systems. We only kept matches with \texttt{parallax\_over\_error}, $\varpi/\delta\varpi$, $> 10$ and more than eight ``good" astrometric observations, as characterized by DR2. Additionally, since these systems are highly variable, we opted not to validate our crossmatch with photometry comparisons. We instead cleaned the crossmatch by removing any matched objects that had another star listed in {\it Gaia} within 5$\dprime$, so as to prevent confusion from lower-resolution studies that have contributed to \hyperlink{cite.ritter_2003}{RK16} measurements. We include the crossmatch for 582 \hyperlink{cite.ritter_2003}{RK16} systems in Table \ref{tab:rk16}.

The \hyperlink{cite.ritter_2003}{RK16}-{\it Gaia} crossmatch reveals a clear dependence of CV period on the location in the CMD, with longer orbital period systems falling closer to or overlapping with the MS and shorter orbital period systems falling closer to the WD sequence as shown in Figure \ref{fig:rk_cmd}. This trend has likely been revealed by {\it Gaia} due to the underlying multi-epoch nature of the survey, allowing for reported measurements to be less dependent on individual observations, which could be serendipitous observations of an active state. Until the time of submission, we were unaware of a simultaneous study, \citet{abril_2020}, that also found a similar trend using multiple CV catalogs, including \hyperlink{cite.ritter_2003}{RK16}. While \citet{abril_2020} identified a resemblant overall trend between $P_{\rm orb}$ and CMD position, their study did not explore a fit to the $P_{\rm orb}$--CMD relationship and the evolutionary implications of this fit. We fit the relationship following a Bayesian methodology with an exponential model containing first-order and cross terms. 

Since the apparent $P_{\rm orb}$-color-$M_{\rm G}$ relation seems to run orthogonal to the MS and the WD sequence, we fit a log-linear model with higher order cross terms. Each variable ($P_{\rm orb}$, $G_{\rm BP}-G_{\rm RP}$, and $M_{\rm G}$) has been scaled by its median to calculate a more stable fit. This scaling allows us to make a direct comparison between coefficients as well:

\begin{equation}
\begin{split}
\hat{P}_{\rm orb} =  \beta_0{\cdot}\exp\big\lbrack&\beta_{X}{\cdot} (G_{\rm BP}-G_{\rm RP}) + \\ &\beta_Y{\cdot}M_{\rm G} + \beta_{XY}{\cdot}(G_{\rm BP} - G_{\rm RP}){\cdot}M_{\rm G} \big\rbrack .
\label{eqtn:porb_hat}
\end{split}
\end{equation}

\vspace{2px}

To test the fit, we used the Monte Carlo Markov Chain (MCMC) software \texttt{emcee} \citep{foreman-mackey_2013}, a \texttt{Python} implementation of the affine-invariant ensemble sampler \citep{goodman_2010}. Since $G_{\rm BP}$, $G_{\rm RP}$, and $G$ rely on different instruments or filters, the model can safely neglect cross-correlations between the errors in $G_{\rm BP}-G_{\rm RP}$ and $M_{\rm G}$. The model includes four free parameters, which are established with uniform priors, given in Table \ref{tab:fit_params}. Where possible, parameters were limited by physical or definitional constraints. A summary of the fit parameters is given in Table \ref{tab:fit_params}. For each run, walkers were initialized with the maximum-likelihood estimate for the model parameters determined with the Broyden-Fletcher-Goldfarb-Shanno algorithm for optimization \citep{kelley_1999}. For sampling, we used 100 walkers, letting each run for 2,500 steps. Convergence was established by checking if the total chain length was at least 50 times as long as the autocorrelation time \citep{goodman_2010}. Average parameter convergence occurred at 52 steps; the first 100 steps of each chain were removed for burn-in. Longer burn-in time was unnecessary, in part because the initial parameter estimates were close to the final answers from the MCMC convergence. The parameter measurements from the MCMC run are included in Table \ref{tab:fit_params}, along with the uncertainties for each parameter, calculated at the 16th and 84th percentile along the parameter distribution.

$M_{\rm G}$ provides the strongest predictor of $P_{\rm orb}$, followed by $(G_{\rm BP}-G_{\rm RP})$ and the crossterm between both variables. As objects become fainter and bluer, $P_{\rm orb}$ shrinks, a relation that fits in well with the prevailing evolutionary models. Figure \ref{fig:rk_cmd}b illustrates the grid of CV periods calculated using this fit given a $G_{\rm BP}-G_{\rm RP}$ color and an $M_{\rm G}$ measurement. This fit has intrinsic noise of nearly 10\%, $\sigma_{\rm int} \approx 0.08$\,hr, that could be a result of $P_{\rm orb}$ uncertainties, which were not reported in \hyperlink{cite.ritter_2003}{RK16}; they could be due to system outbursts influencing the measured $G$, $G_{\rm BP}$, and $G_{\rm RP}$ statistics in {\it Gaia}, or they could be intrinsic to the relationship itself. Additionally, these measurements are limited to the boundaries provided by the observed data, which are not necessarily physical. We indicate the areas for which this fit is unreliable (beyond the bounds of the observed data) in white.

\begin{deluxetable}{cccc}
\tablenum{2}
\tablecaption{$\hat{P}_{\rm orb}$($M_G$, $G_{\rm BP}-G_{\rm RP}$) Parameter Fits \\ (Eqtn. \ref{eqtn:porb_hat})
\label{tab:fit_params}}
\tablehead{
\colhead{$\Theta_i$} & \colhead{Median Fit} & \colhead{Priors} & \colhead{$\tau$}}
\setlength{\tabcolsep}{10pt}
\startdata
$\beta_0$ & $3.351_{-0.064}^{0.064}$ & $|\beta_0| <$ 15.0 & 49.0\\
$\beta_X$ & $0.776_{-0.014}^{0.015}$ & $|\beta_X| <$ 15.0 & 45.8\\
$\beta_Y$ & $-1.416_{-0.025}^{0.025}$ & $|\beta_Y| <$ 15.0 & 50.5\\
$\beta_{XY}$ & $-0.385_{-0.016}^{0.016}$ & $|\beta_{XY}| <$ 15.0 & 47.8\\
\enddata
\tablecomments{We report the median fit for the parameters, $\Theta$, sampled from the model fit after burn-in along with upper and lower 95\% confidence limits. The number of steps required to converge is $\tau$. There is intrinsic scatter in the fit, $\sigma_{\rm int} = 0.08$.}
\end{deluxetable}

\subsection{Gaia Variability of the Short-Period RK16 Sample} \label{subsec:rk_vari}

\noindent To find new spCVs, we selected all objects with $P_{\rm orb}<2.15$\,hr from the \hyperlink{cite.ritter_2003}{RK16}-{\it Gaia} crossmatch. 39 spCVs in the \hyperlink{cite.ritter_2003}{RK16} sample have released light curves in the {\it Gaia} DR2 Variability Catalog, all without an assigned class in {\it Gaia}. We predict MAD$^\prime$ and IQR$^\prime$ for these objects to confirm that our RF model, primarily trained on pulsational variables, is performing well on semi-stochastic systems like CVs. We find that our model predicts with greater accuracy at lower MAD$^\prime$ and IQR$^\prime$, but since we are only making cutoffs on the lower limits of variability, this is sufficient for our current analysis, allowing us to predict IQR$^\prime$ and MAD$^\prime$ for CV systems.

\begin{figure}
  \begin{center}
  \includegraphics[width=240px]{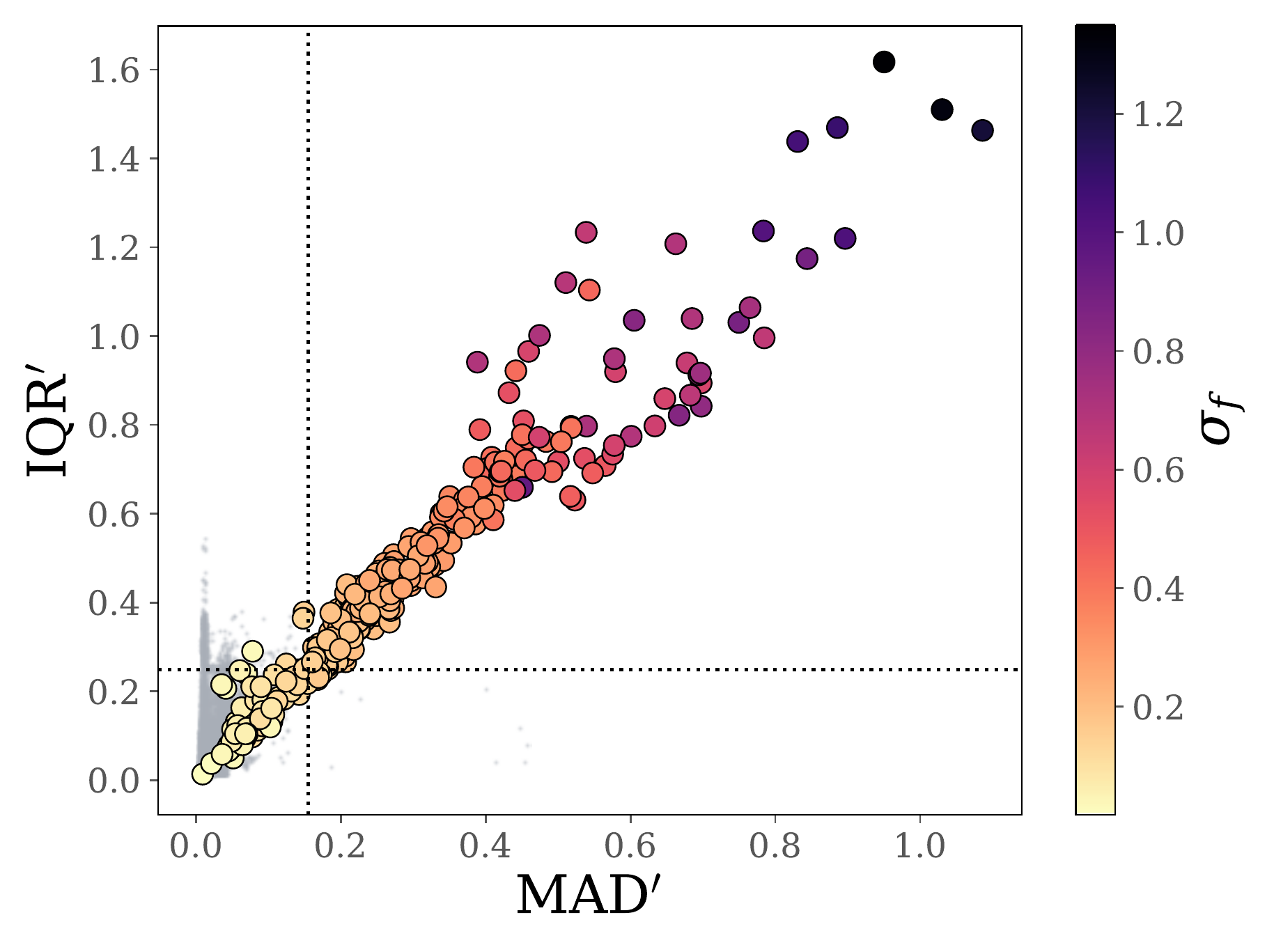}
  \caption{The predicted variability metrics for the spCV \hyperlink{cite.ritter_2003}{RK16} sample are shown here in the colored circles, with color corresponding to $\sigma_f$ (Eq. \ref{eqtn:sigma_f}). The cutoffs for each variability metric (gray dotted lines) were selected to remove 99\% of the degeneracy with the detection of variability from the more quiescent background, which are shown in the gray points on the bottom left.
  }
  \label{fig:var_thresholds}
  \end{center}
\end{figure}

Figure \ref{fig:var_thresholds} shows the IQR$^\prime$ and MAD$^\prime$ predictions for the spCV \hyperlink{cite.ritter_2003}{RK16} sample colored according to $\sigma_f$, with the 100\,pc background shown in gray. We use the comparison between \hyperlink{cite.ritter_2003}{RK16} and the stellar background to define a variability cutoff for both measurements that rejects the lowest 25th percentile of the \hyperlink{cite.ritter_2003}{RK16} sample. This cutoff retains the CVs with the highest variability, but eliminates $>99\%$ of the relatively quiescent 100\,pc background population. The cutoff limits of IQR$^\prime > 0.23$\,mag and MAD$^\prime > 0.14$\,mag are illustrated with the dotted gray lines. This cutoff is approximately equivalent to thresholding at $\sigma_f=0.118$, but owing to the nonparametric nature of RF, this is not an exact equivalence.

\section{The Search for \\ Candidate Cataclysmic Variables} \label{sec:cv_search}

\noindent Section \ref{sec:cv_fit} shows that spCVs have a well-defined CMD space and that when thresholded together, MAD$^\prime$ and IQR$^\prime$ can predict highly variable sources, like spCVs, whether they are periodic or semi-stochastic in nature. To search for spCVs, we query the {\it Gaia} Archive for any objects with 70\,min $< \hat{P}_{\rm orb}$($M_G$, $G_{\rm BP}-G_{\rm RP}$) $<$ 2.5\,hr.  

To minimize detections of spurious sources, we would like to employ a selection of quality cuts, tuned from some of the recommended recipes or \citet{gaiacollaboration_2018a}. Unresolved binary systems are expected to have both excess astrometric and photometric noise in DR2, which does not incorporate a binary solution when measuring parallaxes. We restrict our sample to objects with $\varpi/\delta\varpi > 5 $ and a five-parameter astrometric solution. However, we find that the UWE and RUWE filters recommended by \citet{lindegren_2018} remove many known CV sources in \hyperlink{cite.ritter_2003}{RK16}. Additionally, we find that the recommended cuts on photometric excess noise, or the $\texttt{phot\_bp\_rp\_excess\_factor}$, also filter out potentially interesting CV sources. 

To some extent this is to be expected, as \citet{gaiacollaboration_2018a} suggest that strict employment of {\it Gaia} quality filters will also remove real sources. RUWE and UWE filters are designed to remove excess astrometric noise induced by nearby sources on the CCD but are also known to remove sources between the MS and WD sequence \citep{mowlavi_2020}, just as we find in this study. Similarly, cuts on $\texttt{phot\_bp\_rp\_excess\_factor}$, which measures excess flux in the $G_{\rm BP}$ and $G_{\rm RP}$ integrated photometry with respect to $G$ photometry, are also designed to remove photometric measurements contaminated by flux from nearby stars, but limits on these are calibrated to single systems. Since we are searching for binary systems, we do not impose limits on these quality flags. Instead, to account for spurious measurements due to nearby contaminants, we remove all sources that have neighboring sources recorded by {\it Gaia} within 1.5$\dprime$. To avoid other areas of known astrometric issues in {\it Gaia}, we further eliminated any objects within 20\degree\xspace of the plane of the Galaxy and objects in the line of sight of the Magellanic Clouds. At the boundaries of spCV space, we also removed sources close to the low-mass subdwarf region to avoid degeneracies in CMD space.

This returned 63,825 objects with confident $\varpi$ measurements. Using our RF model, we calculate IQR$^\prime$ and MAD$^\prime$ for these objects and find a sample of 3,253 candidate spCVs with $13 < G < 21$\,mag (above the thresholds described in Section \ref{subsec:rk_vari}) that we list in Table \ref{tab:cand_sample}.

\begin{figure}
  \begin{center}
  \includegraphics[width=240px]{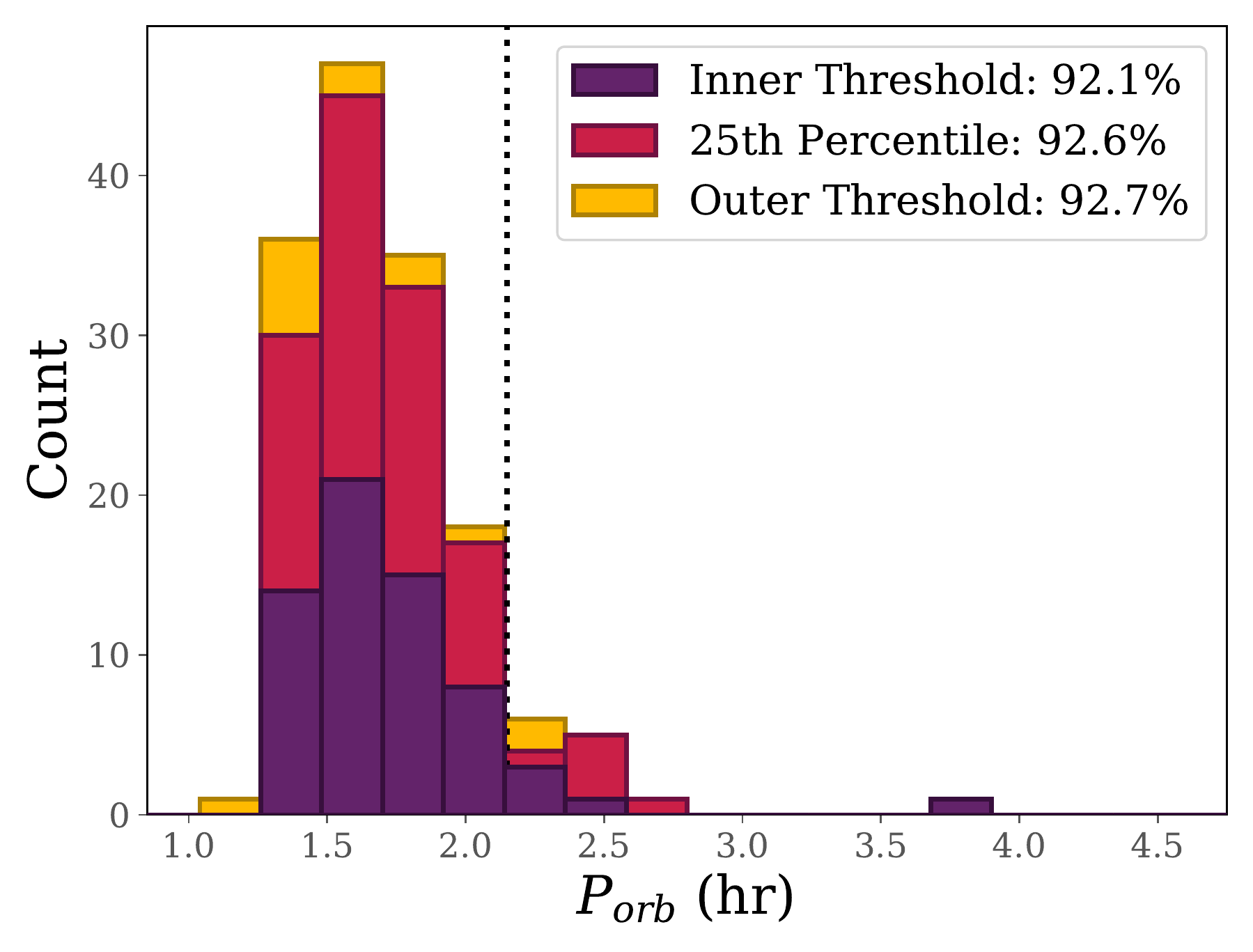}
  \caption{The overlap between \hyperlink{cite.ritter_2003}{RK16} and the candidate sample illustrates that different variability cutoffs lead to similar spCV recovery fractions. Three distributions of \hyperlink{cite.ritter_2003}{RK16} objects are binned by literature $P_{\rm orb}$ across a range of variability cuts, with the lower limit of the period gap shown by the gray dotted line. The red histogram indicates \hyperlink{cite.ritter_2003}{RK16} objects recovered using the 25th percentile variability cutoff that we use for the remainder of the paper. Thresholding instead on the innermost limit which only selects candidates of variability higher than the 100\,pc background recovers the purple histogram. Thresholding more flexibly by selecting all candidates that are more variable than the least variable objects in \hyperlink{cite.ritter_2003}{RK16} recovers the yellow histogram. Agnostic of cutoff choice, all three thresholds recover similar fractions of spCVs.}
  \label{fig:threshold_comparison}
  \end{center}
\end{figure}

\begin{deluxetable*}{ c c c c c c }
\tablenum{3}
\tablecaption{Candidate spCV Sample
\label{tab:cand_sample}}
\tablehead{\colhead{{\it Gaia} \texttt{source\_id}} & \colhead{$G$} & \colhead{$G_{\rm BP}-G_{\rm RP}$} & \colhead{$\hat{P}_{\rm orb}$} & \colhead{$\hat{M}_2$} & \colhead{MAD$^\prime$} \vspace{-5px} \\ 
\colhead{} & \colhead{\deemph{[mag]}} & \colhead{\deemph{[mag]}} & \colhead{\deemph{[hr]}} & \colhead{\deemph{[$M_\odot$]}} & \colhead{}
}
\setlength{\tabcolsep}{10pt}
\startdata
1099223590090176896 & 18.1469 & 0.7794 & $2.3909^{0.2153}_{0.1955}$ & $0.2^{0.00}_{0.00}$ & 1.0303 \\
1563999425873420800 & 18.5760 & 0.1756 & $1.6471^{0.0922}_{0.0874}$ & $0.105^{0.015}_{0.012}$ & 0.9502 \\
1879049845562942592 & 18.6303 & 0.4797 & $1.3076^{0.1084}_{0.0996}$ & $0.069^{0.00}_{0.058}$ & 0.9155 \\
2904668102904871552 & 18.6579 & 0.2052 & $2.1273^{0.112}_{0.1065}$ & $0.194^{0.006}_{0.023}$ & 0.9006 \\
5084805635638179584 & 16.6937 & 0.5885 & $2.0005^{0.1613}_{0.1481}$ & $0.167^{0.033}_{0.028}$ & 0.8857 \\
2395305769240905600 & 18.3096 & 0.7692 & 1.9347$^{0.1833}_{0.1659}$ & 0.154$^{0.192}_{0.029}$ & 0.8010 \\
3521773745637847552 & 19.2306 & 0.5121 & $1.580^{0.1275}_{ 0.1174}$ & $0.096^{ 0.019}_{0.016}$ & 0.8005 \\
6659603532710863232 & 16.7320 & 0.4895 & $2.1876^{0.1563}_{0.1449}$ & $0.2^{0.00}_{0.024}$ & 0.7846 \\
\multicolumn{6}{c}{...}
\enddata
\tablecomments{A truncated table of the candidate spCV sample. The full data are available in a machine-readable format on the web version of this paper. $\hat{P}_{\rm orb}$ is estimated using Equation \ref{eqtn:porb_hat} and $\hat{M}_2$ is estimated using the \hyperlink{cite.knigge_2011a}{Knigge11} donor sequence.}
\end{deluxetable*}

Our candidate sample recovers 144 objects from \hyperlink{cite.ritter_2003}{RK16}-Gaia crossmatch, with $>92\%$ having orbital periods below the period gap. Setting limits on MAD$^\prime$ and IQR$^\prime$ to reject the lower 25th percentile of the \hyperlink{cite.ritter_2003}{RK16} spCV sample is somewhat arbitrary, and does reject even known CV systems. A looser thresholding would recover more of \hyperlink{cite.ritter_2003}{RK16}. To this point, we investigate a range of limiting cutoffs for the variability metrics by matching candidate spCV samples with different variability cutoffs back to \hyperlink{cite.ritter_2003}{RK16}, however even the loosest limits on variability lead to similar spCV/CV recovery fractions, if not similar recovery counts, as shown in Figure \ref{fig:threshold_comparison}. We explore the efficacy of the variability cutoffs further in the next section.

\section{Previous Characterization \\ of Sample Members} \label{sec:previous_char}
\noindent To avoid biasing our sample by the selection functions of previous studies at the time of construction, we determined spCV candidacy using color, $M_{\rm G}$, and variability metrics alone, without including any previous classifications or discovery notes. To check for previous classifications, we crossmatched our candidate catalog (Table \ref{tab:cand_sample}) with the SIMBAD Astronomical Database \citep{wenger_2000}, the AAVSO Variable Star Index \citep[VSX;][]{watson_2006}, and classified spectra from the Sloan Digital Sky Survey \citep[SDSS;][]{kent_1994}. Since many archival classifications were made before {\it Gaia}, we contextualize these objects with astrometric information. 

\subsection{Exploring Variability Cutoffs with \\ SIMBAD Characterization}
\noindent In the previous section, we discussed how different variability cutoffs lead to similar spCV recovery fractions among recovered densities of known CV systems. However, more flexible variability cutoffs also lead to the recovery of other types of variable systems (VSs), both stellar and galactic. Since spCVs display flickering and emit bright, stochastic outbursts from their accretion disk when active, on average spCVs should have higher variability than other VSs. Intuitively, stricter variability cuts should lead to a greater recovery fraction of CVs among other variable types. 

\begin{deluxetable*}{ c | c | c | c }[t!]
\tablenum{4}
\tablecaption{Recovery Fractions in SIMBAD
\label{tab:simbad_recovery}}
\tablehead{
\colhead{{\bf $R_{xmatch}$}} & \colhead{{\bf Inner Threshold} \deemph{(175 candidates)}} & \colhead{{\bf 25th Percentile} \deemph{(3261)}} & \colhead{{\bf Outer Threshold} \deemph{(63825)}}}
\startdata
{\bf 1.5$\dprime$} & 71 CVs/75 objects \deemph{(95\%)} & 155/179 \deemph{(87\%)} & 173/487 \deemph{(36\%)} \\
{\bf 2.5$\dprime$} & 73/77 \deemph{(95\%)} &\setlength{\fboxrule}{1.2pt}\fcolorbox{highlight}{white}{\bf 165/187 (88\% recovery fraction) }& 184/535 \deemph{(34\%)} \\
{\bf 5$\dprime$} & 77/81 \deemph{(95\%)} & 174/212 \deemph{(82\%)} & 191/684 \deemph{(28\%)} \\
{\bf 10$\dprime$} & 77/82 \deemph{(94\%)} & 174/261 \deemph{(67\%)} & 193/1094 \deemph{(18\%)} \\
\enddata
\tablecomments{In order to optimize the variability metric cutoff to recover the highest possible spCV fraction, we investigate the role of the cutoff as a function of crossmatch radius with the SIMBAD database. We find that the 25th percentile limit has the optimal recovery fraction, given the priority to discover new systems.}
\end{deluxetable*}

We pick three different variability cutoffs to explore: an inner threshold that only selects CV candidates with variability higher than the entire 100\,pc background (defined in Section \ref{subsec:vari_metrics}); a 25th percentile threshold on the \hyperlink{cite.ritter_2003}{RK16} sample that rejects the least variable CVs in \hyperlink{cite.ritter_2003}{RK16} while eliminating $>99\%$ of the background; and an outer threshold that selects any candidates with objects more variable than the least variable objects in \hyperlink{cite.ritter_2003}{RK16}. We crossmatch these three different candidate samples with SIMBAD, using a range of cone search radii to allow for a range of positional uncertainty in the reported SIMBAD coordinates. Our recovery fractions of known objects and ratios of CV recovery to the recovery of other classes are reported in Table \ref{tab:simbad_recovery}. 

We find that the strictest variability cut (inner threshold) leads to near-perfect CV recovery at all search radii, but also doesn't allow for the discovery of many new systems: 44\% of the candidates are previously known in a 2.5$\dprime$ crossmatch with SIMBAD. As a note, we consider an object recovered or classified in SIMBAD when it is labeled with a specific classification: CV, CV flavor, or otherwise. If objects are labeled as a star with no further classification or are only noted in the database for being observed at a specific wavelength, we do not consider them as previously classified in SIMBAD and do not include them in recovery counts. The 25th percentile cut still recovers a majority of CV systems among previously classified objects, with an 88\% recovery fraction when crossmatched to SIMBAD with a 2.5$\dprime$ cone-search radius. The 25th percentile cutoff also allows for much greater discovery space than the inner threshold as only 6\% of the candidate sample is already known in SIMBAD. We discuss the classifications of the 22 non-CV objects recovered in SIMBAD in Section \ref{subsec:update_classes}. The outer threshold offers the largest opportunity for discovery of new systems, but has a significantly lower recovery fraction of 34\% in a 2.5$\dprime$ crossmatch with SIMBAD. Using these results, we define our candidate CV catalog using the 25th percentile cutoff limit on variability.

\subsection{Recovery of Known CVs in AAVSO}

\begin{figure*}
  \begin{center}
  \includegraphics[width=450px]{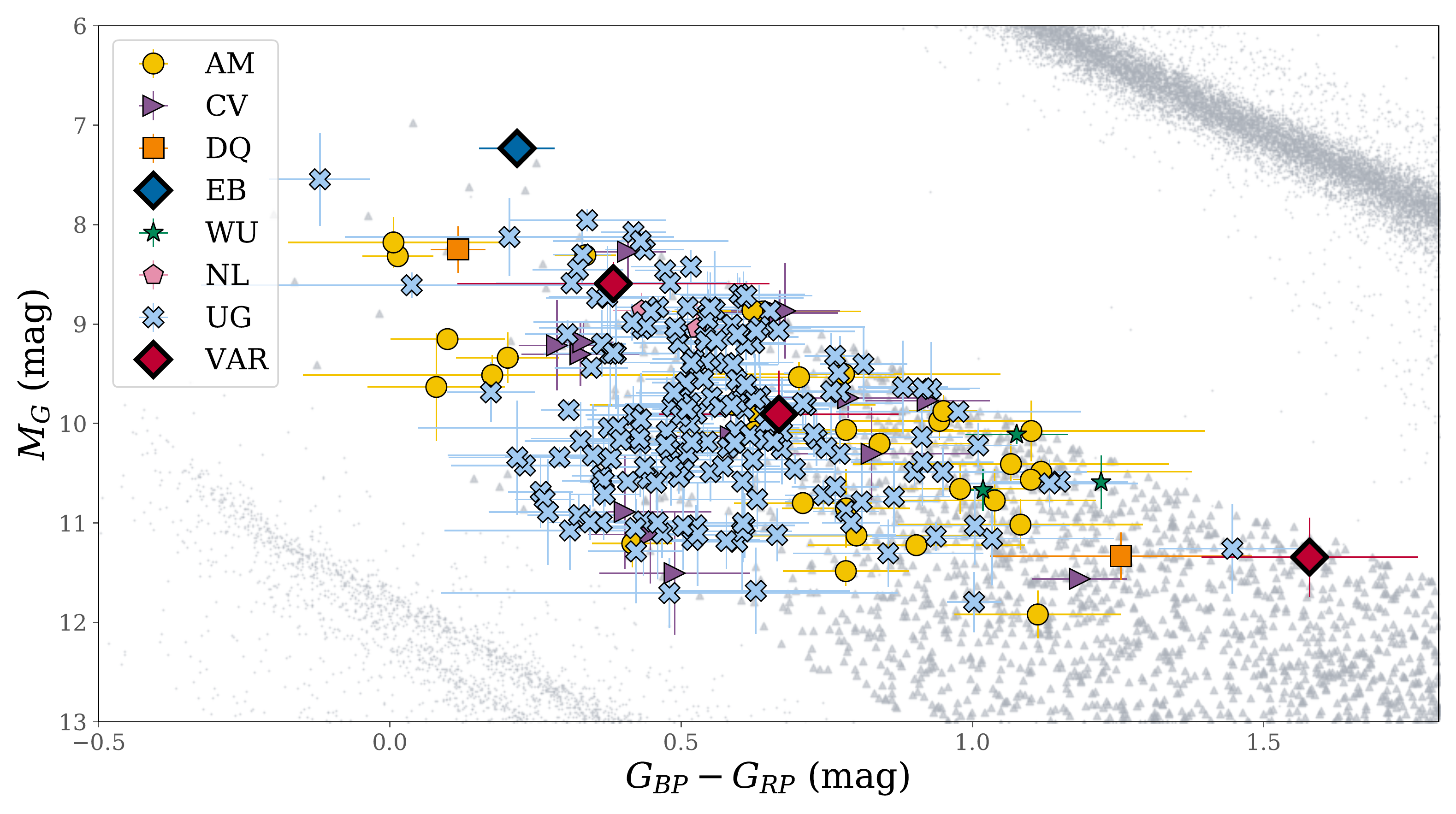}
  \caption{239 objects characterized in AAVSO, placed on the same background CMD (in gray pixels) as Figure \ref{fig:var_thresholds}. We show our full sample of candidate CVs in gray triangles, and overlay the members of the sample that were previously characterized in AAVSO in an assortment of eight colored shapes. The four objects that are classified as general variables (VAR) or as an eclipsing binary (EB) are shown in the highlighted diamond shapes. These objects are likely still CV systems, as they fall within the CV neighborhood on the CMD even within their confidence intervals. Similarly, the W\,Uma eclipsing variables (WU) are likely misclassified CVs, since W\,UMa stars generally fall above the MS. The CVs are shown here as their most general CV subtype: AM\,Herculis-type variables (AM), cataclysmic variables of unspecified type (CV), DQ\,Herculis type (DQ), nova-like systems (NL), and dwarf novae, also known as U\,Geminorum-type variables (UG).}
  \label{fig:aavso_cvs}
  \end{center}
\end{figure*}

\noindent We crossmatch the candidate CV catalog with the AAVSO VSX\footnote{We obtained the VSX on 2/24/20 from the Vizier online database.} and find a nearly perfect CV recovery fraction even at the 25th percentile cutoff limit. In a 2.5$\dprime$ crossmatch we recover 239 variables, and only eight of these variable objects were not classified as CVs or a subtype of CVs in VSX. Of those eight objects, three were only generally classified as variable objects or eclipsing systems and lacked a specific characterization. The remaining were classified as follows: one RR Lyrae star and one $\delta$\,Scuti star, both of which were also recovered in the SIMBAD crossmatch, as discussed in Section \ref{subsec:update_classes}, and three W Ursae Majoris type (W\,UMa) systems. We show the CVs recovered in VSX on the {\it Gaia} CMD in Figure \ref{fig:aavso_cvs}. We highlight the three variable stars and EB with a thicker border around the marker. From the location of these systems on the CMD, even sliding their CMD position along their confidence bands, it is highly likely that these seven systems (the W\,UMa stars, the unclassified variables, and the EB) could be further classified as CVs. We do not find a qualitatively obvious place for CV subtypes on the CMD, though this is explored in much greater detail, unbounded by limits on $P_{\rm orb}$ in \citet{abril_2020}. While we find that UG systems do seem to have a specific cutoff in CMD space, this is also generally the cutoff of all previously classified CV systems (see Figure \ref{fig:rk_cmd}), and systems below this point have smaller, fainter donor stars, owing to where they fall on the CMD-$P_{\rm orb}$ sequence. As previous CV surveys have been shown to be magnitude limited, it is difficult to say that this cutoff is physical. The rightmost CV on this plot proves that UG types extend beyond the locus seen in the data, and further studies are necessary to prove if spCV subtypes can be neatly tied to CMD position.

VSX also provides period measurements for 143 of these systems. We calculate period predictions for these objects using the fit in Eqtn. \ref{eqtn:porb_hat}, since these systems have periods measured by alternate studies from \hyperlink{cite.ritter_2003}{RK16}, and compare the observed periods with the predicted periods in Figure \ref{fig:aavso_periods}. The fit does a reasonable job of predicting the periods, with some evidence for overpredicting periods of true CVs. The fit does a poor job of predicting the three objects that were classified as W\,UMa variables. This could be because these stars are not CV systems and therefore the fit would not perform well, or because these systems might be poorly sampled and the incorrect period has been measured. VSX does warn that some $P_{\rm orb}$ measurements could be multiples of the true period, but does not indicate uncertainty in $P_{\rm orb}$ for any of these objects.

\subsection{Updating Previous AAVSO and SIMBAD Classifications with {\it Gaia} DR2} \label{subsec:update_classes}

\begin{figure}
  \begin{center}
  \includegraphics[width=240px]{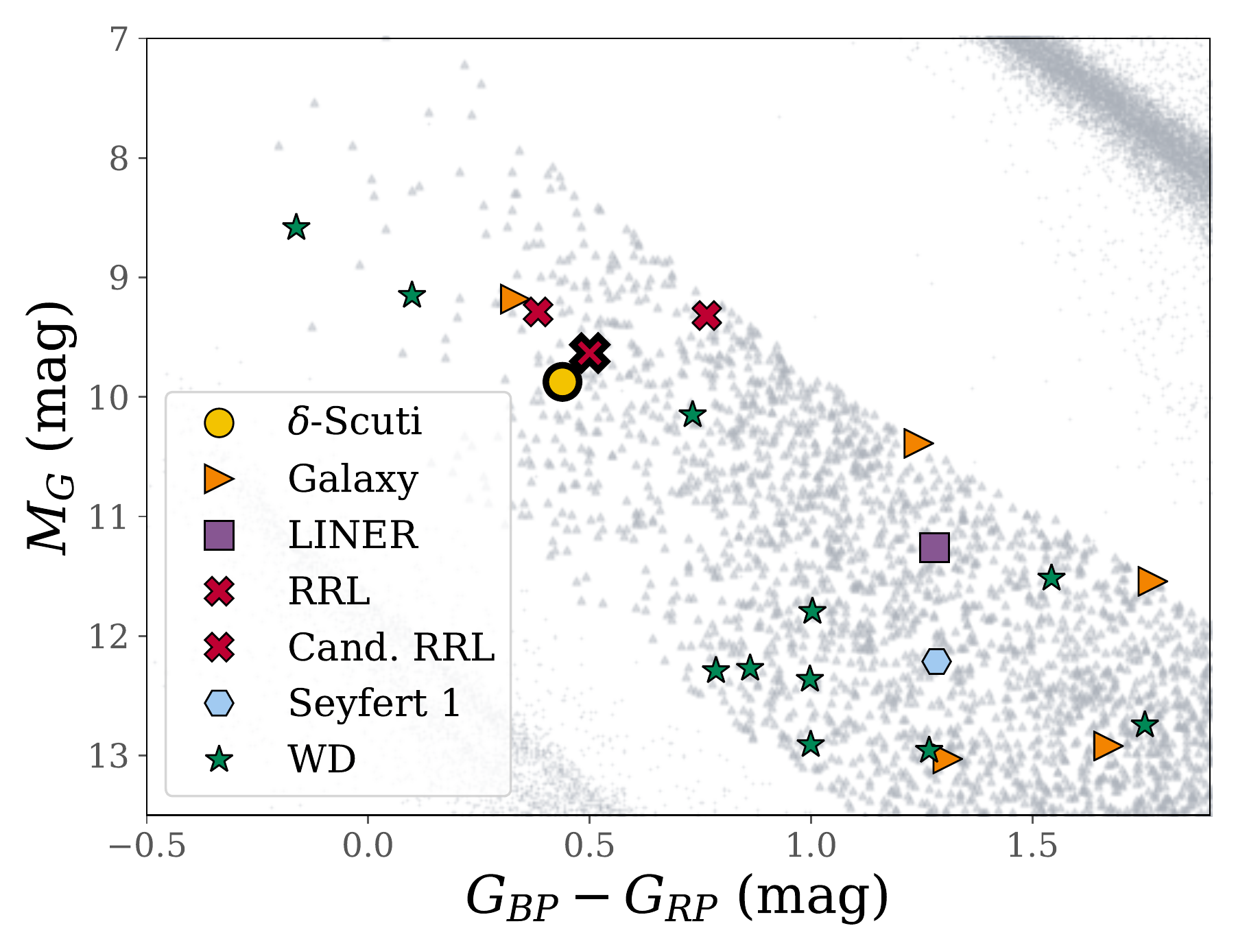}
  \caption{There are four pulsational variable stars (RRL, $\delta$\,Scuti) recovered in our crossmatch with SIMBAD and VSX. One of these objects is labeled as a candidate RRL (Cand. RRL) from its light curve. In this figure, we include the MS and WD CMD sequence from the 100\,pc sample in the gray pixels. From the position of the variable stars/systems on the CMD it is likely that they have been misclassified. The same argument holds true for the 11 recovered WDs. This leaves the seven recovered galaxies and active galactic nuclei (Galaxy, LINER, Seyfert 1) which need further investigation to determine whether they have been misclassified.}
  \label{fig:misclassified_simbad}
  \end{center}
\end{figure}

\begin{figure}
  \begin{center}
  \includegraphics[width=240px]{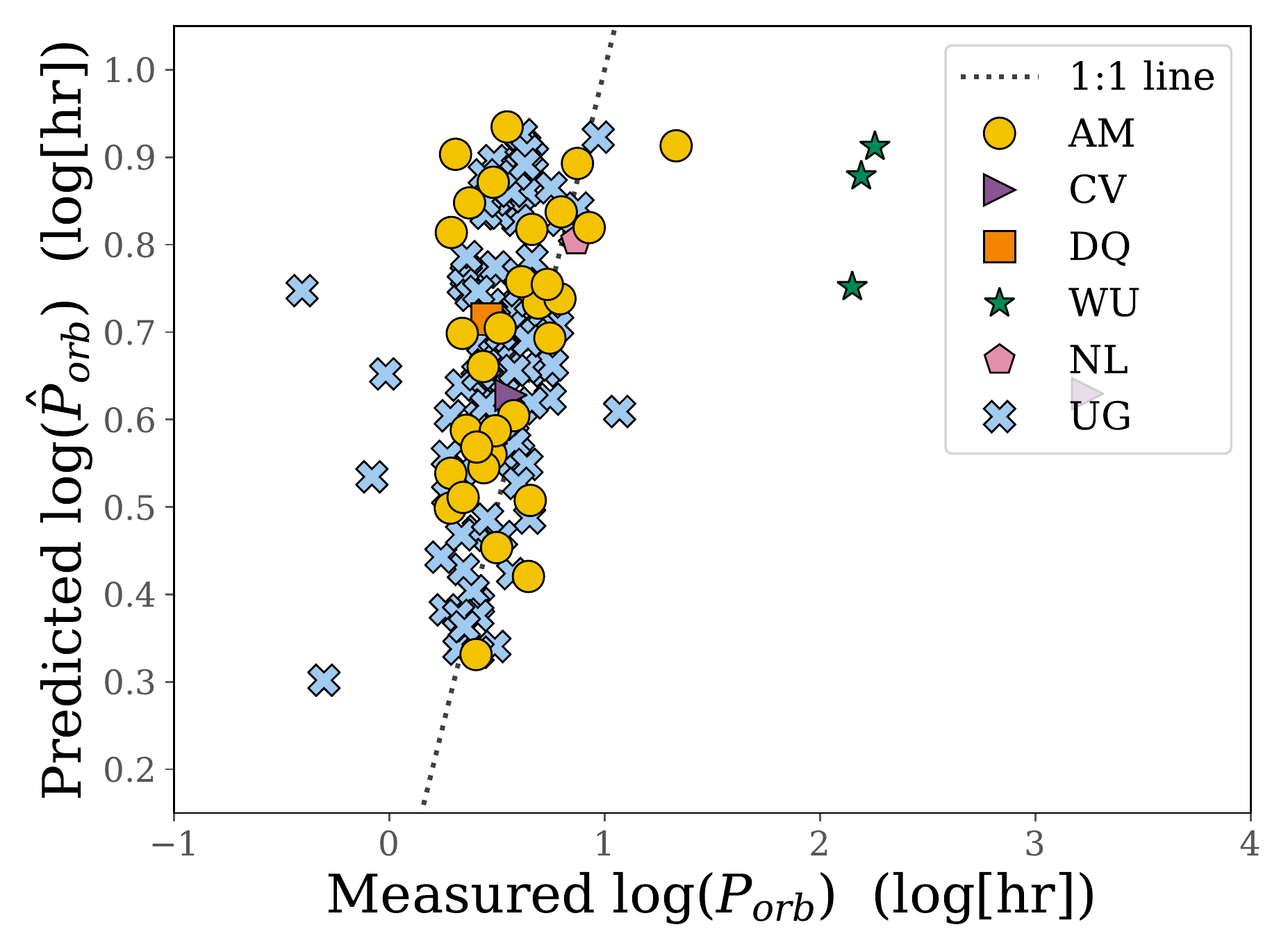}
  \caption{143 systems, represented by six CV subtypes, in the VSX crossmatch with our sample have measured periods. Our model fit to $P_{\rm orb}$ does a reasonable job of predicting period for most CV subtypes, but performs poorly on the three UW types with measured periods. These types display much longer periods than would be expected from their position on the CMD, which places them right below the CV period gap, as shown in Figure \ref{fig:aavso_cvs}.}
  \label{fig:aavso_periods}
  \end{center}
\end{figure}

\noindent Crossmatching with both AAVSO and SIMBAD recovered a total of 22 objects in the candidate sample that were classified as non-CV objects. We show where these objects fall on the {\it Gaia} CMD in Figure \ref{fig:misclassified_simbad}, with the RR Lyrae stars (RRL) and $\delta$\,Scuti variables that were located in both catalogs highlighted in a darker stroke. 

Three of these 22 objects are classified as RRL, and one is classified as a $\delta$\,Scuti pulsator. RRL and $\delta$\,Scuti variables are typically located above the MS of stars (see Fig. 7 in \hyperlink{cite.rimoldini_2019}{Rim18}), which is shown in the diagonal line of gray pixels in the upper-right corner of the figure. Without distance information, there is degeneracy in the colors of pulsational variables (PVs) with other objects like CVs, MS stars, and WDs. Placing the objects on the {\it Gaia} CMD allows us to set further constraints on their classifications, and from their positions alone in Figure \ref{fig:misclassified_simbad}, these objects could be misclassified CVs. We investigate each of the individual objects below.

One of the SIMBAD RRL, V1032 Oph, is classified as an eclipsing CV in VSX\footnote{VSX OID 21518}, and appears to be a SIMBAD misclassification, since it is not listed in the SIMBAD reference paper linked on the webpage \citep{samus_2003}. The other RRL, CRTS J120105.7-164504, is located in both SIMBAD and VSX, and is classified as a type-c RRL (RRc) by \citet{drake_2014}. Their analysis relies on pre-{\it Gaia} derived distances and mentions a confusion rate of a few percent between RRc and contact binaries. From the CMD position, it could be that this is one of those few percent. Finally, the candidate RRL, CRTS J224159.9-662512, is classified instead as a CV by \citet{drake_2014}. 

The $\delta$\,Scuti variable, LINEAR 9345642, is listed in both \hbox{SIMBAD} and VSX, and is classified by \citet{palaversa_2013}. \citet{palaversa_2013} find that the object has a period of 1.533\,hr, which is in the physical range that we would expect for an spCV. This object is classified as a PV with a confidence score of 2/5 and could be alternately classified by the general category in \citet{palaversa_2013} which includes CVs. This a classification worth exploring given the object's {\it Gaia} distance. However, if the $\delta$\,Scuti is correct, it is noticeable that two PVs, the CRTS RRc and the LINEAR $\delta$\,Scuti, would both fall below the MS, which would make these objects orders of magnitude fainter than is typical of their classes \citep[see Fig. \ref{fig:var_thresholds} in][]{gaiacollaboration_2019}. We investigate both of these objects further in Section \ref{subsec:peculiars}.

WDs typically live in the WD sequence, which is shown in the diagonal double line of gray pixels in the lower-left corner, and W\,UMa stars are typically located slightly above the MS. CVs occupy a unique position on the CMD, one that is not known to be shared with other Galactic variables, which limits detection degeneracies with other systems. Owing to the brighter than expected $M_G$ values of the 11 WDs recovered in the SIMBAD sample and the three W\,UMa stars in the VSX sample, we suggest that they too are likely misclassified CVs. We recommend that future classifiers for variable stars and systems include distance information, now largely available due to {\it Gaia}, as classification attributes to help distinguish between physically disparate variable objects with similar light curve and color information.

The remaining seven objects are detections of galaxies or active galactic nuclei (LINER, Seyfert 1) and need further investigation to determine whether they are misclassified. Since these objects all have robust measurements in {\it Gaia} DR2, with more than 11 observations used in their astrometric solutions and position uncertainties ranging between 1.4\% and 0.6\%, it is surprising that they might be extended sources. However, \citet{bailer-jones_2019} finds that there is an extragalactic contamination rate in {\it Gaia} DR2 of $\sim$0.07\%. To infer this contamination rate, \citet{bailer-jones_2019} classify galaxies and quasars directly from {\it Gaia} DR2 covariates and find that the SIMBAD object classified as a Seyfert 1 has a 66\% probability of being a galaxy, given its {\it Gaia} observables. None of the other objects were found to likely be galaxies in that study. Assuming that all extragalactic objects are correctly classified in SIMBAD, the contamination of possible extragalactic objects in our sample is 3.7\%.

\begin{figure*}[!ht]
  \begin{center}
  \includegraphics[width=500 px]{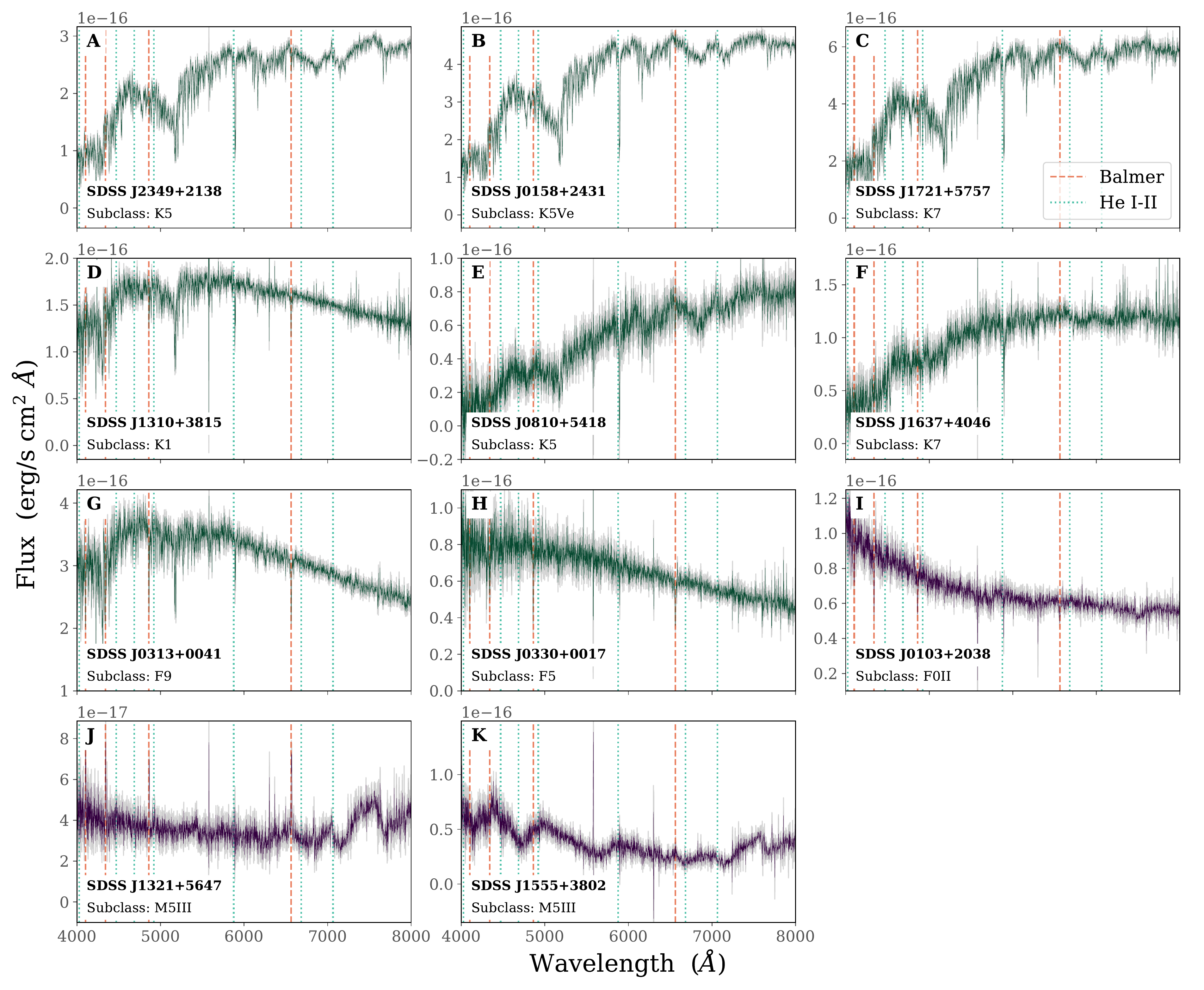}
  \caption{Spectra that meet our CV search criteria without literature classification archival SDSS spectra. The M-dwarf secondary star is visible in the red portion of the {\it SDSS J1321} spectrum along with Balmer emission, making it a likely CV (panel J). {\it SDSS J0103} could be a nova-like CV with a noisy continuum, as its spectrum does not appear to be the F star that it was automatically assigned by the SDSS pipeline (panel I), and {\it SDSS J1555} shows an excess of blue light at certain wavelengths despite the M-dwarf characteristics on the red side of the spectrum (panel K). All three of these spectra also fall within the $u-g-r$ color space of WD--MS binaries, while the remaining spectra are puzzling as they display taxonomical MS characteristics despite their CMD position as is shown in Figure \ref{fig:sdss_cmd}.}
  \label{fig:misclassified_sdss}
  \end{center}
\end{figure*}

\subsection{SDSS Spectral Types} \label{subsec:sdss_spectra}

\noindent In a 2.5$\dprime$ crossmatch with the 16th Data Release of SDSS \citep{ahumada_2020}, 84 objects had previous optical spectroscopic observations. The crossmatch is included in Table \ref{tab:sdss_xmatch}. 78\% of these spectra were marked as CVs from previous studies \citep[][]{szkody_2002, szkody_2003, szkody_2004, szkody_2005, szkody_2006, szkody_2007, szkody_2009, szkody_2011}. 34 objects have measured $P_{\rm orb}$, and 32/34 are below the orbital period gap.  The 54 CVs characterized by the SDSS mission have measured H$\alpha$ and H$\beta$ equivalent widths (EWs) in the literature.

\begin{figure*}[!p]
  \begin{center}
  \includegraphics[width=500 px]{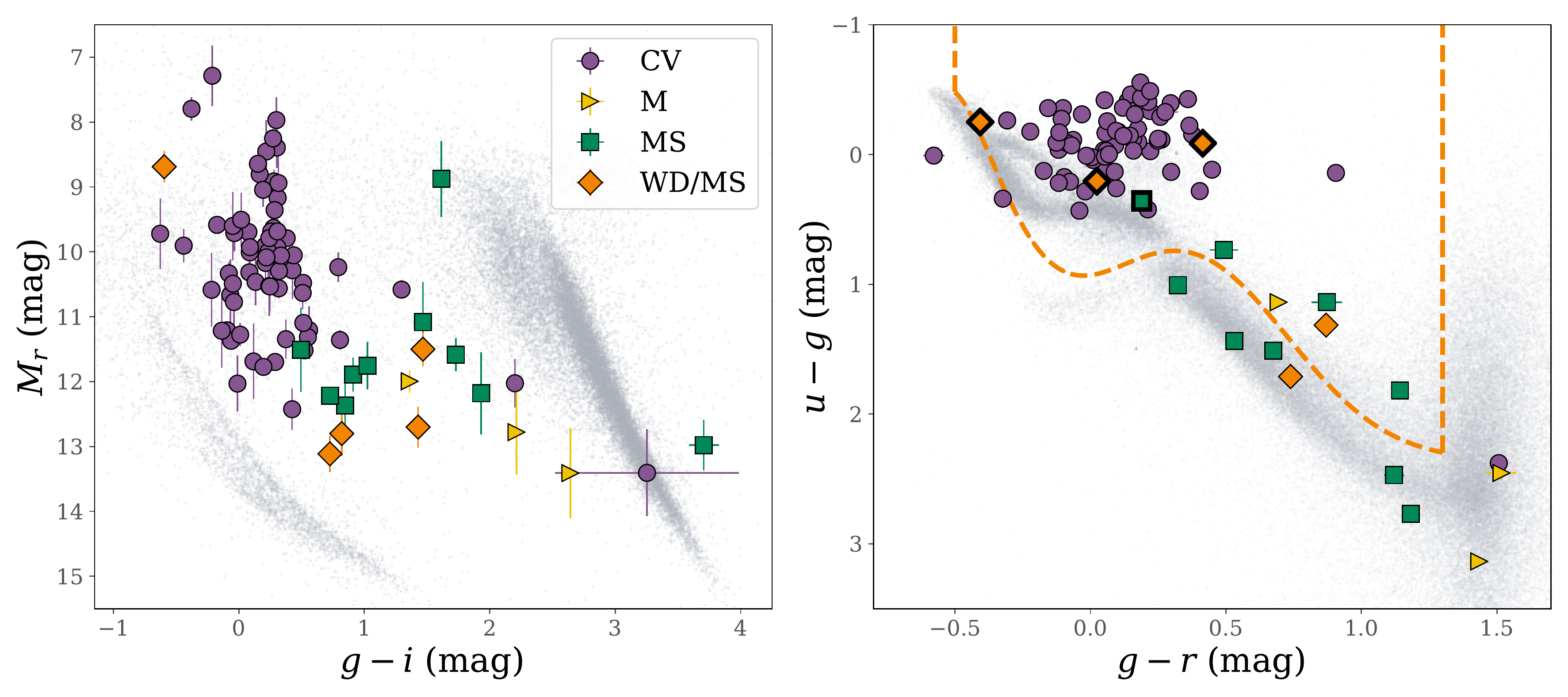}
  \caption{{\it Left panel:} The SDSS CMD for candidate spCVs. The legend indicates how these objects have been classified in the literature \citep{szkody_2002, szkody_2003, szkody_2004, szkody_2005, szkody_2006, szkody_2007, szkody_2009, szkody_2011, rebassa-mansergas_2010, west_2011} or by the automatic SDSS classification algorithm. While CVs and WD--MS binaries are expected to fall between the MS and WD sequence, there are some surprising MS objects (spectra shown in Fig. \ref{fig:misclassified_sdss})  including M dwarfs (M, Fig. \ref{fig:sdss_mdwarfs}) that fall off the MS. We recommend these objects for further follow-up observations as they may be mischaracterized WD--MS binaries. The gray pixels in this panel are the same 100\,pc background used throughout this analysis, constrained to objects that were observed by SDSS, which is about 26\% of that original sample. {\it Right panel:} \citet[Eqs. 2--4 of][]{rebassa-mansergas_2016} defines a $u-g-r$ two-color space that encloses WD--MS binaries, indicated by the dashed orange line. We find that four of the MS objects and one of the M dwarfs fall into this color space, justifying the case that these objects are likely not single MS stars. While this color space encompasses the majority of previously confirmed CVs and WD--MS binaries, there are still three CVs that fall outside the sample, and one WD--MS binary. These confirmed binaries are close to other MS stars, warranting further investigation of those MS objects. The background sample, shown in the gray pixels in this panel, is composed of a random subset of SDSS stars (type = 3 in the SDSS database) with clean photometry, and the confirmed SDSS WDs from \citet{gentilefusillo_2019}.}
  \label{fig:sdss_cmd}
  \end{center}
\end{figure*}

11 objects in the crossmatch are not characterized in the literature, and we show these objects in Figure \ref{fig:misclassified_sdss}. The SDSS database includes an assigned class that is fit to the spectrum, and we indicate the automatically assigned classes in each panel of the figure, however these classifications require further inspection before confirmation. {\it SDSS J1321+5647} (Fig. \ref{fig:misclassified_sdss}, panel J) shows clear signatures of being a CV with visible Balmer and He~I/He~II emission. Additionally, the M-dwarf secondary of the system is visible in the red half of the spectrum which will allow for future characterization of the donor star. Upon inspection {\it SDSS J0103+2038} (Fig. \ref{fig:misclassified_sdss}, panel I) shows characteristics of being a quiescent nova-like CV system, and we recommend further investigation. {\it SDSS J1555+3802} (Fig, \ref{fig:misclassified_sdss}, panel K) exhibits M-dwarf characteristics in the red side of the spectrum and excess luminosity in the blue side, providing evidence that this system might is likely a WD--MS binary, and possibly a CV, owing to its position on the {\it Gaia} CMD. All three of these objects fall into the $u-g-r$ two-color space for WD--MS binaries, defined by \citet[][]{rebassa-mansergas_2016} as is shown in \ref{fig:sdss_cmd}, along with {\it SDSS J1637+4046} (Fig. \ref{fig:misclassified_sdss}, panel F). Including these three systems that display possible CV characteristics, we find a recovery fraction of 82\% in SDSS. This is a significant improvement over previous studies that rely on light curves alone: CRTS searches of light curves have an SDSS recovery fraction of $\sim 23$\% \citep{breedt_2014}.

\begin{deluxetable*}{cccccc}[b]
\tablenum{5}
\tablecaption{SDSS-{\it Gaia} Crossmatch
\label{tab:sdss_xmatch}}
\tablehead{\colhead{{\it Gaia} \texttt{source\_id}} & \colhead{SDSS ID} & \colhead{Plate-MJD-Fiber} & \colhead{Lit. Type} & \colhead{H$\alpha$} & \colhead{$P_{\rm orb}$}
\vspace{-5px} \\ 
\colhead{} & \colhead{} & \colhead{} & \colhead{} & \colhead{\deemph{[\AA]}} & \colhead{\deemph{[hr]}}
}
\setlength{\tabcolsep}{12pt}
\startdata
4461280391188698368 & 1642+1347 & 2210-53535-0592 & CV & 67.0 & 1.07 \\
583340808875713280 & 0859+0536 & 1192-52649-0343 & Polar & 42.0 & 1.1 \\
712480889099565696 & 0903+3300 & 1272-52989-0188 & CV & 153.0 & 1.3 \\
1332378466733219456 & 1625+3909 & 1172-52759-0212 & DN & 100.0 & 1.31 \\
1563999425873420800 & 1307+5351 & 1039-52707-0069 & Polar & 7.6 & 1.33 \\
3876618514794039040 & 1015+0904 & 5334-55928-0884 & Polar & 30.0 & 1.33 \\
1289860214647954816 & 1502+3334 & 1648-53171-0408 & CV & 121.0 & 1.4 \\
2507796391561705728 & 0155+0028 & 0403-51871-0423 & CV & 13.0 & 1.43 \\
2465053942183130240 & 0137-0912 & 0662-52178-0541 & CV & 38.0 & 1.43 \\
772038105376131456 & 1131+4322 & 1366-53063-0231 & DN & 117.0 & 1.53 \\
\multicolumn{6}{c}{...}
\enddata
\tablecomments{We show a truncated table of the crossmatch. The full data are available in a machine-readable format on the web version of this paper.}
\end{deluxetable*}

In contrast, the spectra of {\it SDSS J1721+5757, J0158+2431, J2349+2138, J0810+5418, J1637+4046, J1310+3815, J0313+0041}, and {\it J0330+0017} (Fig. \ref{fig:misclassified_sdss}, panels A--H) all display the typical shape and absorption features of MS F, G, and K stellar types within the measurement uncertainties. This is surprising, as it is clear that these objects are unusually blue in {\it Gaia} for their luminosity, removing them from the well-defined MS. Since {\it Gaia} measures photometry as a mean measurement across multiple observations, we expect the colors in the {\it Gaia} CMD to represent CVs in quiescent states. 

To see if objects might have moved on the CMD since the epoch of the SDSS observations owing to outbursts, we place these objects on an SDSS CMD and on a $u-g-r$ two-color diagram in Figure \ref{fig:sdss_cmd}. On the CMD, in the left panel, all objects are noticeably bluer than the MS in SDSS as well, with the exception of CV {\it SDSS J2345+3429}, which could fall below the MS within its confidence interval, and {\it SDSS J0810+5418} which is redder than the MS, possibly due to extinction. The spectrum of this object is typical of a K dwarf, and the default spectral fit found in SDSS is K5. The $u-g-r$ two-color diagram in the right panel provides further insight into the objects that appear too blue to be the MS stars they have been labeled or characterized to be in the right panel. Eqs. 2--4 of \citet{rebassa-mansergas_2016} provide empirical bounds to the two-color $u-g-r$ space occupied by WD--MS binaries. Both contact types, like CVs (which are seen in the locus near the single WDs in this figure) and non-contact binaries are included in these bounds. A further star labeled as an MS star by SDSS, indicating that its bluer CMD position is likely pointing to a WD companion that is not detected easily in the spectrum.

\begin{figure}
  \begin{center}
  \includegraphics[width=230px]{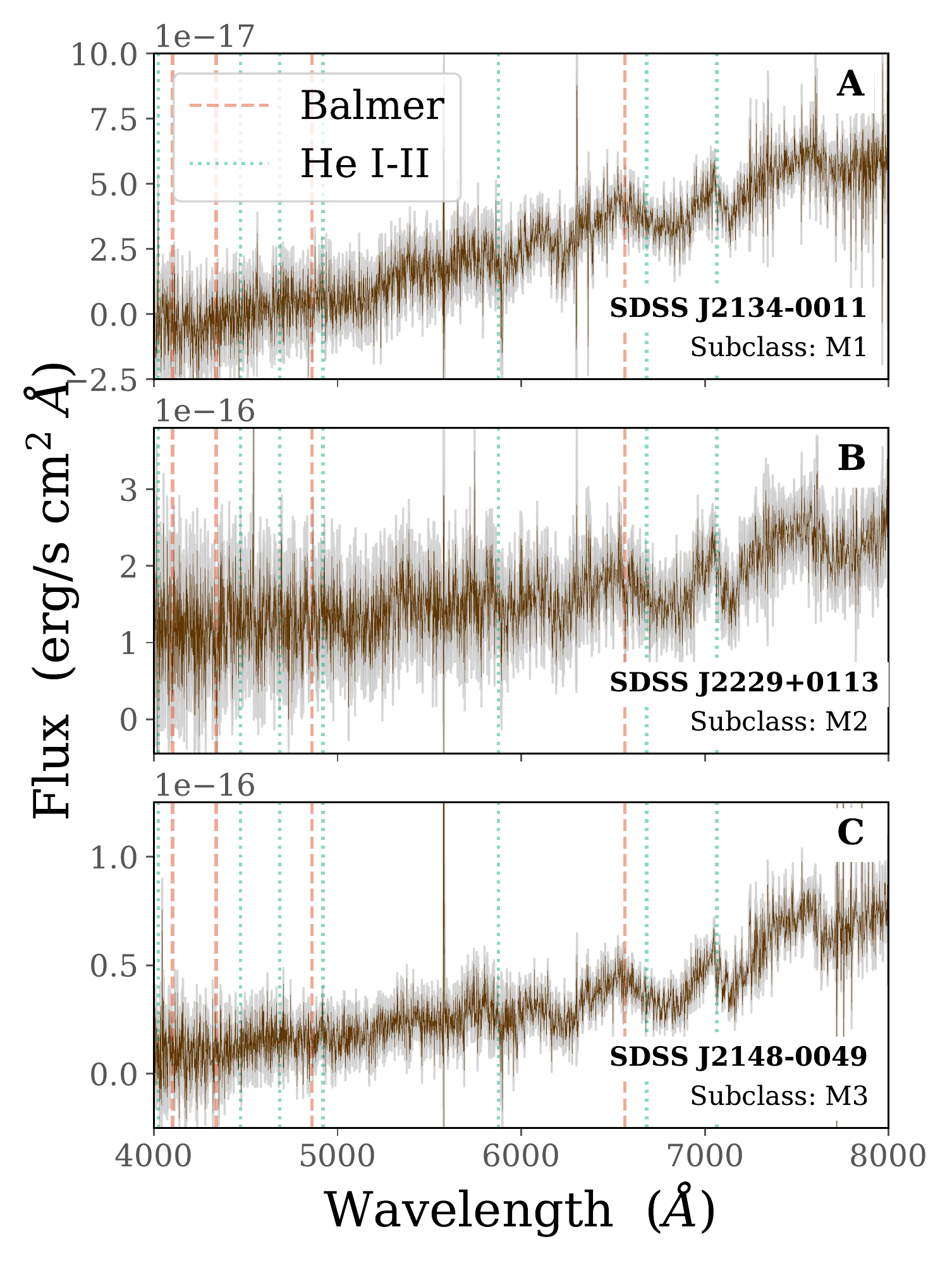}
  \caption{Three objects in the 2.5$\dprime$ crossmatch with SDSS are characterized as M dwarfs in the literature. These objects all show typical M dwarf absorption features and there is no evidence of a WD in their spectra. However, the spectra of confirmed WD--MS binaries in Figure \ref{fig:sdss_binaries} illustrate that a WD is not always immediately apparent in the binary spectrum. The panels in this figure are noticeably similar to panel C in Figure \ref{fig:sdss_binaries} and further analysis of these objects is warranted.}
  \label{fig:sdss_mdwarfs}
  \end{center}
\end{figure}

Three objects, {\it SDSS J2134-0011, J2229+0113}, and {\it J2148-0049}, were characterized as MS M dwarfs in the literature and we show their spectra in Figure \ref{fig:sdss_mdwarfs}. These objects also have the signature spectral shape and absorption that is indicative of M-dwarf spectra, but like the F and K stars from Figure \ref{fig:misclassified_sdss}, these stars are surprisingly blue for M dwarfs and are located far from the MS in Figure \ref{fig:sdss_cmd}. Blue excess can be indicative of a WD companion. Additionally, {\it SDSS J2229+0113} (panel B) falls into the defined color space for WD--MS binaries in $u-g-r$ space. 

\begin{figure}
  \begin{center}
  \includegraphics[width=230px]{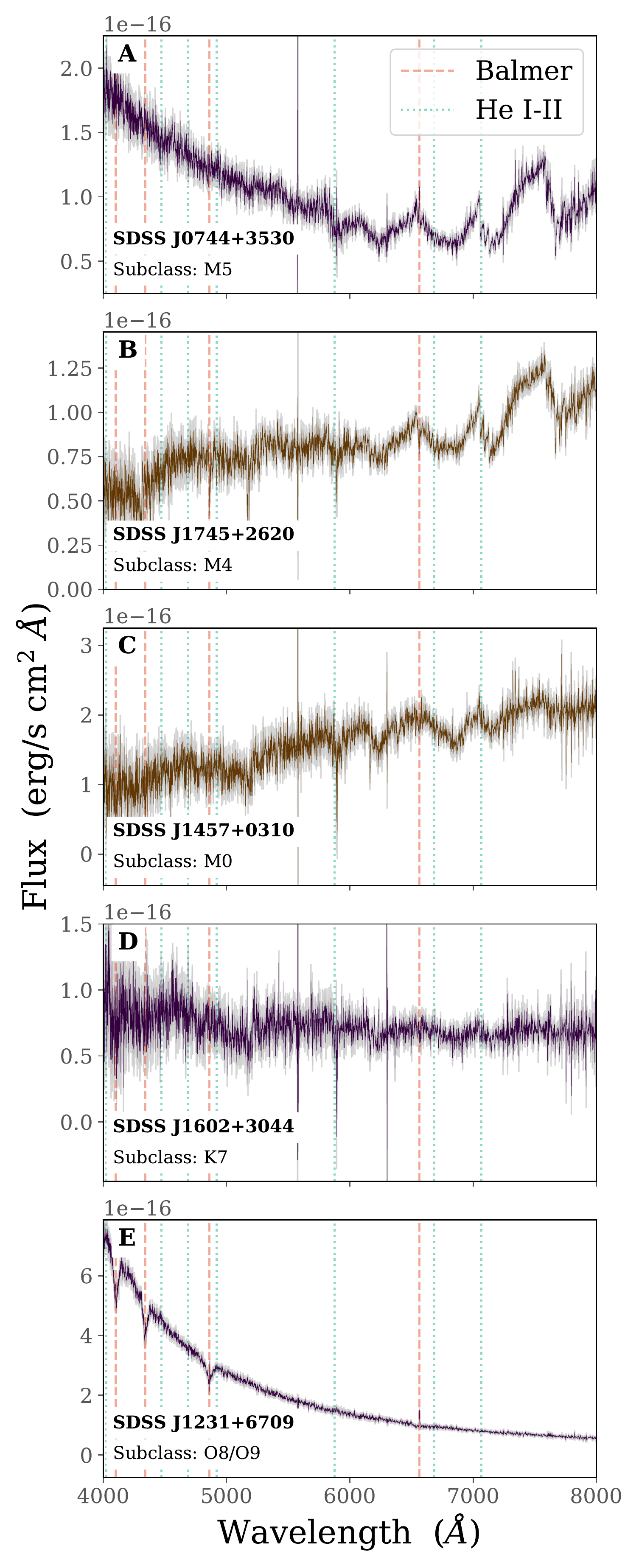}
  \caption{5 objects are previously characterized as WD--MS binaries in the literature \citep{rebassa-mansergas_2010}. The stars in panels A and B show characteristic M-dwarf qualities toward the red portion of their spectra, which are also visible in the spectra of {\it SDSS J1321} and {\it SDSS J1555} in Figure \ref{fig:misclassified_sdss} (panels J and K, respectively). This furthers the case that some of the MS spectra shown in Figure \ref{fig:misclassified_sdss} might be WD--MS binary systems.}
  \label{fig:sdss_binaries}
  \end{center}
\end{figure}

Five spectra have been previously characterized as WD--MS binaries by \citet{rebassa-mansergas_2010}; we show their spectra in Figure \ref{fig:sdss_binaries}. These spectra are not characterized as CVs by \citet{rebassa-mansergas_2010}, but three objects ({\it SDSS J0744+3530, J1602+3044}, and {\it J1231+6709,}; Fig. \ref{fig:sdss_binaries}, panels A, D, and E, respectively) fall tightly within the color space of confirmed CVs in both the three-color diagram and the SDSS CMD shown in Figure \ref{fig:sdss_cmd}. These objects might be quiescent CV systems in the process of crossing the period gap, which means that an accretion disk would not be visible in the spectrum. We discuss the two-color characterization further in Section \ref{subsec:off_ms}.

\section{Spectroscopic Confirmation of \\ New Cataclysmic Variables} \label{sec:new_spectra}

\begin{figure*}[]
  \begin{center}
  \includegraphics[width=500 px]{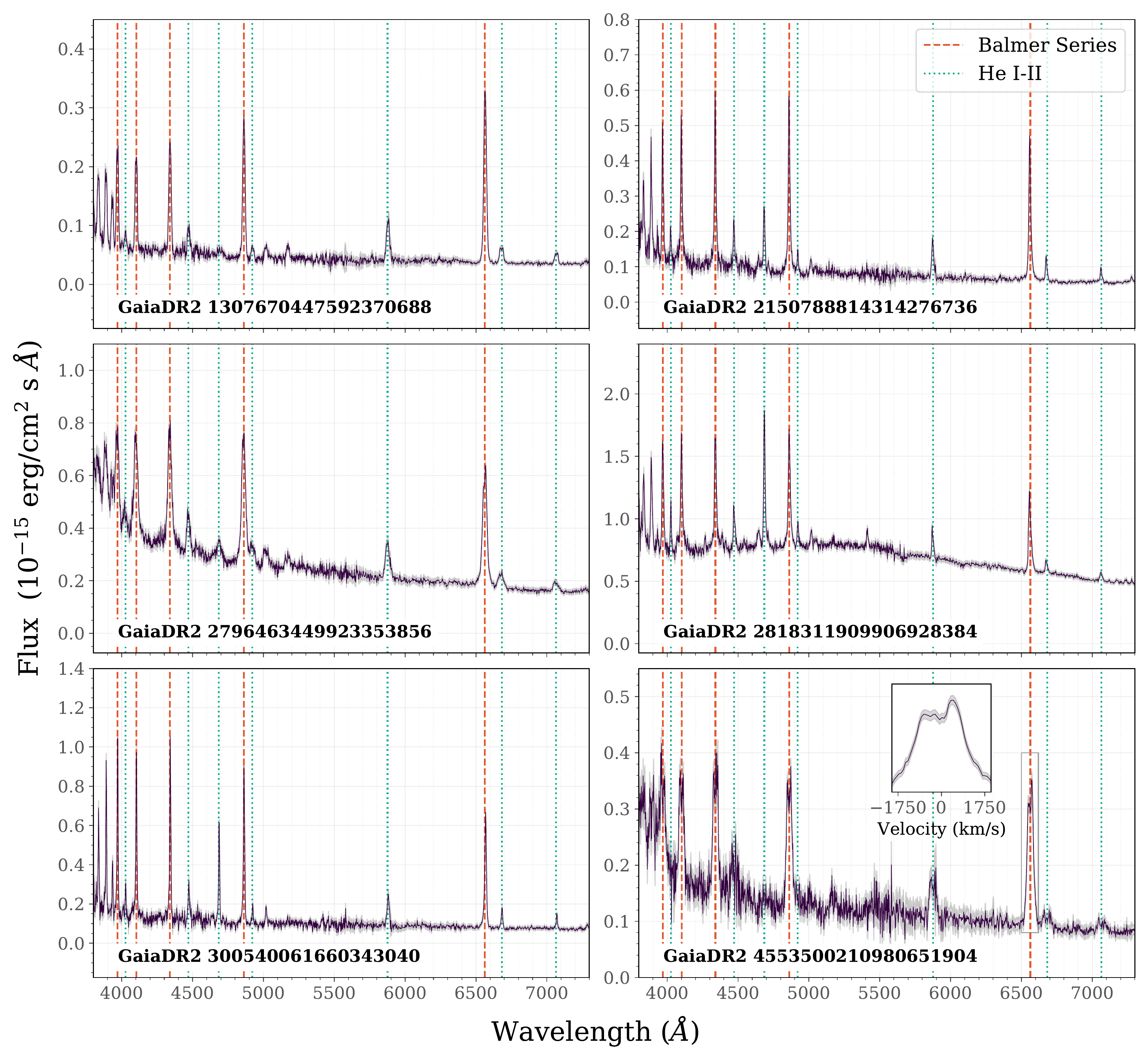}
  \caption{Lick spectra of six new CV systems in active states. Three systems show strong He~II emission lines, indicative of a magnetic CV or SW\,Sex system. {\it Gaia} DR2 4553500210980651904 has prominent central absorption feature in all the Balmer lines, likely caused by a relatively high inclination of the system. While eclipsing systems have double features that extend almost halfway to the continuum, the doubling here indicates that this system is not face-on. An inset displays the double peak seen in H$\alpha$ emission from this system, with the terminal axis displayed in Doppler velocities.}
  \label{fig:lick_lines}
  \end{center}
\end{figure*}

\noindent We obtained spectra of nine objects at Lick Observatory, confirming their CV status. We report on their spectra here. 

\begin{figure}
  \begin{center}
  \includegraphics[width=245px]{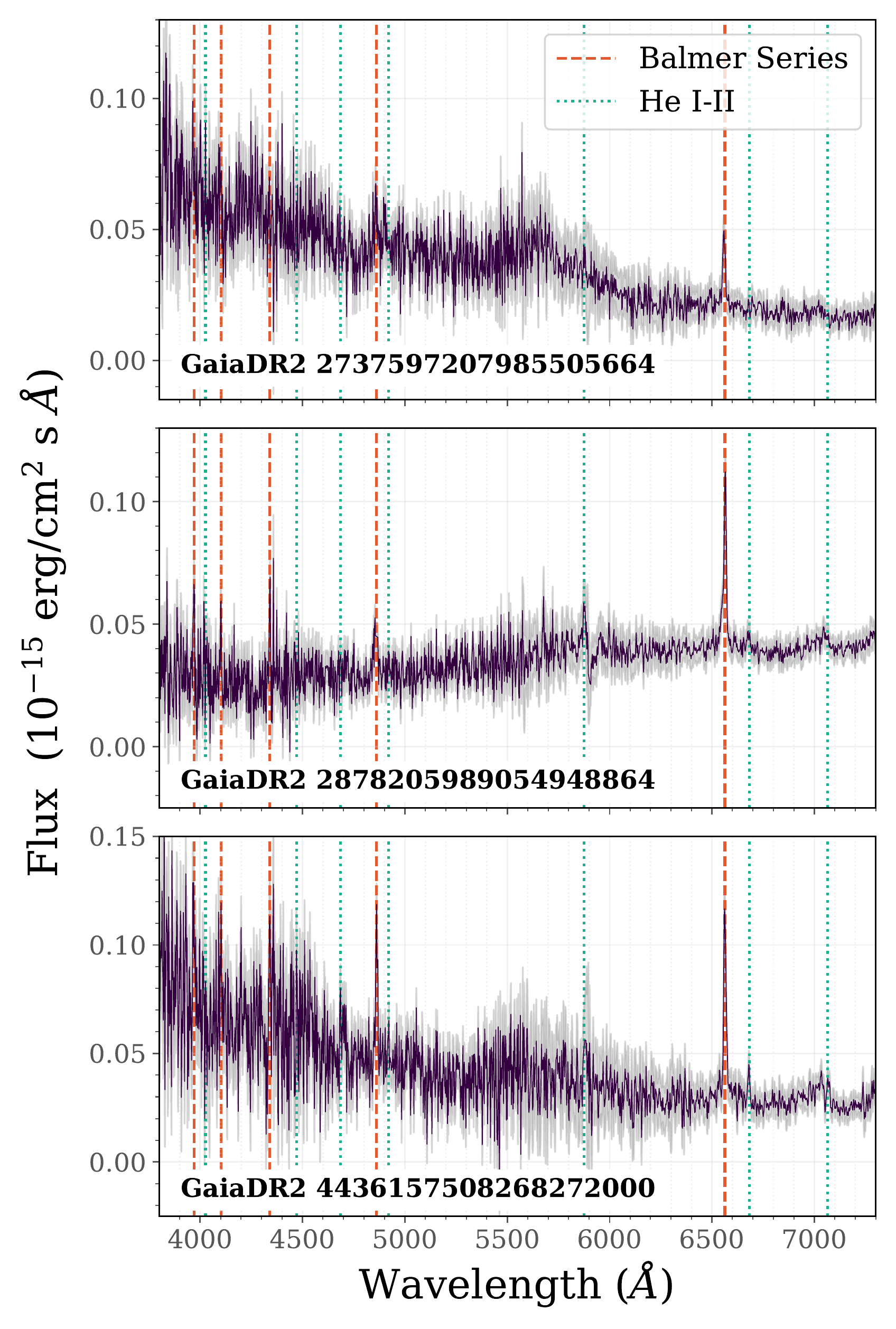}
  \caption{Lick Shane spectra of three new CV systems displaying prominent H$\alpha$ emission, characterizing them as CVs. The continuum in these systems is too noisy to measure any significant He~II emission, but their spectral shapes are typical of quiescent dwarf novae.}
  \label{fig:lick_nolines}
  \end{center}
\end{figure}

\subsection{Kast Spectrograph}
\noindent We observed nine candidate systems with the Kast double spectrograph \citep{miller_1993} mounted on the Shane 3\,m telescope at Lick Observatory on the nights of 6 June 2019, 31 July 2019, and 1 August 2019. Most spectra presented here were obtained with the long slit at or near the parallactic angle so as to reduce the differential light loss caused by atmospheric dispersion \citep{filippenko_1982}.

We briefly summarise the principal steps in our reduction strategy \citep[we generally follow the methods of][]{silverman_2012}, which are implemented using IRAF5 routines and publicly available Python and IDL programs. First, standard preparation steps including bias removal, cosmic ray rejection, and flat-field correction are performed. Following extraction, 1D spectra are wavelength-calibrated using comparison-lamp spectra typically taken in the afternoon prior to each observing run. The spectra are then flux-calibrated using spectra (taken during each observing run with the appropriate instrumental setup) of bright spectrophotometric standard stars at similar air-masses. Finally, atmospheric (telluric) absorption features are removed and overlapping (i.e., red-side and blue-side spectra from Kast) are combined by scaling one so that it matches the other over the common wavelength range. We consider spectra at this stage to be ``science ready.''

\subsection{9 New CV Systems}
\noindent All spectra exhibit evidence of accretion and display the strong Balmer emission lines characteristic of CVs. We show the spectra in Figures 12 and 13, labeled by their {\it Gaia} source identifier. While specific CV subtype assignment requires further observation --- dwarf novae need time-resolved confirmation of the system varying by more than a few magnitudes and Polars need spectropolarimetric confirmation of their polarized signals --- we sort the spectra by the presence (Fig. \ref{fig:lick_lines}) or absence (Fig. \ref{fig:lick_nolines}) of strong He~II emission lines. We discuss some of the noticeable characteristics below.

Strong He~II emission lines are evidence of either a magnetic WD or a nova-like system. The systems shown in Figure \ref{fig:lick_lines} all display strong He~II emission, making them good candidates for magnetic CV subtypes. The prominent central absorption indicated by the double feature in the Balmer emission lines of {\it Gaia} DR2  4553500210980651904 likely indicate a relatively high accretion disk inclination. We obtained two spectra of {\it Gaia} DR2  4553500210980651904 on consecutive nights and we discuss the multi-epoch spectra in Section 6.1.4 below.

While the systems in Figure \ref{fig:lick_nolines} do not feature strong He~II emission lines, they all display the prominent H$\alpha$ and H$\beta$ lines that characterize CV systems. {\it Gaia} DR2 2737597207985505664 seems to have faint signs of a cyclotron hump in its spectrum, but owing to the continuum noise, it cannot be confirmed from this observation.

The spectra in Figures 12 and 13 confirm the efficacy of our variability metrics to select even semi-stochastic highly variable systems. When combined with additional information, like CMD position, the variability metrics in Section \ref{subsec:vari_metrics} allows for the selection of specific variables, like spCVs with high confidence.

\subsection{Equivalent-Width Measurements}

\noindent In Table \ref{tab:lick_ew}, we report EWs for any prominent H$\alpha$ and H$\beta$ emission where available for each spectrum. The EWs were calculated by fitting a Voigt profile to the line emission above the continuum, which was determined by-eye for each spectrum. Uncertainties were calculated by taking the average of 3,000 Monte Carlo trials within the flux uncertainty of the line profile. We find that all CVs show sufficiently strong H$\alpha$ lines.

\begin{deluxetable}{ccc}

\tablenum{6}
\tablecaption{EW Measurements
\label{tab:lick_ew}}
\tablehead{
\colhead{{\it Gaia} \texttt{source\_id}} & \colhead{H$\alpha$} & \colhead{H$\beta$} \vspace{-5px} \\ 
\colhead{} & \colhead{\deemph{[\AA]}} & \colhead{\deemph{[\AA]}}}
\setlength{\tabcolsep}{8pt}
\startdata
1307670447592370688 & 17.91 & 14.77 \\
2150788814314276736 & 12.99 & 14.34 \\
2737597207985505664 & 14.46 & -- \\
2796463449923353856 & 30.45 & 25.51 \\
2818311909906928384 & 17.51 & 14.34 \\
2878205989054948864 & 12.06 & -- \\
300540061660343040 & 9.13 & 7.07 \\
4436157508268272000 & 14.96 & -- \\
\specialrule{.025em}{.1em}{.1em}
4553500210980651904$^{\dagger}$ & 39.99 & 37.71 \\
4553500210980651904$^{\ddagger}$ & 48.67 & 48.81 \\
\enddata
\tablecomments{$\dagger$ denotes the first night 4553500210980651904 was observed. $\ddagger$ denotes the follow-up observation one night later.}
\end{deluxetable}

\subsection{Multi-Epoch Spectroscopy of \\ {\it Gaia} DR2 4553500210980651904}

\begin{figure}
  \begin{center}
  \includegraphics[width=245px]{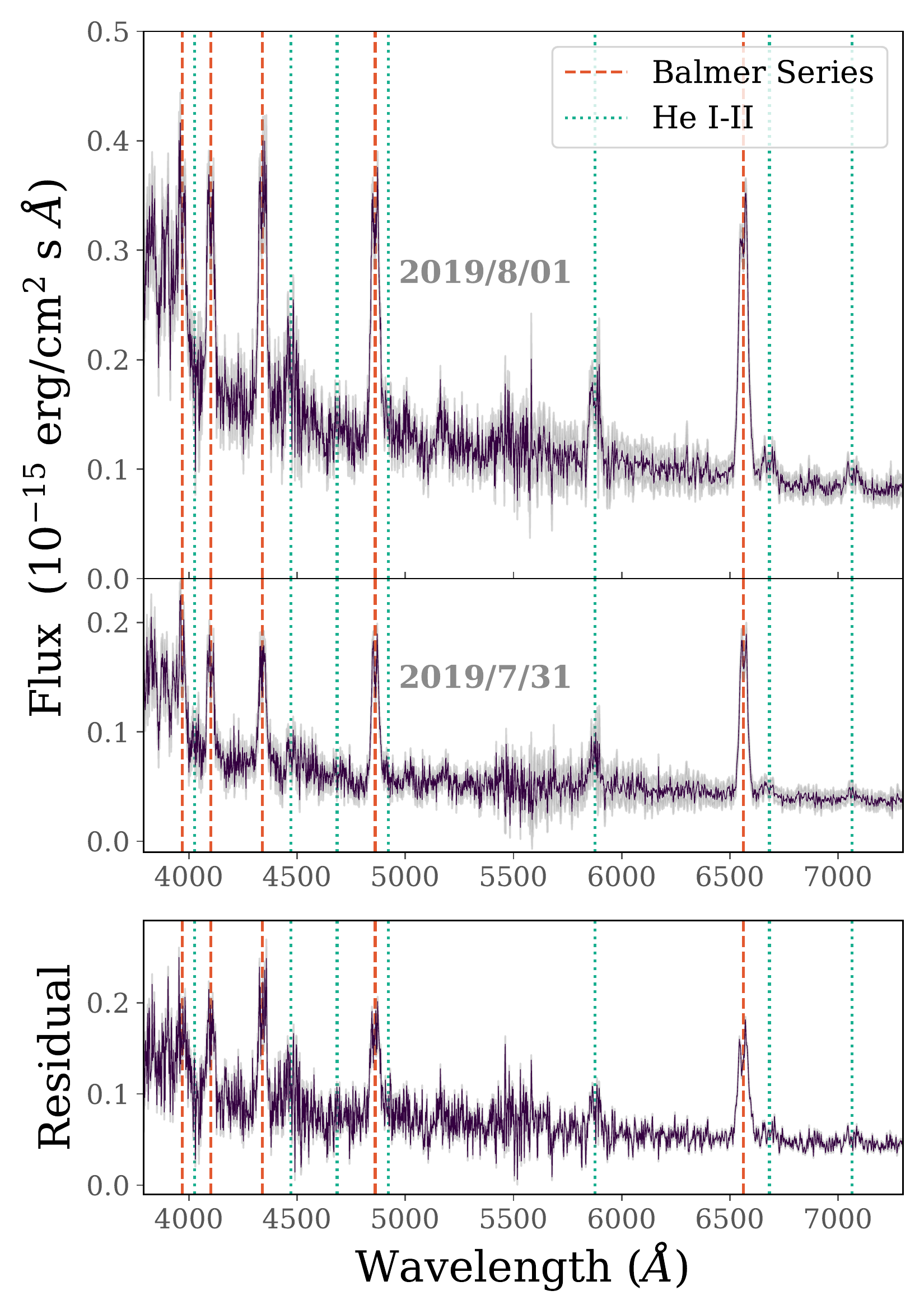}
  \caption{Multi-epoch spectroscopy of the new CV, GaiaDR2  4553500210980651904, observed on two consecutive nights, shown to scale in each panel. The bottom panel shows the residual of the subtraction of the 7/31 spectrum from the 8/01 spectrum in the top panel. The amplitude of Balmer lines in the residual is almost as high as the emission amplitudes in the lower spectrum, indicating that the flux emission from this system nearly doubled overnight. This could be due to differences in accretion between the two nights.}
  \label{fig:lick_double}
  \end{center}
\end{figure}

\noindent We observed {\it Gaia} DR2 4553500210980651904 for 20\,min of integration on two consecutive nights, 31 July 2019 and 1 August 2019. We show a comparison of the two spectra in Figure \ref{fig:lick_double}. While both spectra display strong Balmer lines, the lines from the second night of observation are nearly twice as strong, as is illustrated by the amplitude of the residual plot in the lower panel. We measure the EWs for the H$\alpha$ and H$\beta$ lines in these spectra, as discussed in the previous section. Following Table \ref{tab:lick_ew}, the lines increase in strength between the first and second night. 

\section{Time-Domain \\ Photometric Characterization} \label{sec:new_lcs}

\noindent Many open-source catalogs like {\it Gaia} DR2 and PTF provide all or nearly all-sky spatial coverage of variable sources. In this section we present an analysis of time-domain photometry where available.

\subsection{Gaia DR2 Variable Catalog} \label{sub_sec:gaia_lcs}
\noindent 21 objects from our candidate list have public light curves in the {\it Gaia} DR2 Variable Catalog \citep{holl_2018}. We share their DR2 source identifiers and a selection of their time-series statistics in Table \ref{tab:vari_xmatch}. On average, these objects vary in $M_{\rm G}$ by 1.34\,mag in {\it Gaia}, with 12 objects varying by more than 1\,mag during the {\it Gaia} observing window, and 9 objects vary less.

\begin{figure*}
  \begin{center}
  \includegraphics[width=500px]{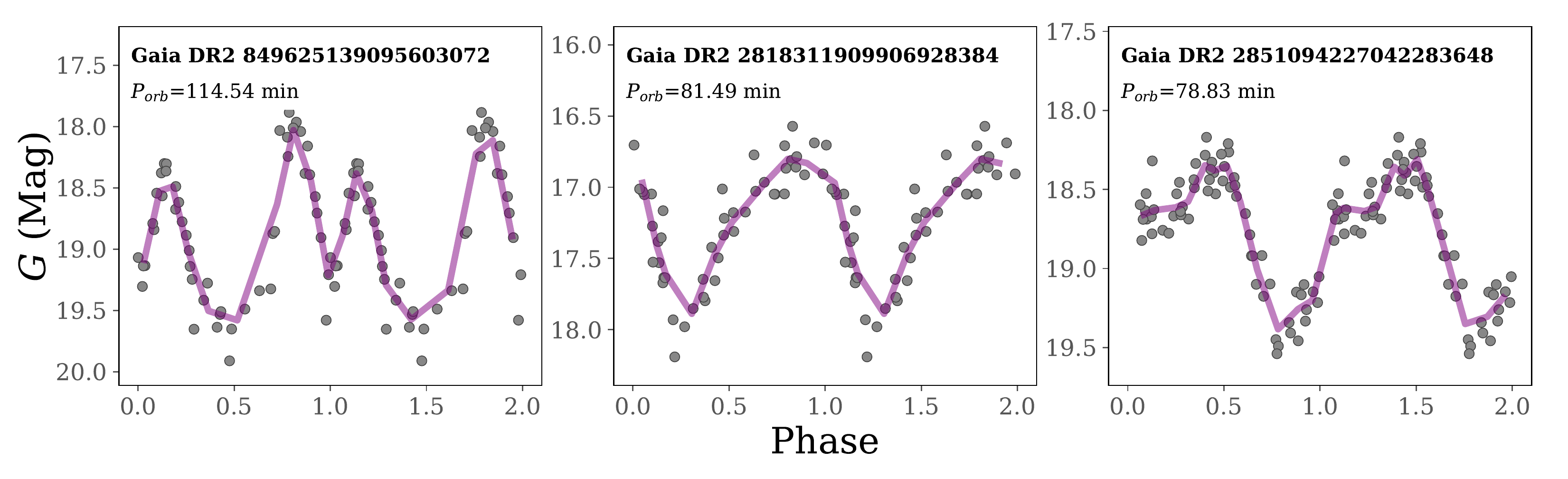}
  \caption{Phase-folded light curves from the {\it Gaia} DR2 Variability Catalog. The gray points symbolize the observed magnitudes at a particular timestamp, and the purple lines show the flux of the phase averaged across four time bins. The $P_{\rm orb}$ recovered with a Lomb-Scargle search is less than 2\,hr for every system. {\it Gaia} DR2 849625139095603072 is the well-characterized AM\,Herculis system, EK\,Uma \citep{morris_1987}. Our L-S search recovers a period that matches the literature measurement of 114.5, despite the sparsity of {\it Gaia} sampling. The other two sources shown here have not been previously characterized, though they both display orbital periods of less than 2\,hr.}
  \label{fig:gaia_lcs}
  \end{center}
\end{figure*}

18 of these objects (86\%) are in the DR2 Short-timescale Variability Catalog \citep{roelens_2018}, which consists of objects that display rapid-timescale variability ($< 1$\,day) in the DR2 light curves composed of data collected the first 22 months of {\it Gaia} observations. The Catalog released 3,018 candidates, identified either by variogram analysis, by peculiar behavior in their light curves, or from prior characterization. 17/18 objects have a characteristic timescale of variability $< 2.15$\,hr in the Short-timescale Variability Catalog, providing further evidence that our method is successful at finding spCVs. However, \citet{roelens_2018} caution that the variogram analysis is only accurate on 23\% of the sample when compared to previously measured periods in the literature, so we only rely on their measurements for determining whether objects might be above or below the period gap.

\begin{deluxetable}{cccc}
\tablenum{7}
\tablecaption{{\it Gaia} DR2 Variability Catalog Members
\label{tab:xmatch_counts}}
\tablehead{\colhead{{\it Gaia} \texttt{source\_id}} & \colhead{IQR} & \colhead{MAD} & \colhead{$G$ Range}}
\startdata
1123290903189100160 & 0.4139 & 0.2635 & 3.4436 \\
2796463449923353856 & 0.3696 & 0.267 & 2.558 \\
3955313418148878080 & 0.4035 & 0.2752 & 2.5405 \\
5041907811522399488 & 0.4899 & 0.3965 & 2.4442 \\
{\bf 849625139095603072} & 0.9211 & 0.686 & 2.0254 \\
{\bf 2818311909906928384} & 0.6696 & 0.4613 & 1.5028 \\
{\bf 2851094227042283648} & 0.6055 & 0.366 & 1.3692 \\
4461280391188698368 & 0.3152 & 0.2319 & 1.2969 \\
4638176586435463296 & 0.5133 & 0.3846 & 1.2649 \\
583340808875713280 & 0.2809 & 0.1886 & 1.0818 \\
3212803625248377088 & 0.4294 & 0.2925 & 1.0542 \\
1101345166494742400 & 0.2578 & 0.2242 & 1.01 \\
1289860214647954816 & 0.2219 & 0.1947 & 0.8445 \\
6298434093495565184 & 0.3779 & 0.1781 & 0.8389 \\
12523059483357312 & 0.3004 & 0.3253 & 0.7906 \\
5107845936158224768 & 0.2097 & 0.1623 & 0.7872 \\
6391968555534104192 & 0.2899 & 0.2254 & 0.7711 \\
1383764490550797824 & 0.2695 & 0.1727 & 0.7402 \\
3567798585117389824 & 0.274 & 0.2271 & 0.6442 \\
4877265084954805504 & 0.202 & 0.1544 & 0.5999 \\
4775330422798414208 & 0.2181 & 0.1737 & 0.4787
\enddata
\tablecomments{The light curves and recovered $P_{\ orb}$ measurements for boldfaced objects are shown in Figure \ref{fig:gaia_lcs}.}
\label{tab:vari_xmatch}
\end{deluxetable}

We run Lomb-Scargle \citep[L-S;][]{lomb_1976,scargle_1982} periodograms using the \texttt{cesium} library implementation \citep{naul_2016} with 8 harmonics over three frequencies on the full set of 21 light curves. We examined the results for each source visually and found three sources with credibly determined periods.

Owing to the {\it Gaia} mission's unique observation windowing, it is challenging to measure periods with high confidence when $< 25$ observations are sampled, so we restrict our search to the 19 objects with more than 25 observations in the $G$ band. Additionally, L-S periodograms will measure a period, whether or not a true period exists, by accepting the frequency of oscillation that has the highest power. To ensure confidence in our measured periods, we fold the light curves to their L-S periods, and only accept periods that display visibly repetitive behavior in phase as shown in Figure \ref{fig:gaia_lcs}.

{\it Gaia} DR2 849625139095603072 is the well-known short-period polar EK\,Ursae Majoris, recovered in our crossmatch with SIMBAD, and observed by SDSS. Though the {\it Gaia} sampling is sparse for this source, L-S still recovers a $P_{\rm orb}$ measurement identical to the known value \citep{morris_1987,beuermann_2009}. To the best of our knowledge, {\it Gaia} DR2 2818311909906928384 and 2851094227042283648 are previously unclassified sources. Both of these candidates show clear photometric variability with a period of less than 2\,hr. 

Four objects in the DR2 Variability Catalog were classified as RRL in \hyperlink{cite.rimoldini_2019}{Rim18}, but three of these objects have been confirmed as CV systems previously by SDSS and CRTS. It is of note that the fourth RRL, {\it Gaia} DR2 3567798585117389824, 
is also listed as an RRL in VSX, and is the RRL highlighted with a thicker stroke in Figure \ref{fig:misclassified_simbad}. {\it Gaia} recovers the same classification and period of 6.53\,hr as \citet{drake_2014}. We discuss this object further in Section 8.4. With the exception of this unusual object, the remainder of these objects fall well below the CV period gap, confirming that our search is correctly constrained to spCV systems.

\subsection{Gaia Alert Light Curves}
\noindent 51 objects have early-release light curves in the {\it Gaia} Alerts Database \citep{wyrzykowski_2012}, found in a 2.5$\dprime$ crossmatch. {\it Gaia} does not provide source identifiers for objects in the {\it Gaia} Alerts Database, but a 1$\dprime$ and 2.5$\dprime$ cone search returns an identical list of crossmatches. 27 of these were previously known CVs, and the remaining 24 are not given a classfication by the Alerts Database, though 20/24 are listed as candidate CVs in the comments. These objects all vary $> 1$\,mag in $M_{\rm G}$ and have a median range of 4.14\,mag. 13 of the sources that were not previously classified have $M_{\rm G}$ variations $> 4$\,mag. In Table \ref{tab:alerts_xmatch} we provide a subset of the time-series statistics for the full crossmatch to the Alerts Database.

Owing to the prevalence of null and untrusted\footnote{This mission labels a timestamp untrusted when the derived flux measurement is unreliable.} detections in the data, we were not able to measure L-S periods for any of these objects with high confidence. Null detections are recorded when {\it Gaia} is predicted by the mission to have observed the location in the sky at a specified timestamp but no observations were recorded, and can occur when CVs at quiescence are fainter than {\it Gaia}'s limiting magnitude. 

\begin{deluxetable*}{cccccc}
\tablenum{8}
\tablecaption{{\it Gaia} Alerts Catalog Members
\label{tab:alerts_xmatch}}
\tablehead{\colhead{{\it Gaia} \texttt{source\_id}} & \colhead{Alert Name} & \colhead{Type} & \colhead{IQR} & \colhead{MAD} & \colhead{$G$ Range}}
\setlength{\tabcolsep}{12pt}
\startdata
1163811724898541440 & Gaia18crs & CV & 0.415 & 0.4003 & 7.93 \\
2465053942183130240 & Gaia18duy & CV & 0.185 & 0.1334 & 5.91 \\
4409581010854168064 & Gaia18cwe & CV & 0.24 & 0.1927 & 5.75 \\
6402626839000977408 & Gaia18bzb & unknown & 0.4575 & 0.3188 & 5.69 \\
3800596876396315648 & Gaia18bwz & CV & 0.285 & 0.2076 & 5.66 \\
321009394756982784 & Gaia19cks & CV & 0.3725 & 0.2817 & 5.47 \\
1707635020719834880 & Gaia18dnv & unknown & 0.38 & 0.2965 & 5.29 \\
772038105376131456 & Gaia18dqq & CV & 0.42 & 0.3113 & 5.12 \\
1367700728748682880 & Gaia19aif & unknown & 0.55 & 0.4151 & 5.09 \\
604649859617712384 & Gaia19cde & CV & 0.37 & 0.2669 & 4.93 \\
3244090159897565952 & Gaia18dzq & unknown & 0.4775 & 0.3336 & 4.82 \\
3212803625248377088 & Gaia19bei & CV & 0.38 & 0.2965 & 4.76 \\
851753282506891776 & Gaia19bwr & CV & 0.545 & 0.4225 & 4.71 \\
4932973490841113216 & Gaia16adj & unknown & 0.46 & 0.341 & 4.71 \\
4797053645827235456 & Gaia18dze & unknown & 0.31 & 0.2372 & 4.71 \\
856604632750452736 & Gaia19dby & CV & 0.45 & 0.3706 & 4.68 \\
4918835764173746432 & Gaia19def & CV & 0.5525 & 0.3484 & 4.66 \\
6475303909056729984 & Gaia19fng & unknown & 0.44 & 0.3707 & 4.63 \\
2806802123399581056 & Gaia16adh & CV & 0.6 & 0.4448 & 4.57 \\
3489472709648571904 & Gaia16aeh & unknown & 0.3775 & 0.2817 & 4.42 \\
6644989197816347776 & Gaia20awo & CV & 0.405 & 0.2669 & 4.4 \\
6396791047893939200 & Gaia18dgr & unknown & 0.62 & 0.4893 & 4.36 \\
581110831791334400 & Gaia19ccf & CV & 0.78 & 0.4448 & 4.27 \\
3549823111895468672 & Gaia19ala & unknown & 0.43 & 0.341 & 4.26 \\
5665509162794290944 & Gaia20air & unknown & 0.5675 & 0.4077 & 4.19 \\
6480394849757928448 & Gaia19ebd & unknown & 0.5625 & 0.4151 & 4.14 \\
3279388782412362624 & Gaia19bel & CV & 0.52 & 0.4003 & 4.13 \\
4581360216423882624 & Gaia18bra & CV & 0.555 & 0.3929 & 4.1 \\
1879049845562942592 & Gaia19bnd & unknown & 2.5525 & 1.3121 & 4.04 \\
5057409070048839040 & Gaia16bce & CV & 0.57 & 0.4003 & 3.97 \\
1647976069553227136 & Gaia18bfc & unknown & 0.3825 & 0.2224 & 3.8 \\
6376903150289173760 & Gaia19atn & unknown & 1.655 & 1.1638 & 3.45 \\
1628286938740022144 & Gaia14aaf & CV & 1.56 & 1.0675 & 3.42 \\
1123290903189100160 & Gaia18dgt & unknown & 0.4175 & 0.3113 & 3.4 \\
2257709040148720640 & Gaia18dno & CV & 1.7 & 0.9933 & 3.24 \\
4846684780367607296 & Gaia19aii & CV & 0.32 & 0.252 & 3.12 \\
2866189598275905152 & Gaia19dam & CV & 0.53 & 0.4003 & 3.11 \\
3741979368899159552 & Gaia20auq & CV & 0.37 & 0.3113 & 3.09 \\
880821067114616832 & Gaia19bkt & CV & 0.3225 & 0.2372 & 3.07 \\
2982847743326395648 & Gaia18cgt & unknown & 0.4325 & 0.3113 & 3.02 \\
4744804750196738560 & Gaia18cfq & CV & 0.73 & 0.5337 & 2.88 \\
300540061660343040 & Gaia18aot & unknown & 0.57 & 0.43 & 2.67 \\
1101345166494742400 & Gaia18dhv & unknown & 0.33 & 0.2372 & 2.28 \\
2737597207985505664 & Gaia20aip & CV & 1.065 & 0.6968 & 2.22 \\
1694471835016084608 & Gaia17bcx & unknown & 0.27 & 0.2076 & 2.19 \\
2756091238377277440 & Gaia15abg & unknown & 1.2325 & 0.3929 & 1.9 \\
2347829338189150080 & Gaia19fcr & unknown & 0.42 & 0.3188 & 1.86 \\
4642096242309238272 & Gaia18deu & unknown & 0.14 & 0.1038 & 1.64 \\
626719772406892288 & Gaia19fdn & unknown & 0.3025 & 0.215 & 1.55 \\
1328757431346007680 & Gaia18cry & CV & 0.1225 & 0.089 & 1.54 \\
3618331314896034048 & Gaia20ahl & CV & 0.225 & 0.1927 & 1.33
\enddata
\end{deluxetable*}

\subsection{PTF Observations}

\noindent PTF, and the subsequent iPTF \citep{kulkarni_2013}, provide public access to an archival set of optical multi-epoch photometric measurements that were collected between 2009 and 2017 at the Palomar 48-inch Oschin Schmidt telescope at Palomar Observatory, thereby limiting the sample to objects visible from that location. The limiting magnitude of the PTF survey was 21 in the {\it g} band, making it an ideal survey to supplement the {\it Gaia} mission, which has the same limiting magnitude in $G$.

A 1.5$\dprime$ crossmatch to the public PTF and iPTF catalogs revealed 249 sources with more than 25 reliable photometric observations, in the $r$ band. We determine observation reliability by limiting our sample to timestamps that were flagged as ``good,'' with calibrations that were flagged ``good,'' and those that are fainter than mag 10. 

\begin{figure*}
  \begin{center}
  \includegraphics[width=500px]{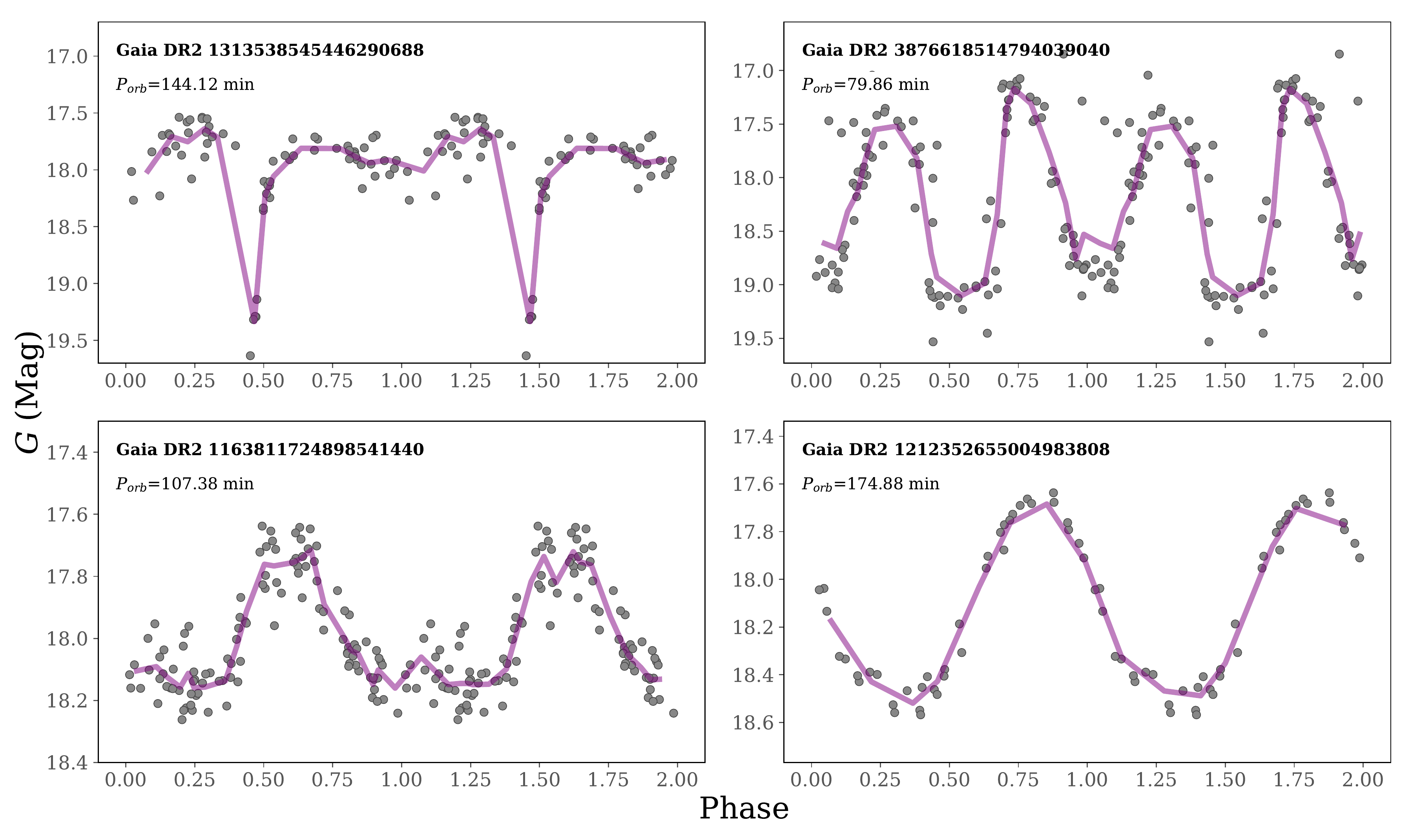}
  \caption{Phase-folded light curves from the PTF catalog. {\it Gaia} DR2 1212352655004983808 is classified as a CV by the SDSS algorithm and is cataloged by \citet{drake_2014}. From the optical light curve, we confirm that this object varies by $\sim $1\,mag and we measure $P_{\rm orb} = 174.88$\,min, agreeing with the period measurement by \citet{drake_2014}, confirming its CV status, and placing the system in the period gap. The large sinusoidal variation indicates that this is likely a magnetic CV. The other three systems are the well-characterized CVs: V1239\,Herculis, GG\,Leo, and QW\,Ser. Our L-S periodogram measurements are in agreement with the literature measurements of $P_{\rm orb}$ for these spCV systems.}
  \label{fig:ptf_lcs}
  \end{center}
\end{figure*}

We use a customized CPU-parallelized version of the conditional entropy algorithm \citep{graham_2013} along with an L-S search, following the specification in Section \ref{sub_sec:gaia_lcs}, to derive period measurements from the PTF light curves. We found that the {\it Gaia} selection function precluded the use of conditional entropy to search for periods in DR2, but that PTF's sampling allowed for corroboration between the methods. Owing to the sparsity of PTF sampling, we recover stable periods for four objects, that we share in Figure \ref{fig:ptf_lcs}. Two of these orbital periods fall above the orbital period gap, but both are within the intrinsic noise of our measured fit.

Three of these objects have been previously characterized as CV systems, and are all in the SIMBAD, AAVSO, and SDSS crossmatches. {\it Gaia} DR2 1313538545446290688 is the eclipsing dwarf nova, V1239\,Herculis. Our period measurement of 144.12\,min agrees exactly with the measurement in the literature \citep{khruzina_2015}. {\it Gaia} DR2 3876618514794039040 is the previously characterized AM\,Her, GG\,Leo. Similarly, the period measurement that we recovered for this system agrees with the literature \citep{burwitz_1998}. {\it Gaia} DR2 1163811724898541440 is the known CV, QW\,Ser, and the period measurement that we recovered agrees with the measurement made by SDSS. \citet{drake_2014} characterized {\it Gaia} DR2 1212352655004983808, also known as CRTS J151500.6+191619, as a post-common-envelope binary system from the light curve. Given the SDSS spectrum and light curve from PTF, we can confirm that this is a CV system owing to its variation of $\sim 1$\,mag, likely a magnetic system given its large sinusoidal modulation. 

\subsection{Characterizing Lick Spectra with Light-Curve Analysis}

\noindent Four of the new CV systems presented in Section \ref{sec:new_spectra} have time-resolved photometric measurements in {\it Gaia} that vary by more than 1\,mag, making all of them dwarf nova candidates.

The optical spectrum of {\it Gaia} DR2 2818311909906928384 is shown in the right panel of the middle row in Figure \ref{fig:lick_lines}. Using L-S on the light curves released in the {\it Gaia} DR2 Variability Catalog, we recover a measurement of $P_{\rm orb}=81.49$\,min, as shown in the middle panel of Figure \ref{fig:gaia_lcs}. This object regularly varies by more than 1\,mag in photometric observations. While this $P_{\rm orb}$ measurement approaches the modeled minimum $P_{\rm min}$ for CVs, it is not the lowest observed measurement of $P_{\rm orb}$ for a CV system. 

{\it Gaia} DR2 2796463449923353856, shown in the left panel of the middle of Figure \ref{fig:lick_lines}, also has time-resolved photometry released in the {\it Gaia} DR2 Variability Catalog, but due to the sparsity of observations, we were unable to recover a measurement of $P_{\rm orb}$ for this system. Across only 31 observations, this object displays a magnitude range of 2.56 in $M_{\rm G}$.

{\it Gaia} DR2 300540061660343040, or {\it Gaia18aot}, was labeled as an unknown source in {\it Gaia} alerts. The booming Balmer emission lines are clearly visible in this system, as shown in the lower-left panel of Figure \ref{fig:lick_lines}, and this system varies over 2.67\,mag in the {\it} {\it Gaia} alerts light curve. This system also has abnormally strong He~II lines, which can indicate a magnetic system. From the sparsity of the {\it Gaia} light curve, it is hard to tell if the system has the typical outburst pattern of a dwarf nova, even though the magnitude range is large over time. This system could be a dwarf nova or an intermediate polar CV.

The optical spectrum of {\it Gaia} DR2 2737597207985505664 is shown in the top panel of Figure \ref{fig:lick_nolines}. The {\it Gaia} Alerts light curve shows variations of more than 2.21\,mag, and comments that this object is a CV candidate. 

\section{Discussion} \label{sec:discussion}

\subsection{Static {\it Gaia} DR2 as a Time-Domain Survey} \label{subsec:alerts_iqr}

\begin{figure}[t!]
  \begin{center}
  \includegraphics[width=240px]{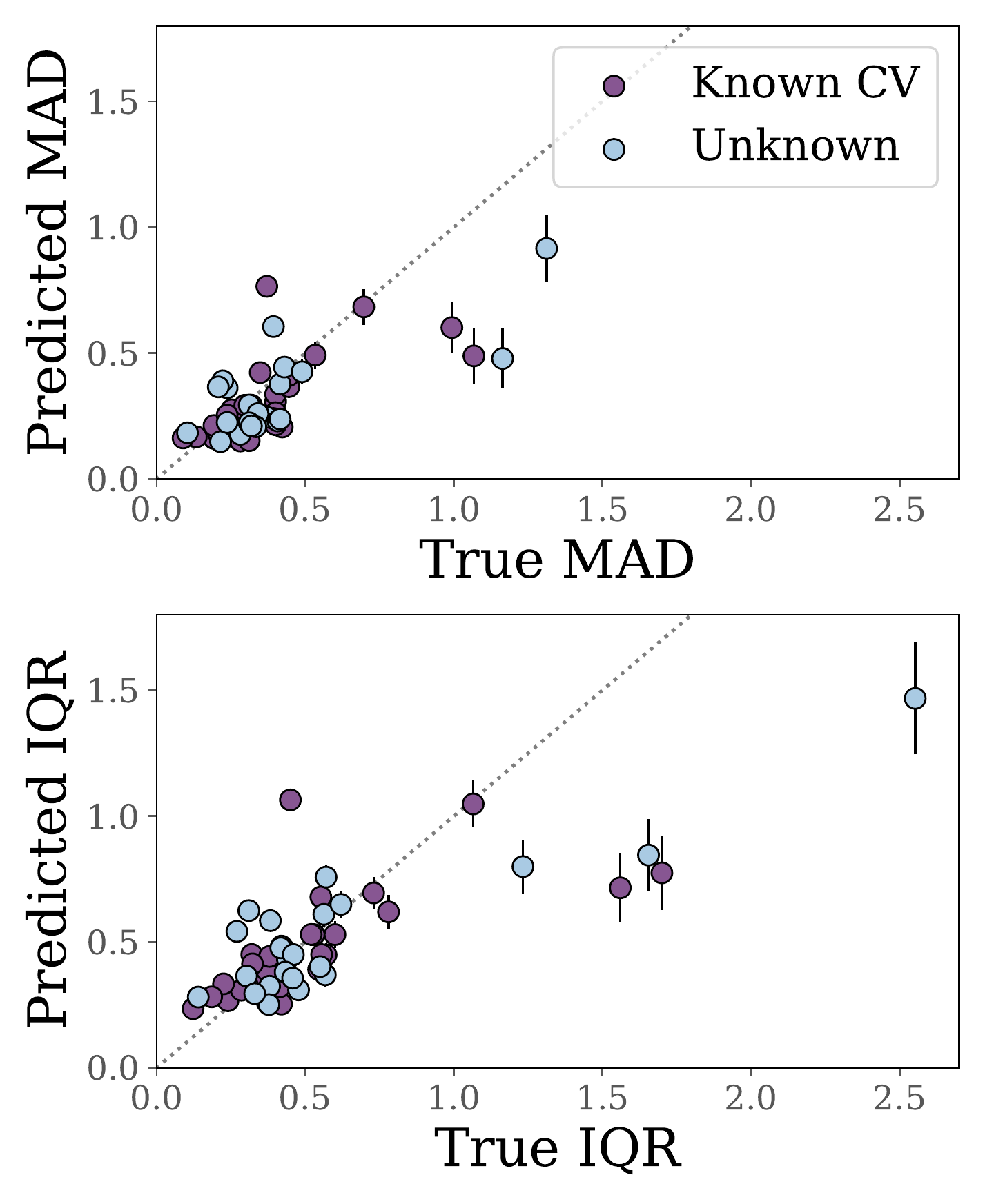}
  \caption{The {\it Gaia} alerts catalog provides an excellent withheld test set for validating our RF model (defined in Section \ref{subsec:rf_model}). RF regression models are known to perform poorly beyond the bounds of the test data. Because several of the alerts light curves have significantly higher true MAD and IQR, the RF model underpredicts their variability. In this study we are only attempting to measure whether an object is variable above a certain threshold, so we are not concerned with exact MAD and IQR predictions. At the lower threshold, the algorithm still performs sufficiently well because even less variable CVs generally have MAD and IQR well within the lower boundaries of the test set: the {\it Gaia} DR2 Variability Catalog.}
  \label{fig:alerts_metrics}
  \end{center}
\end{figure}

\noindent This paper seeks to answer whether the RMS uncertainty on {\it Gaia}'s time-averaged measurements can reliably estimate the time-resolved variability of each object. We find that our variability metrics, when measured on {\it Gaia}'s static covariates, successfully predict MAD$^\prime$ and IQR$^\prime$, recovering variable objects of arbitrary class with nearly complete accuracy in crossmatches to SIMBAD and AAVSO. While our metrics were trained and initially tested on the $\sim$500,000 objects in the {\it Gaia} DR2 Variability Catalog, which is largely composed of highly periodic, pulsational objects like RR Lyrae stars, Cepheids, and $\delta$\,Scuti variables, we find that they also perform well on semi-stochastic systems with energetic outbursts like spCVs.

Figure \ref{fig:alerts_metrics} illustrates the relationship between MAD and MAD$^\prime$ in the upper panel, and between IQR and IQR$^\prime$ in the lower panel for all 51 candidate spCVs in the {\it Gaia} Alerts crossmatch. Since these light curves were not released in the {\it Gaia} DR2 Variability Catalog, they were ``unseen" by the RF algorithm, and are useful as a withheld test set of our method. Additionally, many of these objects display outbursts, leading to the early release of their time-resolved photometry in the Alerts Catalog, which means they provide an alternate test to the sources we withheld from the DR2 Variability Catalog. The RF predictions of MAD$^\prime$ and IQR$^\prime$ are less stable toward the edges of the training dataset, and since the population of alerts light curves extends to both higher MAD and IQR, on average, than the objects included in the DR2 Variability Catalog, MAD$^\prime$ and IQR$^\prime$ consistently underpredict variability at higher true values. This is not of concern for this study because all of these objects still register as highly variable within the range of predictive values available to the RF algorithm, and therefore similarly variable objects would not be excluded in threshold cuts on the lower bound of MAD$^\prime$ and IQR$^\prime$.

In contrast, the relationship between predicted and true MAD and IQR is much closer to unity where the range of detected variability is lower, and this is because CV systems are more variable when compared to the overall training set of the full {\it Gaia} DR2 Variability Catalog, so even the smallest values of MAD$^\prime$ and IQR$^\prime$ selected for this study are not close to the lower boundaries of the training set. While it is likely that our lower threshold eliminated some quiescent CV systems, we maintained a threshold that allowed for strict recovery when crossmatched to known systems.

\subsection{Implications of the CV $P_{\rm orb}$--CMD Relation} \label{subsec:donor_seq}

\noindent A CV loses angular momentum as the donor mass ($M_{2}$) evolves to smaller masses. The mass of the WD primary member ($M_1$) remains relatively stable over time \citep[][\hyperlink{cite.knigge_2011a}{Knigge11}]{townsley_2002}, while the mass of the donor star shrinks, and the system is driven out of thermal equilibrium causing its radius ($R_2$) to slightly inflate as it evolves. Assuming that accretion temperatures stay relatively stable throughout its evolution, a CV will appear fainter in optical observations as it evolves to tighter $P_{\rm orb}$. It has been shown, too, in theory and observation, that the color of a CV in quiescence is determined by the WD member \citep{townsley_2002}. 

\begin{figure}
  \begin{center}
  \includegraphics[width=250px]{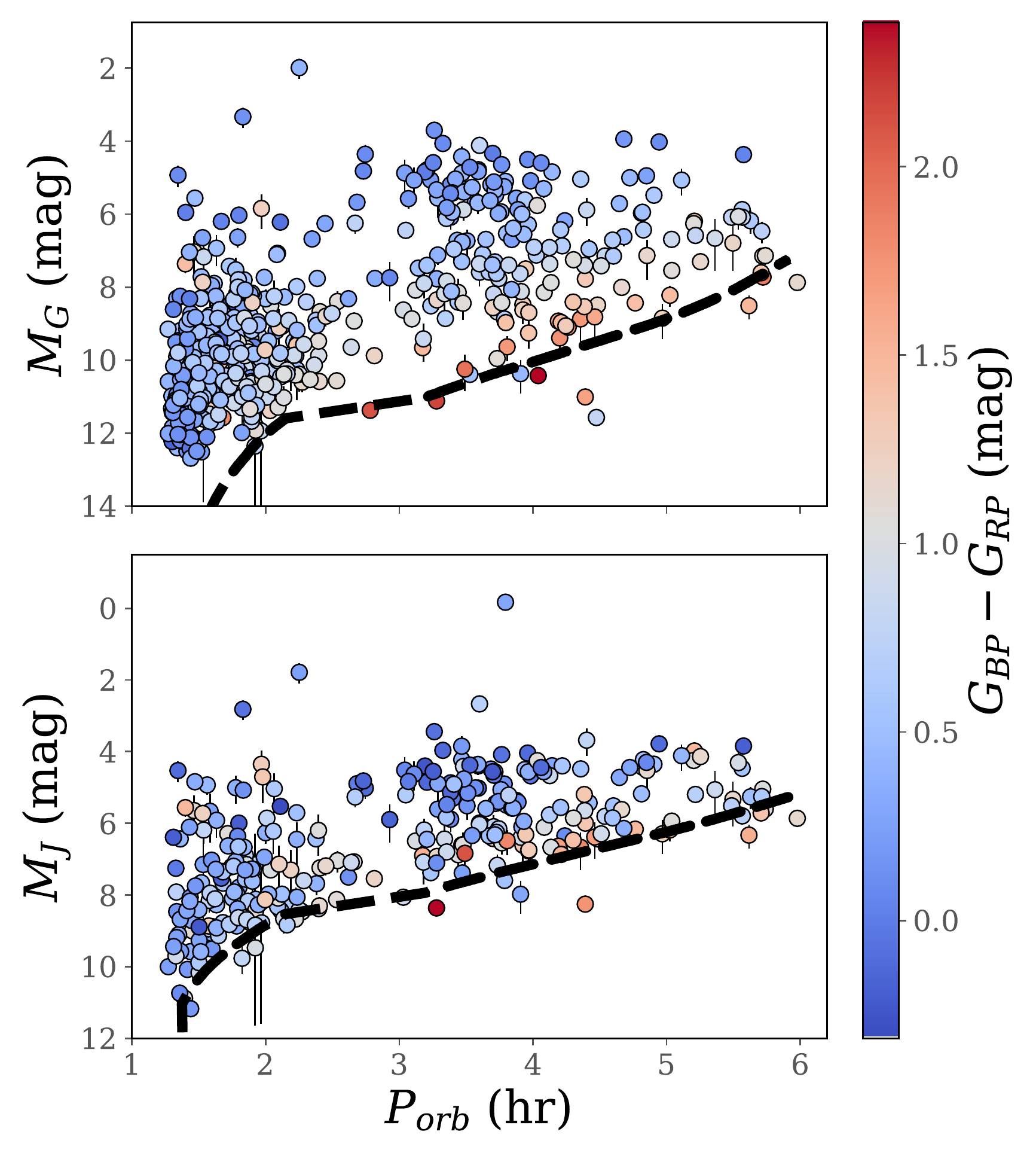}
  \caption{\hyperlink{cite.ritter_2003}{RK16} crossmatch plotted as a function of absolute optical and NIR magnitudes: $M_{\rm G}$ in the upper panel and $M_{\rm J}$ in the lower panel. This is a variant of the CMD in Figure \ref{fig:rk_cmd}, configured to show how the data add a new dimension to the semi-empirically derived donor sequence determined by \hyperlink{cite.knigge_2011a}{Knigge11} (shown in the dashed black line). The donor sequence was intended to provide a lower limit for absolute optical and NIR magnitudes as a function of $P_{\rm orb}$ as derived from the theoretical binary system parameters. These sequences generally correctly predict the lower limit, with some exceptions for redder sources. The scatter along $P_{\rm orb}$ is largely dependent on $G_{\rm BP} - G_{\rm RP}$ color, with redder systems having lower luminosity than bluer systems in the same period bin.
  }
  \label{fig:donor_scatter}
  \end{center}
\end{figure}

The general position that CVs occupy on the CMD has been predicted from theory \citep{townsley_2002}. As a pre-CV system transitions to a CV and accretion turns on, the system, with colors contributed from the WD and MS members in addition to the accretion disk, is blue enough to move left of the MS on the CMD. This locates the longest period CVs at the position that we see them in Figure \ref{fig:rk_cmd}. As the system undergoes AML, $M_{2}$ shrinks and $R_{2}$ slightly inflates, leading to a fainter system with a lower $M_{\rm G}$ \citep{warner_1995}. However, it is speculated that as $P_{\rm orb}$ tightens, the accretion disk heats up, leading to a system that is simultaneously bluer in $G_{\rm BP}-G_{\rm RP}$ as it evolves. Additionally, as the donor star dims, the color from the WD primary dominates \citep{townsley_2002}. This could be the highly generalized physical model behind the relationship that we see in Figure \ref{fig:rk_cmd}. More specifically, the second-order fit to the \hyperlink{cite.ritter_2003}{RK16} data finds a much steeper slope between $P_{\rm orb}$ and the CMD above the period gap, which would be in general agreement with the disrupted magnetic braking models \citep{mcdermott_1989, knigge_2011}. In this model, AML rates are thought to be orders of magnitude faster above the gap, which would mean that systems would transition more rapidly to fainter $M_{\rm G}$ and bluer $G_{\rm BP}-G_{\rm RP}$ colors above the gap.

However, this description does not take into account that the majority of systems previously detected in the period gap are Polars, which generally lack accretion disks owing to the strong magnetic fields of the WD member, and are shown to be redder than other CVs due to the presence of cyclotron emission \citep{krisciunas_1998,szkody_2002}. This means we should see a noticeably different location on the CMD for objects inside the period gap, and we do not find this to be the case. The first panel of Figure \ref{fig:rk_cmd} highlights the CMD location of \hyperlink{cite.ritter_2003}{RK16} systems within the period gap. We can see that they fall neatly into the $P_{\rm orb}$ sequence. To illustrate this further, we rerun the fit to the $P_{\rm orb}$--CMD relation removing any \hyperlink{cite.ritter_2003}{RK16} sources inside the period gap, and we find no noticeable change. There are not enough sources within the gap with a reliable {\it Gaia} crossmatch to run a fit within the gap alone, and this might point to the reason that we see no obvious difference in the fit when sources in the gap are included or excluded. More detections of sources inside the gap with robustly measured $P_{\rm orb}$ are necessary to confirm whether they occupy a noticeably different space on the CMD that would be expected from the $P_{\rm orb}$--CMD sequence.

Since CV evolutionary models rely on different AML rates above and below the period gap, which has been predicted to define a set of different evolutionary tracks with unique CMD positions \citep{townsley_2002}, we also try to fit separate models for the $P_{\rm orb}$--CMD sequence to the data above and below the gap separately. We find that there is too much local scatter to obtain a reliable fit in either regime, and that a fit across the entire \hyperlink{cite.ritter_2003}{RK16} crossmatch, constrained to $P_{\rm orb} < 8$\,hr, is more predictive when applied to a withheld test set. Using this fit, our candidate sample has a total of 176 objects with $\hat{P}_{\rm orb} > 2.15$\,hr within 95\% confidence. Further work is required to confirm the true underlying $P_{\rm orb}$ of these objects, but if these are within the gap, this provides a novel method for locating quiescent CV systems.

We re-examine the semi-empirical donor sequence from \hyperlink{cite.knigge_2011a}{Knigge11}, which provides a lower limit for the absolute optical and near-infrared (NIR) magnitudes that a CV system would emit at a given $P_{\rm orb}$, which are two of the observational covariates in the $P_{\rm orb}$--CMD sequence. Using the \hyperlink{cite.ritter_2003}{RK16} crossmatch, we verify that the \hyperlink{cite.knigge_2011a}{Knigge11} sequence accurately predicts the lower limits for $M_V$ (close in central wavelength to the $G$ passband) and $M_J$ with a much larger sample of CVs with measured parallax distances, as is shown in Figure \ref{fig:donor_scatter}. We find that there is a clear dependence on CV color in creating the upward scatter above the \hyperlink{cite.knigge_2011a}{Knigge11} semi-empirical sequence, shown with the black dashed line. For a given $P_{\rm orb}$, fainter objects are cooler and brighter objects are hotter. This relationship is more obvious in the optical data, and it is harder to determine below the period gap in the smaller subset with measured Two-Micron All-Sky Survey \citep[2MASS;][]{kleinmann_1994} NIR magnitudes. The semi-empirical donor sequence suggests that system luminosity essentially depends on three system parameters: the masses of both the primary and donor stars, and the mass-transfer rate, which should dictate the temperature and size of the accretion disk.

\hyperlink{cite.knigge_2011a}{Knigge11} estimates that 23\% of $M_J$ in CVs is due to the donor star. Scatter in the sequence could indicate different accretion-disk temperatures for a given $P_{\rm orb}$, possibly due to different magnetic field strengths in WD primaries, or could indicate that there is some scatter in the luminosity from the donor star itself. This could mean that the relationship between $M_{2}$ and $P_{\rm orb}$ has some scatter as a CV evolves to tighter orbits. 

\subsection{Sensitivity to Quiescent CVs} \label{subsec:quiescent_cvs}

\begin{figure*}[!ht]
  \begin{center}
  \includegraphics[width=500px]{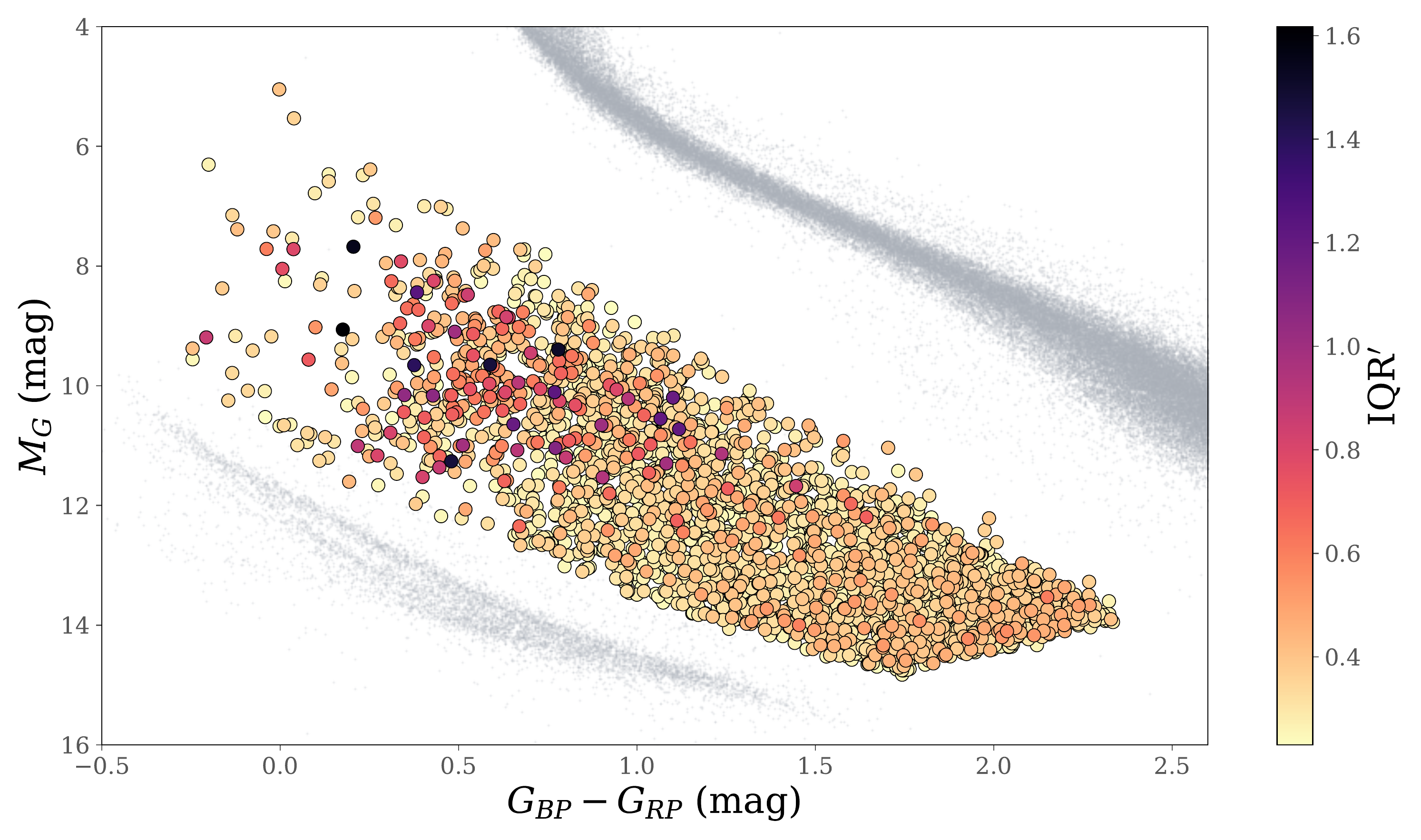}
  \caption{Estimated IQR$^\prime$ for the $> 3,200$ members of the spCV candidate sample. Were we to assume that the light curves of our sample sources were well approximated by a sine wave, IQR$^\prime$ would provide a measure of $\sim70$\% of the magnitude variability. However, since we know that spCVs are highly active and emit energetic flares, which we confirm to be the case on the full set of {\it Gaia} Alerts crossmatched sources, IQR does a better job of describing the scale of the nonflaring CV variability --- that is, the variability of the sample in quiescent states.}
  \label{fig:cand_iqr}
  \end{center}
\end{figure*}

Traditionally, when CVs are discovered and identified from multi-epoch photometry alone, the discovery can be biased toward CVs in outburst and highly active states \citep{thorstensen_2012}. As discussed above in Section \ref{subsec:rf_model} and illustrated using a withheld sample of {\it Gaia} Alerts sources in Section \ref{subsec:alerts_iqr}, we cannot estimate the full amplitude range of variability on the objects in our candidate sample due to the semi-stochastic nature of CV outbursts. However, as {\it Gaia} DR2 reports time-averaged measurements of photometry, this search is conducted on the average behavior of CVs, allowing us to locate CVs that might spend the majority of the {\it Gaia} observing window in quiescence. Figure \ref{fig:cand_iqr} shows the estimated IQR$^\prime$ from our RF model for the full candidate sample. In light curves that are well approximated by sine waves, like certain types of RRL or $\delta$\,Scuti stars, IQR describes about 70\% of the magnitude variability. However, only some components of CV emission can be approximated by traditional wave functions, like sines or sawtooths, since CV light curves can also display energetic flares and periods of superoutburst.

\begin{figure}[]
  \begin{center}
  \includegraphics[width=240px]{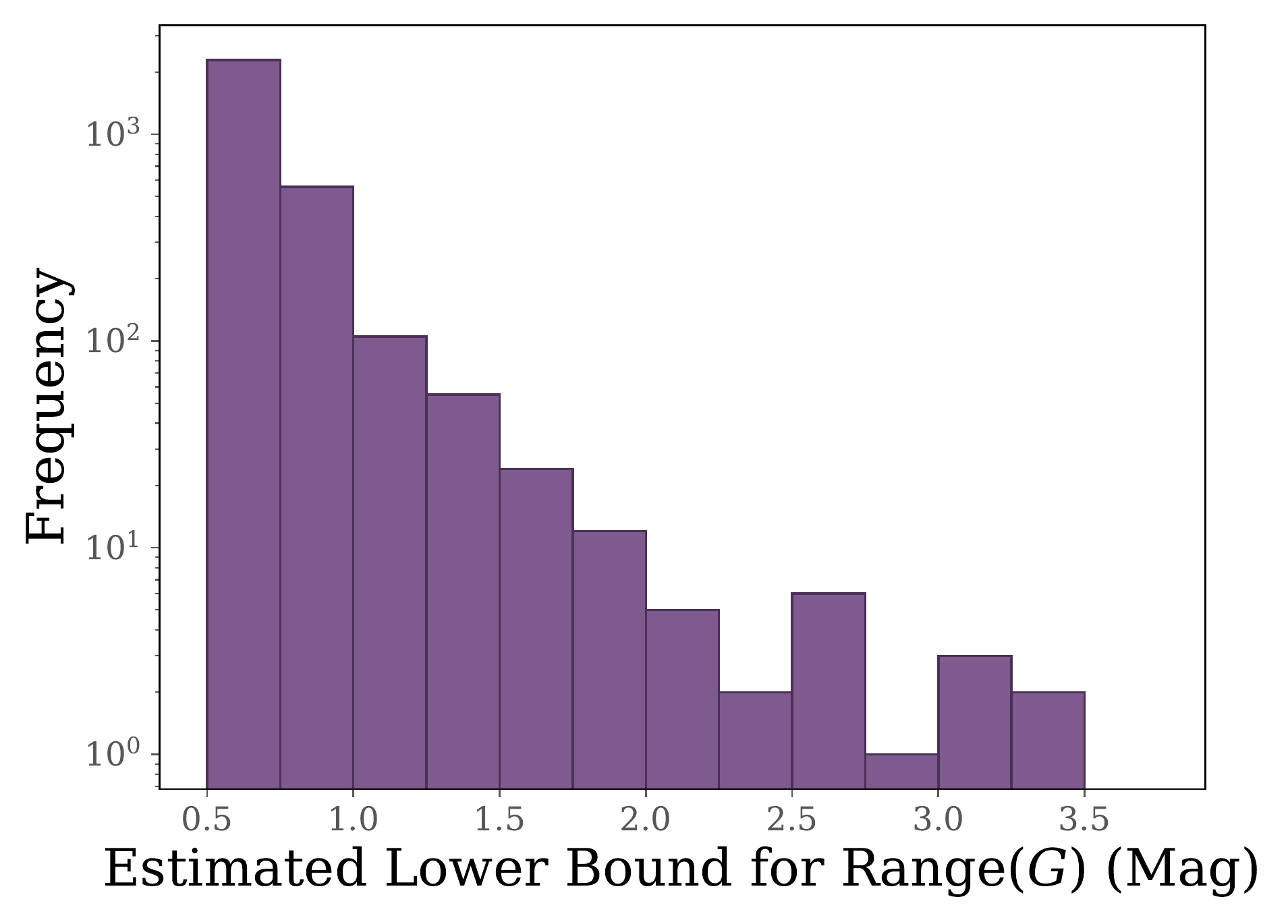}
  \caption{Estimated lower bound for $G$ magnitude ranges of the candidate sample. Since the relationship calculated for IQR$^\prime$ here relies on a single outlier in the {\it Gaia} Alerts crossmatch, this figure is illustrating the lowest possible magnitude ranges expected from the candidate sample.}
  \label{fig:lower_range}
  \end{center}
\end{figure}

Our sample has an average IQR$^\prime$ of 0.32\,mag, which would be equivalent to a variability range of $\sim 0.5$\,mag in a sine curve. Comparing this to the 27 previously characterized CVs in our sample that have a robust crossmatch in {\it Gaia} Alerts, we find that true IQR describes only 11.6\% of the variability on average, and for the vast majority of systems (24/27), the true IQR captures $<$25\% of the variability. This is exacerbated when relying on estimated IQR$^\prime$, as explained in Section \ref{subsec:alerts_iqr}, which tends to underestimate the underlying IQR and MAD for semi-stochastic systems like CVs. We find that IQR$^\prime$ captures $<$25\% of the variability for 26/27 systems. In the system with the best approximation, IQR$^\prime$ describes $\sim$47\% of its magnitude range. If we naively assume this upper limit of range capture as the best way to estimate the variability of $G$ range, we find that candidates in our sample vary at minimum by $\sim 0.5$\,mag. We show the distribution of the estimated lower bound for Range($G$) in Figure \ref{fig:lower_range}. Given that using the median estimation of range capture from the {\it Gaia} Alerts rather than the minimum estimation returns an expected Range($G$) of four times the magnitude variation, the ranges shown in Figure \ref{fig:lower_range} are likely largely underestimating the variability we would expect from this set of spCVs.

\subsection{Nearby Sources as LISA Verification Candidates} \label{subsec:lisa_cands}

\begin{figure*}
  \begin{center}
  \includegraphics[width=500px]{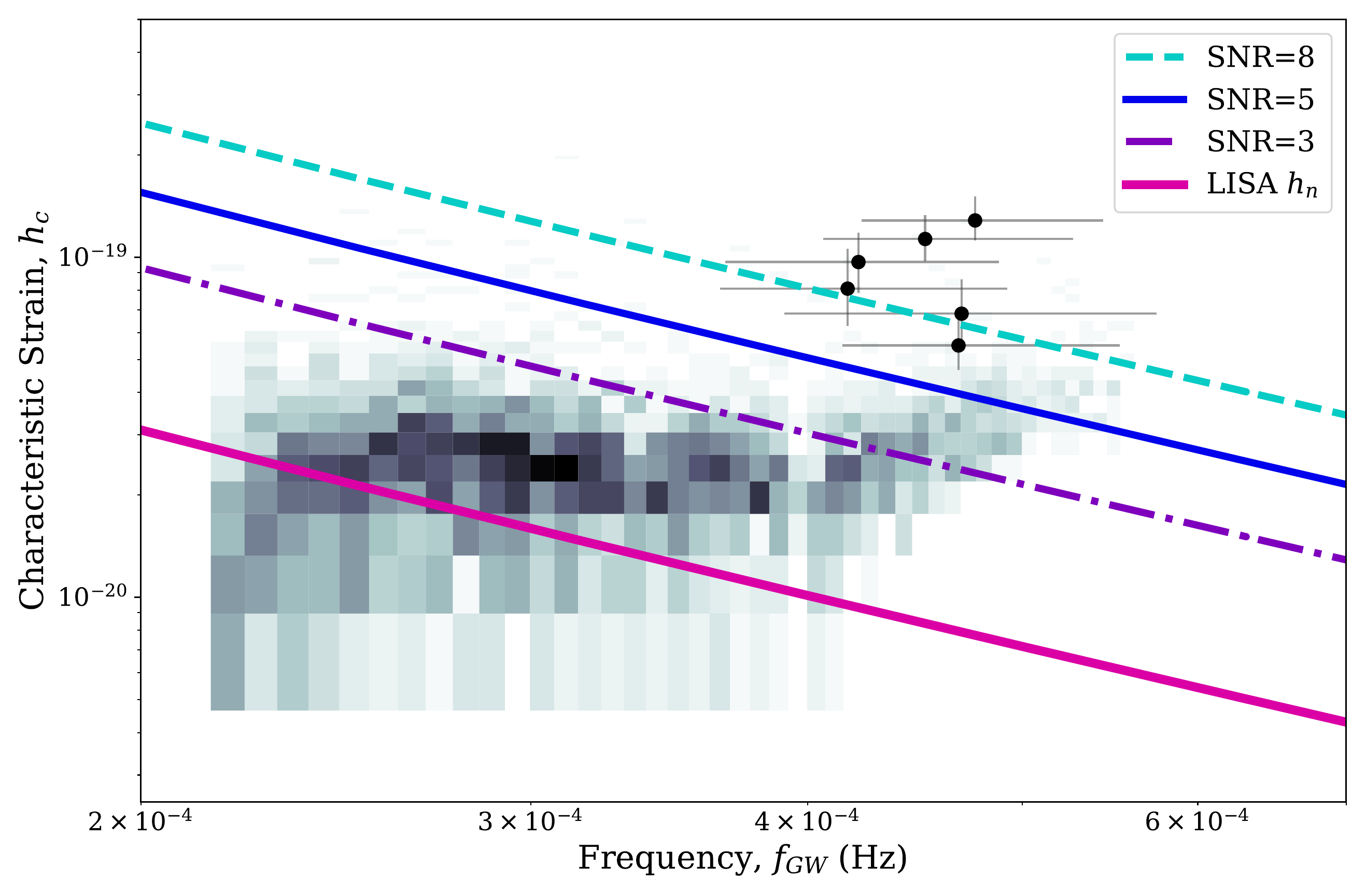}
  \caption{Estimated {\it LISA} detectability for the $> 3,200$ members of the spCV candidate sample. The 2D histogram illustrates the density of the candidate sample, shown beneath the {\it LISA} noise floor, $h_n$, and the curves indicating where the {\it LISA} SNR will be $>3$, 5, or 8 over the 4\,yr mission. Estimating $\hat{P}_{\rm orb}$ using Eq. \ref{eqtn:porb_hat} and $\hat{M}_{2b}$ using \hyperlink{cite.knigge_2011a}{Knigge11}, we predict that six objects might be coherent to the {\it LISA} mission within 4\,yr, and recommend follow-up observations of these objects as candidate {\it LISA} verification binaries.}
  \label{fig:lisa_sample}
  \end{center}
\end{figure*}

\noindent Using inferred $\hat{P}_{\rm orb}$ from the CMD location and the \hyperlink{cite.bailer-jones_2018}{BJ18} distances, we can estimate the detectable GW signature over our entire candidate CV ensemble. To establish the binary chirp mass, $\mathcal{M} = (M_1 M_2)^{3/5} (M_1 + M_2)^{-1/5}$ \citep{kupfer_2018}, we rely on the \hyperlink{cite.knigge_2011a}{Knigge11} semi-empirical $M_2-P_{\rm orb}$ relation to estimate the posterior distribution of the secondary donor mass, $M_2$. For the mass of the WD primary, we use the measurement of $\langle M_1 \rangle=0.79 M_\odot$ with an intrinsic dispersion, $\sigma_{\rm int} = 0.16\,M_\odot$ which is found to be stable above and below the period gap (\hyperlink{cite.knigge_2011a}{Knigge11}). As is recommended, we reject all systems with $M_2 \leq 0.05\, M_\odot$. Also discussed in Section \ref{subsec:donor_seq}, these masses might be underestimated for systems that are bluer than expected from the \hyperlink{cite.knigge_2011a}{Knigge11} donor sequence for their estimated $\hat{P}_{\rm orb}$.

The characteristic strain, $h_c$, over an observing time, $T_{\rm obs}$, is given by $h_c = (f T_{\rm obs})^{1/2} \mathcal{A}$, where $f=2/P_{\rm orb}$. The GW amplitude is given by
$$
\mathcal{A} = \frac{2 (G \mathcal{M})^{5/3}}{c^4 d} (\pi f)^{2/3}.
$$

For each source, we draw samples from our posterior distribution for $\hat{P}_{\rm orb}$ and from the \hyperlink{cite.bailer-jones_2018}{BJ18} distance posteriors to determine estimates of $M_1$, $M_2$, $\mathcal{M}$, and $\mathcal{A}$. To determine the detectability in the nominal {\it LISA} observing time, $T_{\rm obs} = 4$\,yr of observations, we compare the samples to the $2.5 \times 10^{6}$\,km baseline sensitivity curve ($S_n(f)$) given by \citet{cornish_2017} to estimate the noise strain, $h_n(f) = (f S_n(f))^2$ \citep{moore_2015}. The posterior median of the samples for the entire ensemble compared to $h_n(f)$ are shown in Figure \ref{fig:lisa_sample}. The expected signal-to-noise ratio (SNR) for an approximately monochromic frequency source, as is expected for these candidates, should scale as $\sim h_c / h_n$. 

We find that six isolated sources will likely be highly coherent in the {\it LISA} bandpass, with SNR $>5$ in 4\,yr of observations and we recommend these sources, listed in Table \ref{tab:lisa_cands}, for further spectroscopic and photometric follow-up observations to confirm their true $P_{\rm orb}$ and to characterize the primary and donor masses. If confirmed as {\it LISA} sources, these will be the first CV systems characterized as {\it LISA} verification binaries that are not double WD systems. 53 additional sources are estimated to have SNR $> 3$ with 95\% confidence; though these signals will be detectable to {\it LISA}, they are not estimated to have a bright enough signal to serve as calibration systems for the mission. The remainder of CV candidates will contribute to the background signal.

\begin{deluxetable}{cccc}
\tablenum{9}
\tablecaption{Candidate {\it LISA} Verification Binaries}
\label{tab:lisa_cands}
\tablehead{\colhead{{\it Gaia} \texttt{source\_id}} & \colhead{$\hat{\rm SNR}$} & \colhead{$\hat{f}_{\rm GW}$} & \colhead{$\hat{h}_c$}}
\startdata
2664442236918225920 & 13.4 & 4.5194e-4 & 1.1307e-19 \\
1742925044711324160 & 12.6 & 5.2747e-4 & 8.3159e-20 \\
1732551977135581952 & 8.6 & 4.6939e-4 & 6.8187e-20 \\
3747709851704210432 & 8.6 & 4.1696e-4 & 8.0780e-20 \\
2402744897410721280 & 7.8 & 5.0066e-4 & 5.5607e-20 \\
1775956676110237440 & 7.0 & 4.6790e-4 & 5.4986e-20
\enddata
\end{deluxetable}

\begin{figure*}[!t]
  \begin{center}
  \includegraphics[width=510px]{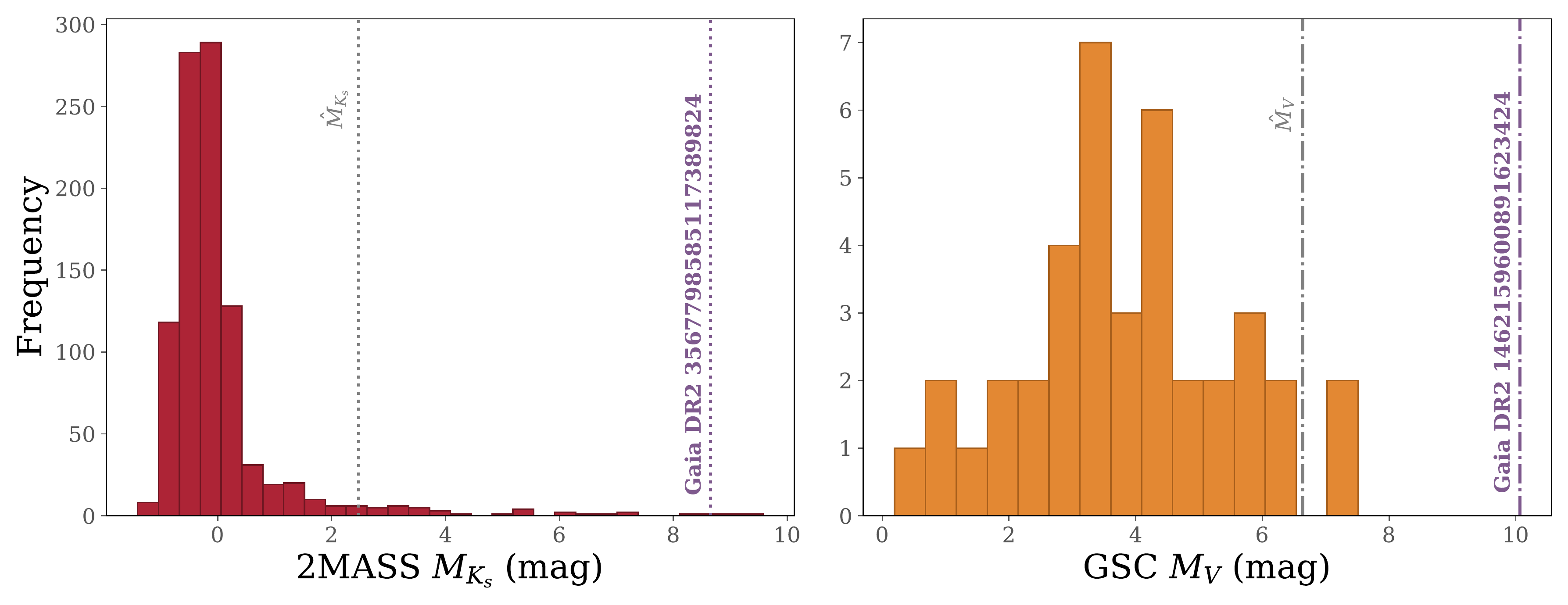}
  \caption{{\it Left panel:} The $M_{K_s}$ distribution of the {\it Gaia} DR2 RR Lyrae Catalog is shown in the red histogram. The purple, dotted line denotes the measured $M_{K_s}$ of the unusual RR Lyrae star, {\it Gaia} DR2 3567798585117389824, more than $8\sigma$ away from the distribution mean. We calculate $\hat{M}_{K_s}$ using the $P$--$L$ relationship from \citet{neeley_2019}, placed in the gray dotted line. {\it Right panel:} Similarly, we show the measured $M_V$ of the unusual $\delta$\,Scuti star, {\it Gaia} DR2 1462159600891623424, in the purple dashed line compared to a distribution of {\it Kepler} and ZTF $\delta$\,Scuti, denoted by the orange histogram and $\hat{M_V}$ calculated from the $P$--$L$ relationship of \citet{ziaali_2019}. Since the measured magnitudes of these objects are much fainter than estimated by their $P$--$L$ relationships, we find these objects likely have mismeasured parallaxes, but given their unusually fast pulse periods for their classes, we conclude they are still unusual pulsational variables.}
  \label{fig:compare_rrl}
  \end{center}
\end{figure*}

\subsection{Peculiar Pulsators} \label{subsec:peculiars}

\noindent In Section \ref{subsec:update_classes} we discuss the mischaracterized objects in the SIMBAD and AAVSO crossmatches. Two objects in particular were classified as pulsational variables, a $\delta$\,Scuti ({\it Gaia} DR2 1462159600891623424) and an RRL ({\it Gaia} DR2 3567798585117389824) from their LINEAR and CRTS light curves \citep{palaversa_2013, drake_2014}, respectively. Through follow-up observations with the Kast spectrograph at Lick Observatory on 24 May 2020, we spectroscopically confirm these characterizations. Despite the careful astrometric cuts on \texttt{parallax\_over\_error} and \texttt{astrometric\_params\_solved} employed in selecting the candidate sample, these two pulsational variables appear to have incorrect parallaxes, causing their measured $M_{\rm G}$ to be too faint and placing them in the incorrect position on the CMD shown in Figure \ref{fig:misclassified_simbad}. These mismeasured parallaxes led to a misclassification of these variables as spCVs using our CMD cuts.

To place these objects in context, we compare them to catalogs of their pulsational class. We rely on the {\it Gaia} DR2 RR Lyrae Catalog \citep{clementini_2019} and compile a $\delta$\,Scuti comparison sample by crossmatching classified $\delta$\,Scuti stars observed by the {\it Kepler} satellite \citep{murphy_2019} and ZTF \citep{chen_2020} to {\it Gaia}. Utilizing the {\it Gaia} crossmatches to 2MASS and the Guide Star Catalog \citep{lasker_2008}, we highlight the unusual faintness of both {\it Gaia} DR2\,1462159600891623424 and {\it Gaia} DR2\,3567798585117389824.

In Figure \ref{fig:compare_rrl}, we show the distributions of $M_{K_s}$ for the {\it Gaia} RR\,Lyrae Catalog, excluding objects within 20${}^{\degree}$ of the galactic plane and those in the Magellanic Clouds, or objects $\varpi/\delta\varpi \leq 5$, using distances from \hyperlink{cite.bailer-jones_2018}{BJ18}. Since the two unusual pulsators have small uncertainties on their astrometric parameters and in their photometric measurements, the \hyperlink{cite.bailer-jones_2018}{BJ18} distances measured for these two objects rely primarily on their individual likelihoods, i.e. their parallax distances, and less on the prior, as is described in Equation 2 of \hyperlink{cite.bailer-jones_2018}{BJ18}. 

We place both of these objects in context with the catalogs mentioned above, according to their classification, and denote them by the labeled, purple dashed lines. \citet{neeley_2019} derive the NIR period-luminosity ($P$--$L$) relationship for RR Lyrae stars using the DR2 RRL Catalog. Using Equation 1 from that study, fit with the parameters specified {\it ibid.} in Table \ref{tab:simbad_recovery} for $K_s$ magnitudes, we calculate the expected $\hat{M_{K_s}}$ for {\it Gaia} DR2\,3567798585117389824, using the literature pulse period of 0.27221\,day \citep{drake_2014}. Similarly, we use the optical $P$--$L$ relationship for $\delta$\,Scuti variables derived by Equation 3 of \citet{ziaali_2019} to calculate $\hat{M_V}$ for {\it Gaia} DR2\,1462159600891623424. Both estimated magnitudes are brighter than the absolute magnitudes measured with {\it Gaia} parallaxes: the RRL, {\it Gaia} DR2\,3567798585117389824, differs in $M_{K_s}$ and $\hat{M_{K_s}}$ by more than 6\,mag. However, we find that even the magnitude estimations dependent on the $P$--$L$ relations for both of these objects lie beyond $2\sigma$ of their distribution means. This is most obvious for {\it Gaia} DR2\,1462159600891623424, which also lies beyond the 98th percentile of the distribution.

After an analysis of the parallactic motions and astrometric covariances of these peculiar objects, we do not find anything unusual in the measured uncertainties when compared to the known CVs in the \hyperlink{cite.ritter_2003}{RK16} crossmatch, and we cannot pinpoint what led to the error in parallax measurement. While the objects have a higher RUWE than recommended by \citet{lindegren_2018}, 2.73 for the RRL and 3.92 for the $\delta$\,Scuti, these are not unusually high RUWE for objects in binary systems, and are within the standard deviation from the mean of our overall spCV candidate sample. As a result, we take these two stars to be indicative of the relative proportion of the sample to the rest of the AAVSO crossmatch (0.8\%) that might have mismeasured parallaxes, and we assume that this is representative of how much of our sample might fall in the incorrect place on the CMD by some orders of magnitude.

\subsection{When MS Spectra Fall Off the Main Sequence} \label{subsec:off_ms}
\noindent In Section \ref{subsec:sdss_spectra}, we show that ten objects in the crossmatch between the candidate spCVs and SDSS have spectra that appear to be MS stars. This is unexpected because these objects fall well below the MS, as is shown in Figure \ref{fig:sdss_cmd}.

\citet{zorotovic_2016} predict a peak of CV systems in the period gap that are detectable as MS--WD binaries with M-type donor stars. Since accretion ceases for the duration of the period-gap crossing, these systems should be redder than CVs and less luminous. Looking at the position on the SDSS CMD in Figure \ref{fig:sdss_cmd} for the confirmed MS--WD binaries, we find that there is overlap in this space between the three M dwarfs, the previously uncharacterized SDSS objects with MS characteristics, and the MS--WD binary sample.

\begin{figure}
  \begin{center}
  \includegraphics[width=240px]{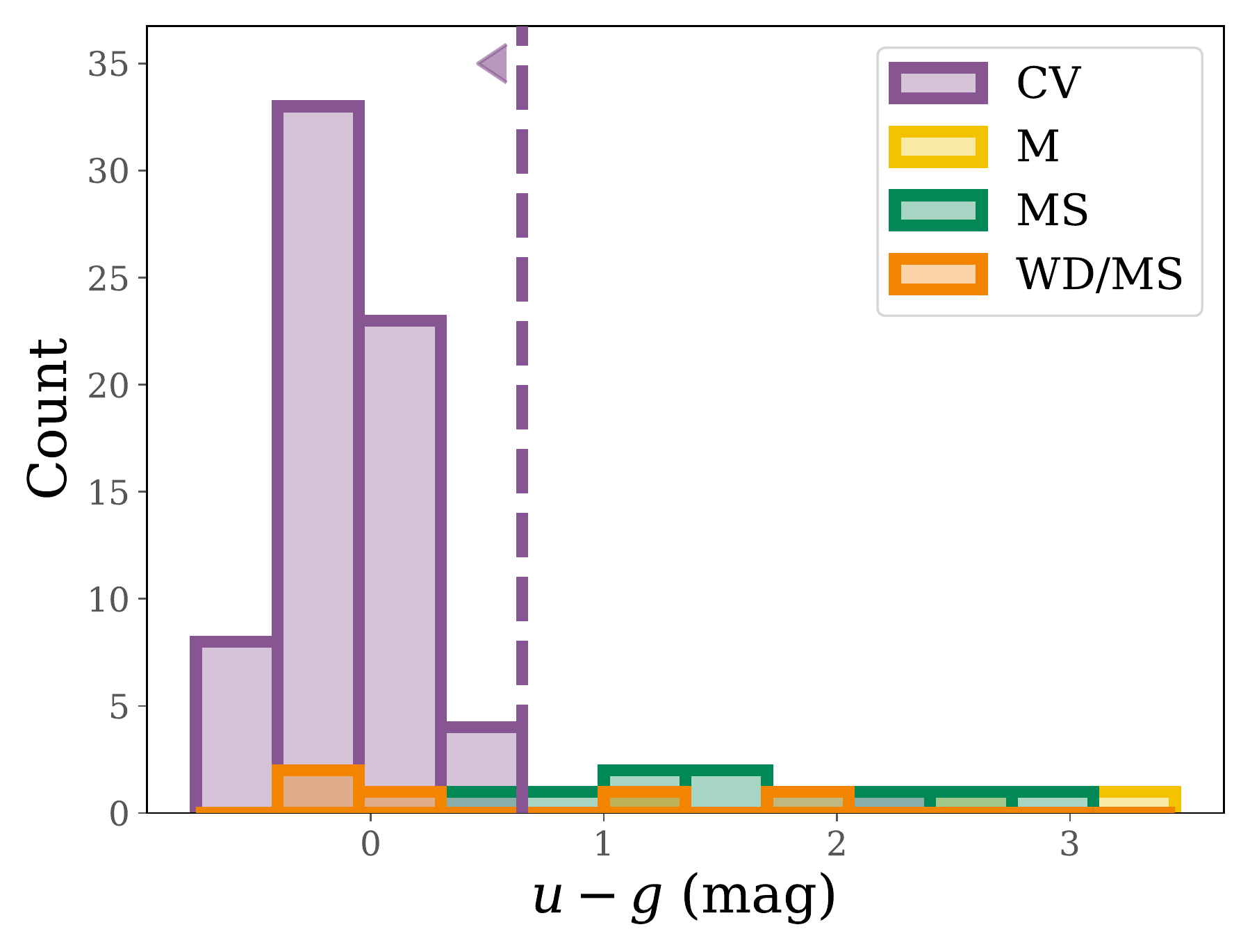}
  \caption{Distributions of objects with SDSS spectra by literature classification. We find ten objects with spectra that appear to be typical of MS stars; however, their position on the CMD, highlighted here in $g-r$ color space, is overly blue, indicating a possible WD companion. Many of these objects share color space with confirmed WD--MS binaries.}
  \label{fig:wsmd_hist}
  \end{center}
\end{figure}

This is highlighted in $u-g$ color space (the ordinate axis of the right panel in Fig. \ref{fig:sdss_cmd}), as is shown in Figure \ref{fig:wsmd_hist}, which furthers the argument that some of these objects may be temporarily detached CVs transitioning across the period gap, as the colors of CVs in quiescence have been shown in theoretical models to be dominated by the WD member \citep{townsley_2002}. Three previously characterized MS--WD binaries fall clearly within the locus of confirmed CVs, which we cut at $u-g < 0.5$\,mag. Making the limiting assumption that only these objects are quiescent CVs, we find a CV recovery fraction of $\sim 86$\% in SDSS. The blue excess is still outstanding for the remaining objects, and they might be CVs in the gap, as we would not expect $\hat{P}_{\rm orb}$ to accurately predict the periods for CVs in the gap, owing to their expected lack of accretion disk, following the discussion in Section \ref{subsec:donor_seq}. They might also be nova-like systems in low states. Follow-up observations will be critical in determining their true classification.

\section{Summary} \label{sec:summary}

\noindent We have shown that stellar variability for semi-stochastic sources in {\it Gaia} DR2 can be predicted from physically-informed transformations of just a few static covariates (color, $\delta f$, $\sigma_f$, $\epsilon_f$), allowing for a nearly unbiased all-sky search for variability, limited only by the DR2 selection function. We combine these variability metrics with CMD position to discover $3,253$ new spCV candidates, and in the process we derive a new fundamental relationship for $P_{\rm orb}$ as a function of CMD position, allowing us to estimate that six candidate CVs will be highly coherent {\it LISA} sources. This paper presents several key results and conclusions, enumerated here.

\begin{enumerate}
 \item We derived a random forest model to predict IQR$^\prime$ and MAD$^\prime$ from physically-informed variability metrics, in order to search for temporal information in the time-averaged measurements provided by {\it Gaia} DR2. We demonstrated that our estimates of IQR$^\prime$ and MAD$^\prime$ predict variability with completeness when proper thresholding is applied, despite the limitation that the variability metrics proposed in this paper assume a Gaussian spread among flux variability in the stars that they describe. In advance of future {\it Gaia} releases, we have shown that these metrics, combined with CMD position, are successful tools in the discovery of new variable Milky Way objects with class candidacy. Additionally, these metrics, relying on only a handful of covariates, allow for a much quicker scanning of large datasets ($> 10^9$ rows), and will fill a useful role in rapidly searching for temporally unusual objects even when the full {\it Gaia} time-resolved photometry is released. By searching for spCVs, we further show that these metrics work for objects that vary semi-stochastically and unevenly, even though they were developed from the {\it Gaia} DR2 Variability Catalog, which largely contains periodic sources.

 \item We have identified a relationship, revealed in {\it Gaia}, that CV period is dependent on CMD position, and derived a linear model for predicting $\hat{P}_{\rm orb}$ given $G_{\rm BP}-G_{\rm RP}$ and absolute $G$ magnitude using the portion of \hyperlink{cite.ritter_2003}{RK16} with measured $P_{\rm orb} \leq 8$\,hr. This model is steeper for objects above the period gap, which is in agreement with the prevailing evolutionary theory that CVs transition more easily to smaller $P_{\rm orb}$ above the gap and evolve to faster orbits more slowly below the gap. We found that, due to the sparsity of objects with measured $P_{\rm orb}$ inside the period gap, the fit parameters were the same, within the intrinsic scatter of the model, regardless of including or excluding objects within the gap. We recommend further follow-up observations to create a larger sample of CVs in the period gap with measured $P_{\rm orb}$ to disentangle these fits.

 \item We spectroscopically confirm nine new CV systems using the Kast spectograph on the Shane 3\,m telescope at Lick Observatory. All of these systems have very strong Balmer lines. Four of these systems have public light curves in the {\it Gaia} DR2 Variability Catalog or the {\it Gaia} Alerts Database. Using an L-S periodogram search, we recover a measurement of $P_{\rm orb} = 81.49$\,min for {\it Gaia} DR2 2818311909906928384, confirming that this object is truly an spCV. While the other {\it Gaia} light curves were too sparse to recover robust measurements of $P_{\rm orb}$, all display a $G$ range of more than a magnitude, possibly indicating that these objects are dwarf novae or magnetic systems.

 \item We show that the $P_{\rm orb}$--CMD relationship in CVs complements previous studies of the semi-empirical donor sequence (\hyperlink{cite.knigge_2011a}{Knigge11}, \citet{knigge_2006}) and provides observational constraints to theoretical predictions of CMD position from the standard evolutionary model \citep{townsley_2002}. We illustrate that the scatter in each $P_{\rm orb}$ bin above the lower limits defined by \hyperlink{cite.knigge_2011a}{Knigge11} follows an optical color sequence, possibly indicating different temperatures in the accretion disks, different magnetic fields in the WD primaries, or different spectral types of the donor stars for systems with similar $P_{\rm orb}$ but disparate optical colors.

 \item Finally, using $\hat{P}_{\rm orb}$ inferred from CMD position, we infer the characteristic strain and GW frequency due to gravitational radiation within these candidate CVs to predict how much SNR these objects will have if observed by the upcoming {\it LISA} mission. We find that six objects will have SNR $> 5$ over the 4\,yr mission.
\end{enumerate}

The {\it Gaia} Early Data Release 3 (EDR3), expected at the end of 2020, will provide $G_{\rm BP}$ and $G_{\rm RP}$ measurements for $>10^8$ sources that lacked these measurements in DR2\footnote{This number was calcuated from the overview table provided by the \href{https://www.cosmos.esa.int/web/gaia/earlydr3}{{\it Gaia} EDR3 Contents Webpage}.}. Additionally, the photometric windowing in DR2 that imposed the necessary limits of $G>13$\,mag used for Equation \ref{eqtn:epsilon_f} is expected to be much reduced, allowing for the identification of closer, and therefore apparently brighter, spCV sources using these methods. However, full characterization of spCVs requires both spectroscopic and temporal photometric follow-up observations. A subset of the brightest southern candidates from this study will be observed by the {\it TESS} mission \citep{ricker_2014} during Cycle 3 at 2\,s cadence, allowing for the measurement of $P_{\rm orb}$ and the study of short-timescale evolution in these systems. Time-resolved spectroscopy will also play a critical role, allowing for a more full characterization and understanding of the role accretion plays in the $P_{\rm orb}$--CMD relationship. 

The RF models and the code for deriving the linear fit to $\hat{P}_{\rm orb}$ (Eqtn. \ref{eqtn:porb_hat}) will be made publicly available on GitHub. Associated datasets, like the full set of spectra from Lick Observatory, will be made available on Zenodo.

\vspace{30 px}

\noindent The majority of this work was completed at U.C. Berkeley, which sits on the territory of xučyun, the ancestral and unceded land of the Chochenyo speaking Ohlone people, the successors of the historic and sovereign Verona Band of Alameda County. E.S.A. along with all of the authors affiliated with U.C. Berkeley acknowledge that we have benefited and continue to benefit from the use and occupation of this land. In this acknowledgement, E.S.A. recognizes the importance of taking actions to support the rematriation of indigenous land, and pledges to take and continue action in support of American Indian and Indigenous peoples.

E.S.A thanks Leigh Smith for initial help in formulating the variability metrics that have been finalized in Section \ref{subsec:vari_metrics}. E.S.A. thanks Anthony Brown for helpful insight into the quality flags in {\it Gaia} DR2, for suggesting the use of the {\it Gaia} Alerts Database, and for helpful insights into the magnitude dependence of {\it Gaia} photometric uncertainty. E.S.A. is grateful to Lorenzo Rimoldini for useful discussions on the {\it Gaia} Variability Database and variability classification. E.S.A. thanks Alcione Mora for assistance in writing ADQL crossmatch queries in the {\it Gaia} Archive, Brigitta Sip{\H{o}}cz for help in writing custom SIMBAD queries in \texttt{astroquery}, and Wren Suess for guidance in downloading Sloan Digital Sky Survey (SDSS) spectra. E.S.A. is grateful to Dan Weisz for his thoughts and insightful comments on a draft of this manuscript, and Courtney Dressing for her expertise on using the Shane Telescope at Lick. E.S.A. thanks Lars Bildsten for sharing a helpful paper on CV quiescence, and David W. Hogg for useful conversations on {\it Gaia} DR2 completeness. E.S.A. is grateful to Sara Jamal, Sal Wanying Fu, and Munazza Alam for helpful conversations on spectral analysis and manuscript preparation. The authors are grateful to the support astronomers and telescope operators at Lick Observatory for their assistance with the Shane telescope and the Kast spectrograph: Matthew Brooks, Dan Espinosa, Elinor Gates, Paul Lynam, Adam Nichols, Donnie Redel, Jeff Roark, Shawn Stone, and Anthony Watson. E.S.A. would like to dedicate this paper to the essential workers and around the world that have kept us safe and healthy by providing access to healthcare, food, mail, and connectivity during the worldwide COVID-19 pandemic.

E.S.A. was supported by a National Science Foundation (NSF) Graduate Research Fellowship, under grant DGE 1752814, and by a Two Sigma Ph.D. Fellowship. J.S.B. was partially supported by a Gordon and Betty Moore Foundation Data-Driven Discovery grant. P.S. acknowledges support from NSF grant AST-1514737. A.V.F. is grateful for financial assistance from the TABASGO Foundation, the Christopher R. Redlich Fund, and the Miller Institute for Basic Research in Science (U.C. Berkeley).

This work was started at the {\it Gaia} DR2 Exploration Lab, hosted by the European Space Astronomy Centre in Madrid, Spain, and benefited from E.S.A.'s participation in the 2019 Santa Barbara {\it Gaia} Sprint, hosted by the Kavli Institute for Theoretical Physics at the University of California, Santa Barbara. This research was supported in part at KITP by the Heising-Simons Foundation and the NSF under grant PHY-1748958. This work has made use of data from the European Space Agency mission {\it Gaia} (\url{https://www.cosmos.esa.int/gaia}), processed by the {\it Gaia} Data Processing and Analysis Consortium (DPAC,
\url{https://www.cosmos.esa.int/web/gaia/dpac/consortium}). Funding for the DPAC has been provided by national institutions, in particular the institutions
participating in the {\it Gaia} Multilateral Agreement. 

A major upgrade of the Kast spectrograph on the Shane 3\,m telescope at Lick Observatory was made possible through generous gifts from William and Marina Kast as well as the Heising-Simons Foundation. Research at Lick Observatory is partially supported by a generous gift from Google.
This research has made use of the SIMBAD database, operated at CDS, Strasbourg, France.
We acknowledge with thanks the variable star observations from the AAVSO International Database contributed by observers worldwide and used in this research.

The Palomar Transient Factory project is a scientific collaboration among the California Institute of Technology, Los Alamos National Laboratory, the University of Wisconsin, Milwaukee, the Oskar Klein Center, the Weizmann Institute of Science, the TANGO Program of the University System of Taiwan, and the Kavli Institute for the Physics and Mathematics of the Universe. LANL participation in iPTF is supported by the US Department of Energy as a part of the Laboratory Directed Research and Development program.

Funding for SDSS-IV has been provided by the Alfred P. Sloan Foundation, the U.S. Department of Energy Office of Science, and the Participating Institutions. 
SDSS-IV acknowledges support and resources from the Center for High Performance Computing at the University of Utah. The SDSS website is www.sdss.org.
SDSS-IV is managed by the Astrophysical Research Consortium for the Participating Institutions of the SDSS Collaboration including the Brazilian Participation Group, the Carnegie Institution for Science, Carnegie Mellon University, Center for Astrophysics (Harvard \& Smithsonian), the Chilean Participation Group, the French Participation Group, Instituto de Astrof\'isica de Canarias, The Johns Hopkins University, Kavli Institute for the Physics and Mathematics of the Universe (IPMU) / University of Tokyo, the Korean Participation Group, Lawrence Berkeley National Laboratory, Leibniz Institut f\"ur Astrophysik Potsdam (AIP),  Max-Planck-Institut f\"ur Astronomie (MPIA Heidelberg), Max-Planck-Institut f\"ur Astrophysik (MPA Garching), Max-Planck-Institut f\"ur Extraterrestrische Physik (MPE), National Astronomical Observatories of China, New Mexico State University, New York University, University of Notre Dame, Observat\'ario Nacional / MCTI, The Ohio State University, Pennsylvania State University, Shanghai Astronomical Observatory, United Kingdom Participation Group, Universidad Nacional Aut\'onoma de M\'exico, University of Arizona, University of Colorado Boulder, University of Oxford, University of Portsmouth, University of Utah, University of Virginia, University of Washington, University of Wisconsin, Vanderbilt University, and Yale University.

\software{\texttt{python} \citep{rossum_2009}, 
          \texttt{jupyter} \citep{kluyver_2016},
          \texttt{pandas} \citep{mckinney_2010},
          \texttt{numpy} \citep{van_2011},  
          \texttt{matplotlib} \citep{hunter_2007},
          \texttt{scipy} \citep{virtanen_2020},
          \texttt{scikit-learn} \citep{pedregosa_2011}, 
          \texttt{astropy} \citep{astropy_2018},
          \texttt{astroquery} \citep{astroquery},
          \texttt{emcee} \citep{emcee},
          and \texttt{pyia} \citep{price-whelan_2018}
          }

\appendix

\section{Feature Selection for Random Forest Regression} \label{app:rf}
\setcounter{figure}{0}
\counterwithin{figure}{section}

In this paper, we use the \texttt{scikit-learn}\footnote{\href{https://scikit-learn.org/stable/modules/generated/sklearn.ensemble.RandomForestRegressor.html}{Version 0.21.3}} implementation of the RF regression algorithm, applying it to the training dataset with, initially, 20 features and 464,436 instances from the {\it Gaia} Variable Star Catalog with no NaN measurements in DR2 photometry.  We limit the number of features used in the split criterion at each node of the tree to ${\rm log}_2(N_{\rm features})$. At training time, we minimized the out-of-bag (OOB) mean-square error (MSE), or the mean prediction error on a training sample that has been (randomly) discarded from the subsample used to create the model. By relying on the minimization of the OOB error, we eliminate the need for a withheld test-set, as we can test from within the training set on trees that have not ``seen" that part of the training set.  

\begin{figure*}
  \begin{center}
  \includegraphics[width=240px]{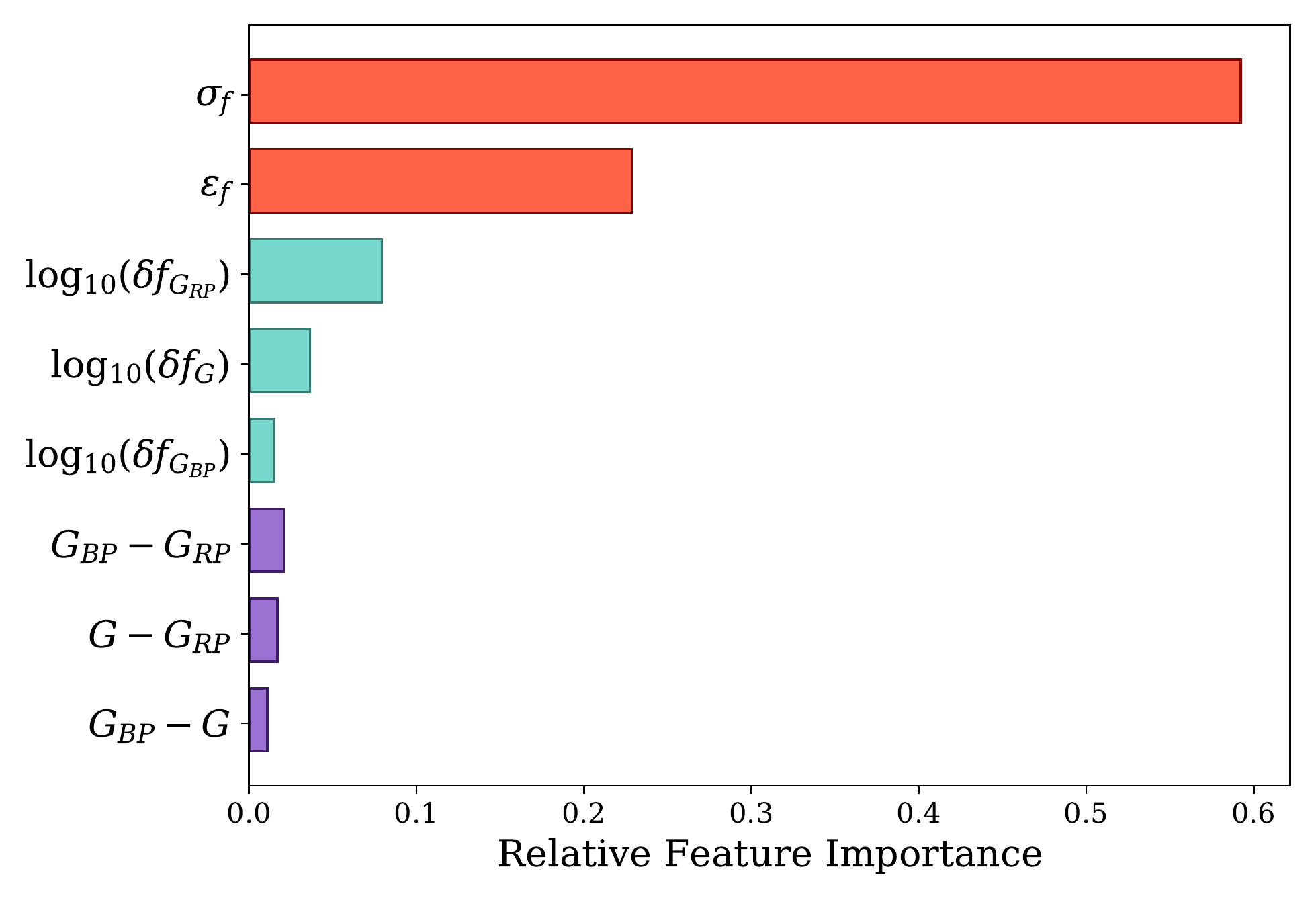}
  \caption{The relative importance of the final eight features used in the RF model to predict IQR$\dprime$ and MAD$\dprime$. The features contributing to $\sim80\%$ of the prediction are the variability metrics, $\sigma_f$ and $\epsilon_f$. Both of these metrics do not account for the distribution uncertainty in $G_{\rm BP}$ and $G_{\rm RP}$, so we include the flux uncertainties for these bandpasses. There is slight color dependency in predicting IQR$\dprime$ and MAD$\dprime$ which may be due to CMD position of variables or to color-dependent systematics in the {\it Gaia} detector.}
  \label{fig:feature_imp}
  \end{center}
\end{figure*}

We investigate the importance of 20 input features, including all of the DR2 photometric covariates and both of the variability metrics described in Section \ref{subsec:vari_metrics}. While the RF model is still reasonably predictive of IQR$\dprime$ and MAD$\dprime$ when relying on the variability metrics alone, the OOB score of the model improves by 7\% when additional features are included. We find the eight features that contribute most significantly are $\sigma_f$, $\epsilon_f$, the {\it Gaia} colors, and scaled flux errors in every passband, and we rerun the featurized RF regression on these covariates alone for our final predictive model. Figure \ref{fig:feature_imp} shows the relative importance of each of these eight features.

Cross-validation provides a lower-variance estimate of the model’s true OOB score. We use the $k$-folds method of cross-validation in this paper, in which the training sample is split into $k$ equal-sized folds. One fold is withheld as a test set, and the machine-learning algorithm (in this case RF regression) is then trained and evaluated on the remaining $k - 1$ folds. This allows for the development of $k$ unique learned models, each with an OOB score on a different withheld set. Averaging these OOB scores provides a cross-validation score. We run a $k$-folds cross-validation on our RF regression model, shuffling the input data and using $k=5$.



\end{document}